\documentclass[11pt,a4paper]{article}

\usepackage{blindtext}
\usepackage{amsmath,amssymb,amsfonts}
\usepackage{hyperref}
\usepackage{float}
\usepackage{microtype}
\usepackage{graphicx}
\usepackage[font=small]{caption}
\usepackage{bm}
\usepackage{latexsym}
\usepackage{epsfig}
\usepackage{psfrag}
\usepackage{color}
\usepackage{textcomp}
\usepackage{pstricks}
\usepackage{fontenc}
\usepackage[sort,compress]{cite}

\usepackage{fancyhdr}

\pdfoutput=1

\def\nn{\nonumber} 
\def\pa{{\partial}}
\def\f{\frac}
\def\l{\left}
\def\r{\right}
\def\d{{\rm d}}
\def\Mpl{M_{_{\mathrm{Pl}}}}
\def\Mp{M_{_{\mathrm{Pl}}}}

\def\cB{{\mathcal B}}
\def\cG{{\mathcal G}}
\def\cI{{\mathcal I}}
\def\cP{{\mathcal P}}
\def\cR{{\mathcal R}}

\def\cT{{\mathcal T}}
\def\cN{{\mathcal N}}

\def\ei{\eta_{\mathrm{i}}}

\def\ee{\eta_{\mathrm{e}}}
\def\Ni{N_{\mathrm{i}}}

\def\fpbh{f_{_{\mathrm{PBH}}}}
\def\ogw{\Omega_{_{\mathrm{GW}}}}
\def\fnl{f_{_{\rm NL}}}

\def\vka{{\bm k}_{1}}
\def\vkb{{\bm k}_{2}}
\def\vkc{{\bm k}_{3}}

\def\vx{{\bm{x}}}
\def\vk{{\bm{k}}}
\def\vp{{\bm{p}}}

\def\ps{\mathcal{P}_{_{\mathrm{S}}}}

\def\pt{\mathcal{P}_{_{\mathrm{T}}}}
\def\ph{\mathcal{P}_h}

\def\ns{n_{_{\mathrm{S}}}}
\def\mpcinv{{\rm Mpc}^{-1}}

\title{\bf
Observational imprints of enhanced scalar power on small scales 
in ultra slow roll inflation and associated non-Gaussianities}
\author{{\bf H.~V.~Ragavendra}$^{1}$\footnote{\footnotesize
Current address: Raman Research Institute, C.~V.~Raman Avenue, 
Sadashivanagar, Bengaluru 560 080, India;\linebreak
E-mail:~\texttt{ragavendra@rrimail.rri.res.in}}~
and {\bf L.~Sriramkumar}$^{2\ddagger}$ 
\vspace{0.2cm}\\
$^{1}$Department of Physical Sciences, 
Indian Institute of Science Education and Research Kolkata,\\ 
Mohanpur, Nadia 741 246, India \\
$^{2}$Centre for Strings, Gravitation and Cosmology, \\
Department of Physics, Indian Institute of Technology Madras, \\
Chennai 600 036, India.
E-mail:$^\ddagger$\texttt{sriram@physics.iitm.ac.in}
}
\date{}

\begin{document}

\maketitle
\thispagestyle{fancy}
\fancyhead{} 
\fancyhead[LO]{Accepted in \textit{Galaxies}}
\fancyhead[RO]{\href{https://www.mdpi.com}{
\includegraphics[scale=0.4]{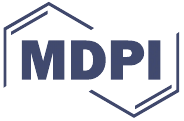}}
}

\vspace{-1cm}
\abstract{The discovery of gravitational waves from merging binary black holes 
has generated considerable interest in examining whether these black holes could 
have a primordial origin. 
If a significant number of black holes have to be produced in the early universe,
the primordial scalar power spectrum should have an enhanced amplitude on small 
scales, when compared to the COBE normalized values on the larger scales that is 
strongly constrained by the anisotropies in the cosmic microwave background. 
In the inflationary scenario driven by a single, canonical scalar field, such 
power spectra can be achieved in models that permit a brief period of ultra slow 
roll inflation during which the first slow roll parameter decreases exponentially.
In this {\it review},\/ we shall consider a handful of such inflationary models 
as well as a reconstructed scenario and examine the extent of formation of 
primordial black holes and the generation of secondary gravitational waves in 
these cases.
We shall also discuss the strength and shape of the scalar bispectrum and the 
associated non-Gaussianity parameter that arise in such situations. 
We shall conclude with an outlook wherein we discuss the wider implications of 
the increased strengths of the non-Gaussianities on smaller scales.}




\section{Introduction}

Without a doubt, the inflationary scenario is the most compelling paradigm 
for overcoming the horizon problem associated with the hot big bang model 
and for simultaneously providing a natural mechanism for generating the 
perturbations in the early universe (see, for example, the 
reviews~\cite{Mukhanov:1990me,Martin:2003bt,Martin:2004um,Bassett:2005xm,
Sriramkumar:2009kg,Baumann:2008bn,Baumann:2009ds,Sriramkumar:2012mik,
Linde:2014nna,Martin:2015dha}).
The simplest of inflationary models involve a single, canonical scalar field 
governed by a smooth potential, which typically leads to a long enough epoch 
of slow roll inflation to overcome the horizon problem.
The perturbations are induced by the quantum fluctuations associated with
the scalar field and they are expected to turn classical during the later
stages of inflation.
The increasingly precise measurements of the anisotropies in the cosmic microwave
background (CMB) over the last couple of decades has led to strong constraints on
the primordial scalar and tensor power spectra on large scales, i.e. over the
wave numbers $10^{-4} \lesssim k \lesssim 10^{-1}\,\mathrm{Mpc}^{-1}$.
The observations by the Planck mission~\cite{Planck:2015sxf,Planck:2018jri} and 
the BICEP/Keck telescopes\cite{BICEP:2021xfz} 
constrain the primordial scalar amplitude~$A_{_{\mathrm{S}}}$, the scalar spectral 
index~$n_{_{\mathrm{S}}}$, and the tensor-to-scalar ratio~$r$ at the pivot scale 
of $k_\ast=0.05\,\mathrm{Mpc}^{-1}$ to be: $A_{_{\mathrm{S}}}=2.1\times10^{-9}$
(a value referred to as COBE normalization), $n_{_{\rm S}} = 0.9649 \pm 0.0042$ 
and $r < 0.036$ at 95\% confidence.
In addition to the constraints on the power spectra, the data from Planck
have also led to bounds on the primordial scalar bispectrum, limiting the 
values of the corresponding non-Gaussianity parameters~$\fnl$.
The Planck data constrain the values of the parameters associated with the 
local, equilateral and orthogonal shapes of the bispectrum to 
be:~$f_{_{\mathrm{NL}}}^\mathrm{local} = -0.9 \pm 5.1$,
$f_{_{\mathrm{NL}}}^{\mathrm{equil}} = -26 \pm 47$ 
and $f_{_{\mathrm{NL}}}^{\mathrm{ortho}} = - 38 \pm 24$~\cite{Ade:2015ava,
Planck:2019kim}.
These constraints suggest that simple slow roll inflationary models which are 
consistent with the data at the level of power spectra are also consistent at
the level of scalar bispectra~\cite{Martin:2013tda,Martin:2013nzq}.

In contrast to the CMB scales, the constraints on the primordial scalar power 
spectrum over the smaller scales are considerably weaker.
With the detection of gravitational waves~(GWs) from merging black hole 
binaries~\cite{LIGOScientific:2018mvr,LIGOScientific:2021usb,
LIGOScientific:2021djp}, there has been interest in the literature to examine 
if these black holes could have a cosmological origin~\cite{Bird:2016dcv,
DeLuca:2020qqa,Jedamzik:2020ypm,Jedamzik:2020omx,Wang:2022nml}.
It has been known earlier that, if the amplitude of the primordial perturbations 
at small scales are adequately high (when compared to the COBE normalized 
amplitude over the CMB scales that we mentioned above), then these perturbations 
can be expected to collapse and form black holes when they reenter the Hubble 
radius during the radiation and matter dominated epochs (for earlier discussions,
see, for example, Refs.~\cite{Carr:1975qj,Khlopov:2008qy,Carr:2009jm}; also see the
recent reviews~\cite{Carr:2016drx,Carr:2018rid,Sasaki:2018dmp,Carr:2020xqk,
Escriva:2022duf,Ozsoy:2023ryl}).
However, in simple models of inflation that permit only slow roll, the amplitude
of the scalar power spectrum will remain roughly at (or less than) the COBE 
normalized value even at smaller scales. 
A strong departure from slow roll during the later stages of inflation is required 
in order to generate power spectra with an enhanced amplitude on smaller scales that
can produce a significant number of primordial black holes~(PBHs).

In this review, we shall focus on single field models of inflation driven 
by the canonical scalar field.
In such cases, it has been shown that a short period of ultra slow roll 
inflation wherein the first slow roll parameter decreases exponentially 
results in increased scalar power over scales that leave the Hubble radius 
just prior to or during the epoch of ultra slow roll (for the original 
discussions, see Refs.~\cite{Tsamis:2003px,Kinney:2005vj}; in this regard,
also see Refs.~\cite{Choudhury:2013jya,Choudhury:2013woa}; 
for recent efforts, see, for example, Refs.~\cite{Garcia-Bellido:2017mdw,
Ballesteros:2017fsr,Germani:2017bcs,Ezquiaga:2017fvi,Bezrukov:2017dyv,
Cicoli:2018asa,Dalianis:2018frf,Bhaumik:2019tvl,Drees:2019xpp,
Dalianis:2020cla,Ragavendra:2020sop}).
We should mention that though the first slow roll parameter remains small 
during the period of ultra slow roll, the second and higher order slow roll
parameters turn large during the phase leading to strong departures from 
slow roll inflation.
Interestingly, it is found that inflationary potentials that contain a (near)
inflection point inevitably lead to a phase of ultra slow roll. 
Besides, it has been found that potentials wherein a bump or a dip is added by 
hand or those that simply contain a sharp change in slope can also lead to an 
epoch of ultra slow roll (in this regard, see, for example, 
Refs.~\cite{Starobinsky:1992ts,Atal:2018neu,Mishra:2019pzq}).
If inflation has to be terminated after the period of ultra slow roll, the 
first slow roll parameter has to steadily rise towards unity. 
The rapid fall and a steady rise in the first slow roll parameter leads to a 
peak in the spectrum of curvature perturbations whose height is determined by 
the smallest value attained by the parameter.
Moreover, in such situations, it can be established that the scalar power 
rises as~$k^4$ before it reaches the peak in the spectrum (for discussions 
in this regard, see Refs.~\cite{Byrnes:2018txb,Cheng:2018qof,Ozsoy:2018flq,
Carrilho:2019oqg,Liu:2020oqe,Tasinato:2020vdk,Motohashi:2019rhu,Ng:2021hll}).
A large enough value for the peak in the scalar power spectrum (but less than 
unity to ensure that the perturbation theory remains valid) can produce copious
amounts of PBHs that can, in principle, constitute all of cold dark matter
today.
We shall restrict our discussion in this review to inflationary models that 
contain a (near) inflection point in the potential and lead to an epoch of
ultra slow roll.

Interestingly, one finds that, when the curvature perturbations are enhanced 
over small scales in order to lead to increased formation of PBHs, they also 
induce secondary GWs 
of significant amplitudes
when these wave numbers reenter the Hubble radius during the 
radiation dominated epoch (for early discussions, see  Refs.~\cite{Ananda:2006af,
Baumann:2007zm,Saito:2008jc,Saito:2009jt}; for recent discussions, see, for
instance, Refs.~\cite{Kohri:2018awv,Espinosa:2018eve,Pi:2020otn,Domenech:2021ztg, 
Balaji:2022rsy}).
In fact, depending on the amplitude and location of the peak in the spectrum
of curvature perturbations, the induced GWs can be of strengths comparable to
the sensitivities of the ongoing and forthcoming GW observatories (for a 
summary of the sensitivity curves and their updated versions, see
Ref.~\cite{Moore:2014lga} and the associated web-page).  
Moreover, recall that, in slow roll inflation, the non-Gaussianity 
parameter~$\fnl$ that reflects the strength of the scalar bispectrum
is of the order of the first slow roll parameter, which is typically 
$\mathcal{O}(10^{-2})$ or less (see, for example, 
Refs.~\cite{Maldacena:2002vr,Seery:2005wm,Chen:2006nt}). 
However, when deviations from slow roll occur, the non-Gaussianities 
can be considerably larger (in this regard, see, for instance, 
Refs.~\cite{Chen:2008wn,Chen:2010xka,Martin:2011sn,Hazra:2012yn,
Ragavendra:2020old,Ragavendra:2020sop}).
The enhanced strengths of non-Gaussianities can play a crucial role on 
the extent of PBHs produced (for very early efforts in this context, see 
Refs.~\cite{Chongchitnan:2006wx,Seery:2006wk,Hidalgo:2007vk}; for recent 
discussions, see Refs.~\cite{Motohashi:2017kbs,Atal:2018neu,
Franciolini:2018vbk,Kehagias:2019eil,Atal:2019erb,DeLuca:2019qsy,
Passaglia:2018ixg,Ezquiaga:2019ftu,Germani:2019zez,Taoso:2021uvl,Riccardi:2021rlf,
Matsubara:2022nbr,Ferrante:2022mui,Pi:2022ysn}) as well as the strengths of the induced 
GWs~\cite{Unal:2018yaa,Cai:2018dig,Cai:2019elf,Ragavendra:2020vud,
Adshead:2021hnm,Ragavendra:2021qdu}.

The plan of this review is as follows.
In the following section, we shall initially arrive at the equations of motion
governing the background and the perturbations using the Arnowitt-Deser-Misner~(ADM) 
formalism~\cite{Arnowitt:1960es}.
We shall then go on to introduce a handful of inflationary models that contain 
a near inflection point which permit an epoch of ultra slow roll inflation.
We shall first describe the evolution of the background in such situations.
We shall then discuss the challenges that arise in ensuring that the scalar 
and tensor power spectra in these models are consistent with the CMB data on 
large scales and describe the manner in which the challenges can be overcome 
by reverse engineering the desired potentials.
In Sections~\ref{sec:pbhs} and~\ref{sec:sgws}, we shall discuss the formation 
of PBHs and the generation of secondary GWs (during the epoch of radiation 
domination) in the various models and scenarios of interest.
In Section~\ref{sec:sbs}, after highlighting a few points related to the third
order action that governs the scalar bispectrum, we shall describe the procedure
for numerically computing the scalar bispectrum.
We shall then go on to present the scalar bispectrum and the associated 
non-Gaussianity parameter that arise in some of the models that we consider.
Lastly, in Section~\ref{sec:outlook}, we shall conclude with an outlook
wherein we discuss the wider implications of the enhanced levels of 
non-Gaussianities on small scales.

Before we proceed further, let us make a few clarifying remarks regarding
the conventions that we shall follow and the notations that we shall use.
We shall work with natural units such that $\hbar=c=1$ and set the reduced 
Planck mass to be $\Mpl=\l(8\,\pi\, G\r)^{-1/2}$.
We shall adopt the signature of the metric to be~$(-,+,+,+)$.
Note that Latin indices will represent the spatial coordinates, apart 
from~$k$ which will be reserved for denoting the wave number. 
Also, an overdot and an overprime will denote differentiation with respect 
to the cosmic and conformal time coordinates~$t$ and~$\eta$, respectively.
Moreover, $a$ will denote the scale factor, and the quantities $H=\dot{a}/a$ 
and $\mathcal{H}=a\,H=a'/a$ will represent the Hubble and the conformal Hubble
parameters.
Further, $N$ will represent the number of $e$-folds, which is defined through 
the relation $\d N/\d t=H$.
A subscript $N$ will denote the differentiation with respect to the number of 
$e$-folds.
Lastly, we shall denote the canonical scalar field that we shall consider 
as~$\phi$, and a subscript $\phi$ will represent differentiation with
respect to the scalar field.


\section{Inflationary models, power spectra and reverse engineered 
potentials}\label{sec:qo}

In this section, we shall consider inflation driven by a canonical scalar field
and first discuss the equations governing the background and the perturbations 
at the linear order.
In contrast to the more common approach of using the zeroth and first order
Einstein's equations to arrive at the governing equations, we shall arrive 
at the equations using the ADM formalism~\cite{Arnowitt:1960es}.
As we shall see later, the approach proves to be helpful when we discuss the 
scalar bispectrum generated in the models of our interest.
We shall then go on to introduce a handful of inflationary models that lead 
to an epoch of ultra slow roll inflation and discuss the scalar and tensor 
power spectra that arise in the models.
We shall illustrate that, in these models, if we desire a sufficiently large 
peak in the power spectrum at small scales, there arises a challenge in 
ensuring that the scalar and tensor power spectra are consistent with the 
constraints from the CMB on large scales.
In order to circumvent this difficulty, we shall discuss the method of reverse
engineering desired potentials from a specific form of the first slow roll 
parameter.


\subsection{Arriving at the equations governing the background and 
the perturbations at the linear order}\label{subsec:adm}

As we mentioned, to arrive at the equations of motion governing the background
and the perturbations, we shall make use of the ADM 
formalism~\cite{Arnowitt:1960es}. 
Recall that, in the ADM formalism, the spacetime metric is expressed in 
terms of the lapse function~$\cN$, the shift vector~$\cN^{i}$ and the spatial 
metric~$\mathsf{h}_{ij}$ as follows:
\begin{equation} 
\d s^{2}=-\cN^{2}\,  \l(\d x^{0}\r)^{2}
+\mathsf{h}_{ij}\,\l(\cN^{i}\, \d x^{0}+\d x^{i}\r)\, 
\l(\cN^{j}\, \d x^{0}+\d x^{j}\r),\label{eq:adm-m}
\end{equation}
where $x^0$ and $x^{i}$ denote the time and the spatial coordinates,
respectively. 
We shall assume that gravitation is described by Einstein's general theory
of relativity.
Since we shall be focusing on the epoch of inflation (to calculate the 
primordial power and bi-spectra), we shall assume that 
gravitation is sourced by a canonical, minimally coupled, scalar field~$\phi$, 
which is described by the potential~$V(\phi)$.
In such a case, the action describing the complete system consisting of 
gravitation and the scalar field can be written in terms of the metric 
variables $\cN$, $\cN^i$ and~$\mathsf{h}_{ij}$ and the field~$\phi$ 
as follows (see, for instance, Refs.~\cite{Maldacena:2002vr,
Seery:2005wm,Chen:2006nt,Martin:2011sn,Arroja:2011yj}):
\begin{eqnarray}
\mathcal{S}[\cN,\cN^{i},\mathsf{h}_{ij},\phi]\!\!\!  
&=&\!\!\! \int \d x^{0} \int \d^{3} \vx\,\cN\,\sqrt{h}\,
\Biggl\{\frac{\Mp^{2}}{2} \l[\f{1}{\cN^2}\,\l(E_{ij}E^{ij} -E^2\r) 
+ ^{(3)}\!\!R\r]\nn\\
& &\!\!\! + \biggl[\f{1}{2\, \cN^2}\, \l(\pa_0\phi\r)^2 
-\f{\cN^i}{\cN^2}\, \pa_0\phi\, \pa_i\phi 
+ \f{\cN^i\,\cN^j}{2\,\cN^2}\, \pa_i\phi\, \pa_j\phi\nn\\
& &\!\!\! - \f{1}{2}\, \mathsf{h}^{ij}\, \pa_i\phi\; \pa_j\phi- V(\phi)\biggr]\biggr\},
\label{eq:adm-a}
\end{eqnarray}
where $\pa_0\phi=\pa\phi/\pa x^0$, $\mathsf{h} \equiv \mathrm{det.}~(\mathsf{h}_{ij})$ 
and $^{(3)}\!R$ is the curvature associated with the spatial metric~$\mathsf{h}_{ij}$. 
The quantity $ E_{ij}$ is given by
\begin{equation}
E_{ij} = \f{1}{2}\, \l(\pa_0 \mathsf{h}_{ij} - \nabla_i \cN_j 
- \nabla_j \cN_i\r),
\end{equation}
with $E = \mathsf{h}_{ij}\,E^{ij}$.
Note that the variation of the above action with respect to the Lagrange 
multipliers~$\cN$ and~$\cN^{i}$ leads to the so-called Hamiltonian and momentum
constraints, respectively.
Upon solving the constraint equations and substituting the solutions back 
in the original action~(\ref{eq:adm-a}), we can arrive at the action 
governing the dynamical variables of interest up to a given order in the
perturbations.


\subsubsection{Equations of motion describing the background and the slow 
roll parameters}

We shall assume the background to be the spatially flat, 
Friedmann-Lema\^itre-Robertson-Walker~(FLRW) universe described by 
the following line element: 
\begin{equation}
\d s^2=-\d t^2+a^2(t)\,\d {\bm x}^2
=a^2(\eta)\, \l(-\d \eta^2+\d {\bm x}^2\r),\label{eq:FLRW}
\end{equation}
where, as we mentioned, $t$ and~$\eta$ denote cosmic and conformal 
time coordinates.
Since we shall be interested in the situation wherein the FLRW universe 
is dominated by the canonical scalar field~$\phi$ described by the 
potential~$V(\phi)$, we can arrive at the action governing the scale factor~$a$ 
and the homogeneous scalar field~$\phi$ upon substituting the above 
line-element in the ADM action~\eqref{eq:adm-a}.
At the leading order---i.e. at the zeroth order, with $\cN_i$ set to zero 
and $\mathsf{h}_{ij}=a^2(t)\, \delta_{ij}$---we find that the complete 
action describing the system is given by 
\begin{equation}
\mathcal{S}_0[\phi(t)] 
=\int \d t \int \d^{3} \vx\,a^3\, \l[-\f{3\,\Mpl^2\,H^2}{\cN}
+\f{\dot{\phi}^2}{2\,\cN}-V(\phi)\r].
\end{equation}
If we now vary this action with respect to the lapse function~$\cN$, we 
arrive at the following constraint equation upon setting~$\cN$ eventually
to unity:
\begin{equation}
H^2=\f{1}{3\,\Mpl^2}\, \l[\f{\dot{\phi}^2}{2}+V(\phi)\r],\label{eq:ffe}
\end{equation}
which is the first Friedmann equation.
On varying the above action with respect to the scalar field (with $\cN$ 
set to unity) leads to the equation of motion for the scalar field 
given by
\begin{equation}
\ddot{\phi}+3\,H\,\dot{\phi} + V_{\phi}=0.\label{eq:eom-sf}
\end{equation}
It is useful to note that the above two equations can be combined to arrive
at the equation
\begin{equation}
\dot{H}=-\f{\dot{\phi}^2}{2\,\Mpl^2}.
\end{equation}

The first slow roll parameter is defined as~\cite{Mukhanov:1990me,Martin:2003bt,
Martin:2004um,Bassett:2005xm,Sriramkumar:2009kg,Baumann:2008bn,Baumann:2009ds,
Sriramkumar:2012mik,Linde:2014nna,Martin:2015dha}
\begin{equation}
\epsilon_1=-\f{\dot{H}}{H^2}=-\f{H_N}{H}
=\f{\dot{\phi}^2}{2\,H^2\,\Mpl^2}=\f{\phi_N^2}{2\,\Mpl^2}.\label{eq:f-srp}
\end{equation}
The higher order slow roll parameters are defined in terms of the first slow
roll parameter~$\epsilon_1$ through the relations 
\begin{equation}
\epsilon_{n+1}=\f{\d\, \mathrm{ln}\,\epsilon_n}{\d N}\label{eq:ho-srp}
\end{equation}
for $n \geq 1$.
For instance, one can show that the second slow roll parameter can be 
written as
\begin{equation}
\epsilon_2=\f{2\,\phi_{NN}}{\phi_{N}}.
\end{equation}
As we shall see, it is the first three slow roll parameters, viz. $\epsilon_1$, 
$\epsilon_2$, and $\epsilon_3$, that determine the amplitude and shape of the 
inflationary power spectrum and bispectrum.


\subsubsection{Scalar and tensor perturbations, equations of motion, 
quantization and power spectra}

Let us now arrive at the action and the equations of motion governing the 
scalar and tensor perturbations using the ADM formalism.
To do so, it turns out to be convenient if we choose to work in a particular 
gauge. 
We shall work in the so-called comoving gauge wherein the perturbation in the 
scalar field, say, $\delta \phi$, vanishes identically~\cite{Maldacena:2002vr}.
In other words, in the gauge of our choice, the scalar field~$\phi$ depends 
only on time.
Let the scalar perturbation be described by the curvature perturbation~$\cR$ 
and let the tensor perturbation be characterized by~$\gamma_{ij}$.
On taking into account these perturbations, it is convenient to express the 
spatially flat FLRW metric~\eqref{eq:FLRW} as~\cite{Maldacena:2002vr}
\begin{equation}
\d s^2 = -\d t^2 
+ a^{2}(t)\; {\rm e}^{2\,{\cal R}(t,\bm{x})}\,
\l[{\rm e}^{\gamma(t,\bm{x})}\r]_{ij}\,
\d x^i\, \d x^j.\label{eq:metric}
\end{equation}
The assumption for the scalar field~$\phi$, the above form of 
the FLRW
metric and the solutions to the Hamiltonian and momentum constraint equations 
allow us to arrive at the action describing the scalar and tensor perturbations
$\cR$ and $\gamma_{ij}$ at a given order~\cite{Maldacena:2002vr,Seery:2005wm,
Chen:2006nt}.
It can be shown that, in the comoving gauge of interest, at the quadratic order,
the actions governing the curvature perturbation~${\cal R}$ and the tensor 
perturbation~$\gamma_{ij}$ can be expressed as~\cite{Maldacena:2002vr,
Seery:2005wm,Martin:2011sn,Sreenath:2013xra}
\begin{subequations}\label{eq:a-p}
\begin{eqnarray}
\mathcal{S}_2[\cR(\eta,{\bm x})]\!\!\!
&=&\!\!\! \f{1}{2}\, \int \d \eta\, \int \d^{3}\vx\,\, z^{2}\,
\l[{\cR'}^2-\l(\pa\cR\r)^{2}\r],\label{eq:a-rr}\\
\mathcal{S}_2[\gamma_{ij}(\eta,{\bm x})]\!\!
&=&\!\! \f{\Mpl^2}{8}\, \int \d \eta\, \int \d^{3}\vx\,\, a^{2}\,
\l[{\gamma_{ij}'}^2-\l(\pa\gamma_{ij}\r)^{2}\r],\label{eq:a-gg}
\end{eqnarray}
\end{subequations}
where $z=a\sqrt{2\,\epsilon_1}\,\Mp$, with $\epsilon_1$ being the first 
slow roll parameter.
The above quadratic actions will evidently lead to linear equations of motion.
In Fourier space, the modes functions, say, $f_k$ and $g_k$, associated with 
the scalar and the tensor perturbations are found to satisfy the differential 
equations
\begin{subequations}\label{eq:de-p}
\begin{eqnarray}
f_k''+2\, \f{z'}{z}\, f_k' + k^2\, f_k \!\!\!
&=&\!\!\! 0,\label{eq:de-fk}\\
g_k''+2\, \f{a'}{a}\, g_k' + k^2\, g_k\!\!\!
&=&\!\!\! 0,\label{eq:de-gk}
\end{eqnarray}
\end{subequations}
respectively.

As we indicated earlier, in the inflationary paradigm, the primordial 
perturbations arise due to quantum fluctuations.
On quantization, the scalar and tensor perturbations~$\cR$ and~$\gamma_{ij}$
can be elevated to be quantum operators.
The operators $\hat{\cR}$ and $\hat\gamma_{ij}$ can be decomposed in terms of 
the corresponding mode functions~$f_k$ and~$g_k$---which satisfy the equations
of motion~\eqref{eq:de-p}---as follows:
\begin{subequations}
\label{eqs:st-m-dc}
\begin{eqnarray}
\hat{\cR}(\eta, {\bf x})\!\!\! 
&=&\!\!\!  \int \f{\d^{3}{\bm k}}{(2\,\pi)^{3/2}}\,
\hat{\cR}_{\bm k}(\eta)\, {\mathrm{e}}^{i\,{\bm k}\cdot{\bm x}}\nn\\
&=&\!\!\!  \int \f{\d^{3}{\bm k}}{(2\,\pi)^{3/2}}\,
\l[\hat{a}_{\bm k}\,f_{k}(\eta)\, {\mathrm{e}}^{i\,{\bm k}\cdot{\bm x}}
+\hat{a}^{\dagger}_{\bm k}\,f^{\ast}_{k}(\eta)\,
{\mathrm{e}}^{-i\,{\bm k}\cdot{\bm x}}\r],\label{eq:s-m-dc}\\
\hat{\gamma}_{ij}(\eta, {\bf x})\!\!\!  
&=&\!\!\!  \int \frac{\d^{3}{\bm k}}{\l(2\,\pi\r)^{3/2}}\,
\hat{\gamma}_{ij}^{\bm k}(\eta)\, {\rm e}^{i\,{\bm k}\cdot{\bm x}}\nn\\
&=&\!\!\!  \sum_{s}\int \frac{\d^{3}{\bm k}}{(2\,\pi)^{3/2}}\,
\l[\hat{b}^{s}_{\bm k}\, \varepsilon^{s}_{ij}({\bm k})\,
g_{k}(\eta)\, {\mathrm{e}}^{i\,{\bm k}\cdot{\bm x}}
+\hat{b}^{s\dagger}_{\bf k}\,\varepsilon^{s\ast}_{ij}({\bm k})\, 
g^{\ast}_{k}(\eta)\, {\mathrm{e}}^{-i\,{\bm k}\cdot{\bm x}}\r].\label{eq:t-m-dc}
\end{eqnarray}
\end{subequations}
In the expression for the operator describing the tensor perturbations, the quantity 
$\varepsilon^{s}_{ij}({\bm k})$ represents the polarization tensor of the GWs 
with their helicity being denoted by the index~$s$.
Moreover, in the above decompositions, the two independent sets of operators 
$(\hat{a}_{\bm k},\hat{a}^{\dagger}_{\bm k})$ and $(\hat{b}_{\bm k}^{s}, 
\hat{b}^{s\dagger}_{\bm k})$ denote the annihilation and creation operators 
associated with the scalar and tensor modes corresponding to the wave vector 
${\bm k}$ and helicity~$s$ (with the latter applying to the case of tensors).
These operators are governed by the following, standard commutation relations:
\begin{subequations}
\begin{eqnarray}
[\hat{a}_{\bm k},\hat{a}_{\bm{k}'}] \!\!\!
&=&\!\!\! [\hat{a}^{\dagger}_{\bm k},\hat{a}^{\dagger}_{{\bm k}'}]=0,\quad\;\;
[\hat{a}_{\bm k},\hat{a}^{\dagger}_{{\bm k}'}]
=\delta^{(3)}\l({\bm k}-{{\bm k}'}\r),\\{}
[\hat{b}_{\bm k}^{s},\hat{b}^{s'}_{{\bm k}'}] \!\!\!
& = &\!\!\! [\hat{b}_{\bm k}^{s\dagger},\hat{b}^{s'\!\dagger}_{{\bm k}'}]=0,\quad
[\hat{b}_{\bm k}^{s},\hat{b}^{s'\!\dagger}_{{\bm k}'}]=\delta^{ss'}\,
\delta^{(3)}\l({\bm k}-{{\bm k}'}\r).
\end{eqnarray}
\end{subequations}
Note that the transverse and traceless nature of GWs leads to the conditions 
$\delta^{ij}\,k_{i}\,\varepsilon_{jl}^s({\bm k})=
\delta^{ij}\,\varepsilon^{s}_{ij}({\bm k})=0$. 
We should point our that we shall work with the normalization 
condition $\delta^{ij}\,\delta^{lm}\,\varepsilon_{il}^{r}({\bm k})\,
\varepsilon_{jm}^{s\ast}({\bm k})=2\,\delta^{rs}$~\cite{Maldacena:2002vr}.

Often, it proves to be convenient to introduce the so-called Mukhanov-Sasaki 
variables to describe the scalar and tensor perturbations.
These variables are defined as $v_k = z\, f_k$ and $u_k = \Mpl\, a\, g_k/\sqrt{2}$
and, in terms of these variables, the equations of motion~\eqref{eq:de-p} 
that govern the scalar and the tensor perturbations reduce to
\begin{subequations}
\label{eqs:ms}
\begin{eqnarray}
v_k''+\l(k^2-\f{z''}{z}\r)\, v_k\!\!\!  &=&\!\!\!  0,\label{eq:ms-f}\\
u_k''+\l(k^2-\f{a''}{a}\r)\, u_k\!\!\!  &=&\!\!\!  0.\label{eq:ms-g}  
\end{eqnarray}
\end{subequations}
The scalar and the tensor power spectra, viz. $\ps(k)$ and $\pt(k)$, are 
defined through the relations
\begin{subequations}
\begin{eqnarray}
\langle {\hat \cR}_{\bm k}(\ee)\,{\hat \cR}_{\bm k'}(\ee)\rangle\!\!\! 
&=&\!\!\! \f{(2\,\pi)^2}{2\, k^3}\, \ps(k)\;
\delta^{(3)}({\bm k}+{\bm k'}),\label{eq:sps-d}\\
\langle {\hat \gamma}_{ij}^{\bm k}(\ee)\,
{\hat \gamma}_{mn}^{\bm k'}(\ee)\rangle\!\!\! 
&=&\!\!\! \f{(2\,\pi)^2}{2\, k^3}\, \f{\Pi_{ij,mn}^{{\bm k}}}{4}\,\pt(k)\,
\delta^{(3)}({\bm k}+{\bm k'}),\label{eq:tps-d}
\end{eqnarray}
\end{subequations}
where the expectation values on the left hand sides are to be evaluated 
in the specified initial quantum state of the perturbations and $\ee$~is 
the conformal time at late times, close to the end of inflation.
The quantity $\Pi_{ij,mn}^{\vk}$ is given by~\cite{Maldacena:2011nz}
\begin{equation}
\Pi_{ij,mn}^{\vk}=\sum_{s}\,\varepsilon_{ij}^{s}(\vk)\,
\varepsilon_{mn}^{s\ast}(\vk).\label{eq:pi-ijmn}
\end{equation}
Typically, the expectation values are evaluated in the vacuum state, say, 
$\vert 0 \rangle$, associated with the quantized perturbations.
The state satisfies the conditions ${\hat a}_{\bm k}\vert 0\rangle=0$ 
and ${\hat b}_{\bm k}^{s}\vert 0\rangle=0$ for all wave numbers~${\bm k}$
and helicity~$s$. 
The initial state is defined at very early times when all the scales of 
cosmological interest are well inside the Hubble radius during inflation 
and the quantum state is referred to as the Bunch-Davies vacuum~\cite{Bunch:1978yq}.
The scalar and tensor power spectra $\ps(k)$ and $\pt(k)$ can be expressed
in terms of the mode functions $(f_k,g_k)$ and the associated 
Mukhanov-Sasaki variables $(v_k,u_k)$ as follows:
\begin{subequations}\label{eqs:stps}
\begin{eqnarray}
\ps(k)\!\!\! 
&=&\!\!\! \f{k^3}{2\, \pi^2}\, \vert f_k(\ee)\vert^2
=\f{k^3}{2\, \pi^2}\, \f{\vert v_k(\ee)\vert^2}{z^2(\ee)},\label{eq:sps}\\
\pt(k)\!\!\! 
&=&\!\!\!  4\,\f{k^3}{2\, \pi^2}\, \vert g_k(\ee)\vert^2
=\f{8}{\Mpl^2}\,\f{k^3}{2\, \pi^2}\, 
\f{\vert u_k(\ee)\vert^2}{a^2(\ee)}.\label{eq:tps}
\end{eqnarray}
\end{subequations}


\subsection{A short list of models permitting ultra slow roll 
inflation}\label{subsec:im} 

A wide variety of models which permit an epoch of ultra slow roll inflation
have been considered in the literature (for a short list of such efforts,
see Refs.~\cite{Garcia-Bellido:2017mdw,
Ballesteros:2017fsr,Germani:2017bcs,Ezquiaga:2017fvi,Bezrukov:2017dyv,
Cicoli:2018asa,Dalianis:2018frf,Bhaumik:2019tvl,Drees:2019xpp,
Dalianis:2020cla,Ragavendra:2020sop,Kawai:2021edk}).
Curiously, many of these models contain a point of inflection, i.e. a point 
where the first and the second derivatives of the potential with respect to 
the scalar field (viz. $V_\phi$ and $V_{\phi\phi}=\d^2V/\d\phi^2$) 
vanish (in this regard, see Appendix~\ref{app:pi}).
We shall consider six of these models and we shall now briefly describe the
models of our interest and the parameters that we shall work with.
For convenience, in our discussions that follow, we shall refer to these 
models as M1 to M6.
In the introductory section, we had mentioned that the CMB observations 
strongly constrain the scalar amplitude, the scalar spectral index and
the tensor-to-scalar ratio at the pivot scale~$k_\ast$.
We shall assume that the pivot scale leaves the Hubble radius~$N_\ast$ 
number of $e$-folds {\it before the end of inflation}.\/
We should point out that, in the absence of detailed modeling
of post-inflationary dynamics, there arises some uncertainty in the choice of 
$N_\ast$ and it is often assumed to lie in the range $50\lesssim N_\ast 
\lesssim 60$~\cite{Liddle:2003as,Planck:2015sxf,Planck:2018jri}.

$\bullet$~{\bf Model~1:}~The first of the models that we shall consider,
which leads to a period of ultra slow roll inflation, is described by a
potential that can be written in the following fashion~\cite{Garcia-Bellido:2017mdw}:
\begin{equation}
V(\phi) = V_0\;
\f{6\,x^2 - 4\,\alpha\,x^3 + 3\,x^4}{(1 + \beta\,x^2)^2},\label{eq:phi4}
\end{equation}
where $x = \phi/v$, with $v$ being a constant.
We shall work with the following choices for the four parameters that 
describe the potential:~$V_0= 4\times10^{-10}\,\Mpl^4 $, $v = 
\sqrt{0.108}\,\Mpl$, $\alpha = 1$ and $\beta = 1.4349$.
For these values of the parameters, the (near) inflection point, say, 
$\phi_0$, is located at $0.39\,\Mpl$.
(For a discussion on the determination of the inflection points numerically 
in the inflationary models being considered, see Appendix~\ref{app:pi}.)
If we choose the initial value of the field to be $\phi_{\mathrm{i}}=3.614\,\Mpl$ 
and $\epsilon_{1\mathrm{i}}=10^{-3}$~[which determines $\phi_{N\mathrm{i}}$,
cf. Eq.~\eqref{eq:f-srp}], we find that inflation lasts for about
$63$~$e$-folds in the model.
Also, in this case, we shall set $N_\ast=50$.

$\bullet$~{\bf Model~2:}~The second potential that we shall consider can 
expressed in terms of the quantity $x = \phi/v$ that we had introduced in
the first model, and is given by~\cite{Bhaumik:2019tvl}
\begin{equation}
V(\phi) = V_0\,\frac{\alpha\,x^2-\beta\,x^4+\gamma\,x^6}{(1+\delta\,
x^2)^2}.\label{eq:phi6}
\end{equation}
We shall consider the following set of values for the six parameters 
involved:~$V_0=1.3253\times10^{-9}\,\Mpl^4$, $v=10\,\Mpl$, 
$\alpha=8.53\times 10^{-2}$, $\beta=0.458$, $\gamma=1$ and $\delta= 1.5092$.
For these values of the parameters and the initial conditions
$\phi_\mathrm{i}=17.245\,\Mpl$ and $\epsilon_{1\mathrm{i}}= 10^{-2}$, 
inflation continues for about $75$~$e$-folds before it is terminated.
Also, the point of inflection is located at $\phi_0=1.72\,\Mpl$.
For this model, we shall choose $N_\ast=55$.

$\bullet$~{\bf Model~3:}~A potential referred to as the critical Higgs
model is given by~\cite{Ezquiaga:2017fvi,Bezrukov:2017dyv,Drees:2019xpp}:
\begin{equation}
V(\phi)=V_0\,
\f{\l[1+a\,\l(\mathrm{ln}\,x\r)^2\r]\,x^4}{\l[1
+c\,\l(1+b\,\mathrm{ln}\,x\r)\,x^2\r]^2},
\end{equation}
where $x=\phi/\mu$ and we shall set $\mu = 1\,\Mpl$.
We shall choose the values of the other parameters that describe the potential
to be $V_0=7.05\times 10^{-8}\,\Mpl^4$, $a = 1.694$, $b = 0.601$ and $c = 2.850$. 
For these values, the point of inflection occurs at $\phi_0 = 0.820\,\Mpl$. 
The initial values of the field and the first slow roll parameter are taken 
to be $\phi_\mathrm{i}=8.00\,\Mpl$ and $\epsilon_{1\mathrm{i}}=10^{-3}$.
In such a case, we achieve about $103$ $e$-folds of inflation.
The pivot scale is set to exit the Hubble radius at about $70$ $e$-folds before 
the end of inflation to achieve the feature at the desired wave number.

$\bullet$~{\bf Model~4:}~The fourth potential that we shall consider is given
by~\cite{Dalianis:2018frf}
\begin{equation}
V(\phi) = V_0\,\l\{\mathrm{tanh}\l(\f{\phi}{\sqrt{6}\,\Mpl}\r) 
+ A\,\sin\l[\f{\mathrm{tanh}\l[\phi/\l(\sqrt{6}\,\Mpl\r)\r]}{f_\phi}\r]\r\}^2
\label{eq:D1}
\end{equation}
and we shall work with the following values of the parameters 
involved:~$V_0 = 2\times10^{-10}\,\Mpl^4$, $A = 0.130383$ and 
$f_\phi = 0.129576$.
For these values of the parameters, we find that a point of inflection 
arises at $\phi_0 = 1.05\,\Mpl$.
If we set initial value of the field to be $\phi_\mathrm{i}=6.1\,\Mpl$ 
and the first slow roll parameter to be $\epsilon_{1\mathrm{i}}=10^{-3}$, 
we obtain about $66$~$e$-folds of inflation in the model.
Also, we shall choose $N_\ast=50$.

$\bullet$~{\bf Model~5:}~A model constructed from supergravity which 
permits a period of ultra slow inflation is described by the 
potential~\cite{Dalianis:2018frf,Dalianis:2020cla}
\begin{equation}
V(\phi) = V_0\,\l[c_0 + c_1\,\tanh\, \l(\f{\phi}{\sqrt{6\,\alpha}}\r) 
+ c_2\,\tanh^2\l(\f{\phi}{\sqrt{6\,\alpha}}\r)
+ c_3\,\tanh^3{\l(\f{\phi}{\sqrt{6\,\alpha}}\r)}\r]^2.\label{eq:pi-tanh}
\end{equation}
We shall work with the following values for the parameters involved:~$V_0
=2.1 \times 10^{-10}\,\Mpl^4$, $c_0=0.16401$, $c_1=0.3$, $c_2=-1.426$, 
$c_3=2.20313$ and $\alpha=1\,\Mpl^2$.
This model too contains a point of inflection and, for the above values for
the parameters, the inflection point is located at $\phi_0 = 0.53\,\Mpl$.
We find that, for the initial values $\phi_\mathrm{i} = 7.4\,\Mpl$ and 
$\epsilon_{1\mathrm{i}}=10^{-3}$, inflation ends after about $68$~$e$-folds. 
Also, in this case, we shall set $N_\ast=50$.

$\bullet$~{\bf Model~6:}~The sixth and last model that we shall consider is
motivated by string theory, and is described by the potential~\cite{Cicoli:2018asa}
\begin{eqnarray}
V(\phi)\!\!\!  &=&\!\!\!  \f{W^2_0}{{\mathcal{V}}^3}\,
\biggl\{\f{C_{\mathrm{up}}}{{\mathcal{V}}^{1/3}} 
- \f{C_W}{\sqrt{\tau_{K3}(\phi)}} 
+ \f{A_W}{\sqrt{\tau_{K3}(\phi)}-B_W}\nn\\
& &\!\!\! +\, \f{\tau_{K3}(\phi)}{\mathcal{V}}\,\l[D_W - 
\f{G_W}{1+(R_W/\mathcal{V})\,\tau^{3/2}_{K3}(\phi)}\r]\biggr\},
\end{eqnarray}
where $\tau_{K3}(\phi) = \exp\,[2\, \phi/(\sqrt{3}\,\Mpl)$,
$W_0 = 9.469\,\Mpl^2$, ${\mathcal{V}} = 10^3$, $C_{\mathrm{up}} = 0$, 
$C_W = 0.04$, $A_W = 0.02$, $B_W = 1.00$, $D_W = 0$,
$G_W = 3.081\times10^{-5}\,{\mathcal{V}}$,
and $R_W = 7.071\times 10^{-4}\,{\mathcal{V}}$.
To achieve the required duration of inflation, we shall set 
$\phi_{\mathrm{i}}=10.0\,\Mpl$ and $\epsilon_{1{\mathrm{i}}}=2\times10^{-3}$. 
These initial conditions lead to about $68$ $e$-folds of inflation. 
We shall set $N_\ast=50$ to compute the power spectra arising in the model.


\subsection{Evolution of the background in ultra slow roll inflation}

We shall solve the equations of motion governing the background and the 
perturbations numerically to arrive at the scalar and tensor power spectra
in a given inflationary model.
To make the numerical computation efficient, as is usually done, we shall
use the number of $e$-folds~$N$ as the independent time variable.
Note that the first Friedmann equation~\eqref{eq:ffe} and the equation of 
motion~\eqref{eq:eom-sf} can be combined to arrive at the following equation
for the scalar field:
\begin{equation}
\phi_{NN} 
+ \l(3 - \f{\phi_N^2}{2\,\Mp^2}\r)\,\phi_{N} 
+ \l(3\,\Mp^2 - \f{\phi_N^2}{2}\r)\, \f{V_{\phi}}{V} = 0.\label{eq:phi-N}
\end{equation}
Given the potential describing the scalar field, the values of the parameters
and the initial conditions (viz. $\phi_{\mathrm{i}}$ and $\epsilon_{1{\mathrm{i}}}$,
with the latter determining $\phi_{N\mathrm{i}}$), we utilize the fifth order 
Runge-Kutta method, with an adaptive step size, to evolve the above 
equation~\cite{Press:2007nrf,Fehlberg:1969rk5}.
Once the solution for the scalar field is at hand, all the other background
quantities, including the slow roll parameters, can be expressed in 
terms of the scalar field and its time derivatives.

\begin{figure}[!t]
\centering
\hskip -46.2pt
\includegraphics[width=9.50cm]{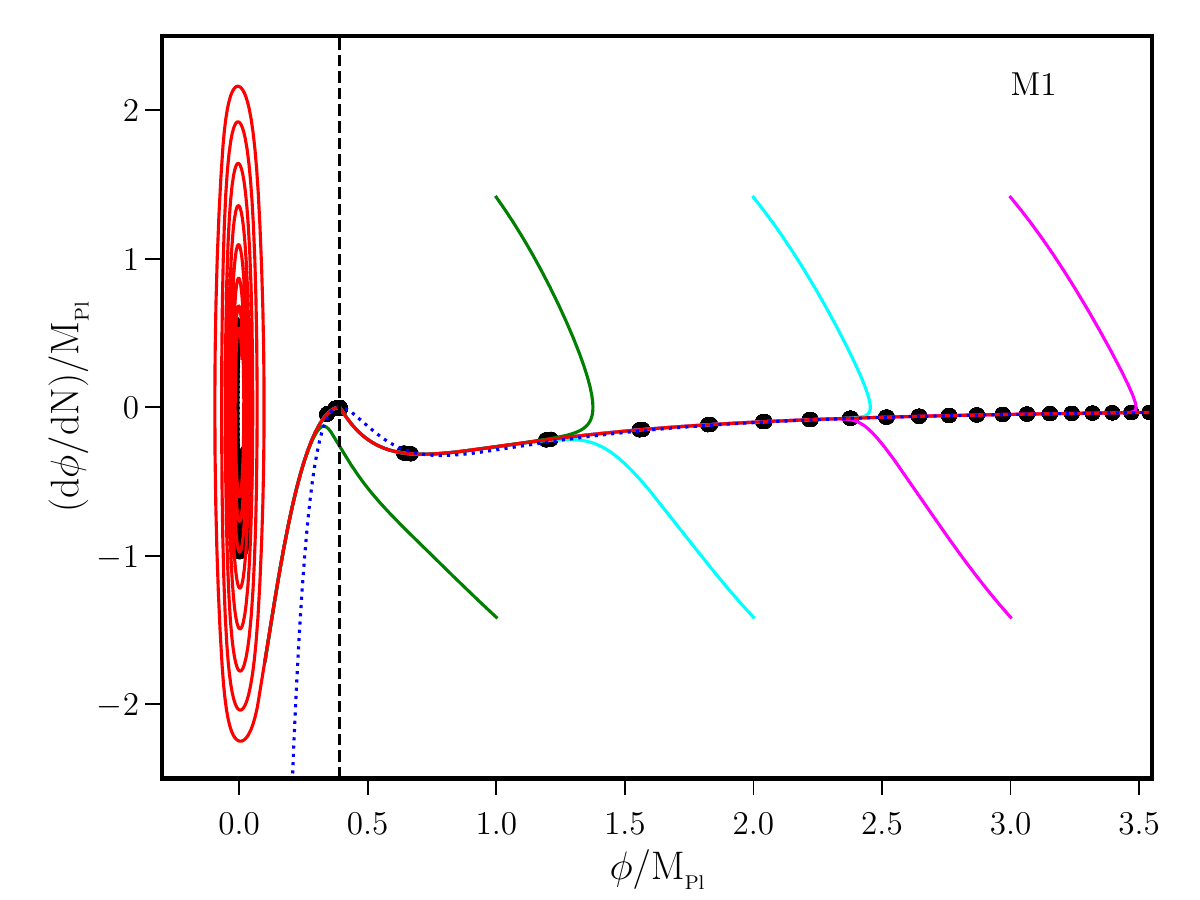}
\includegraphics[width=9.50cm]{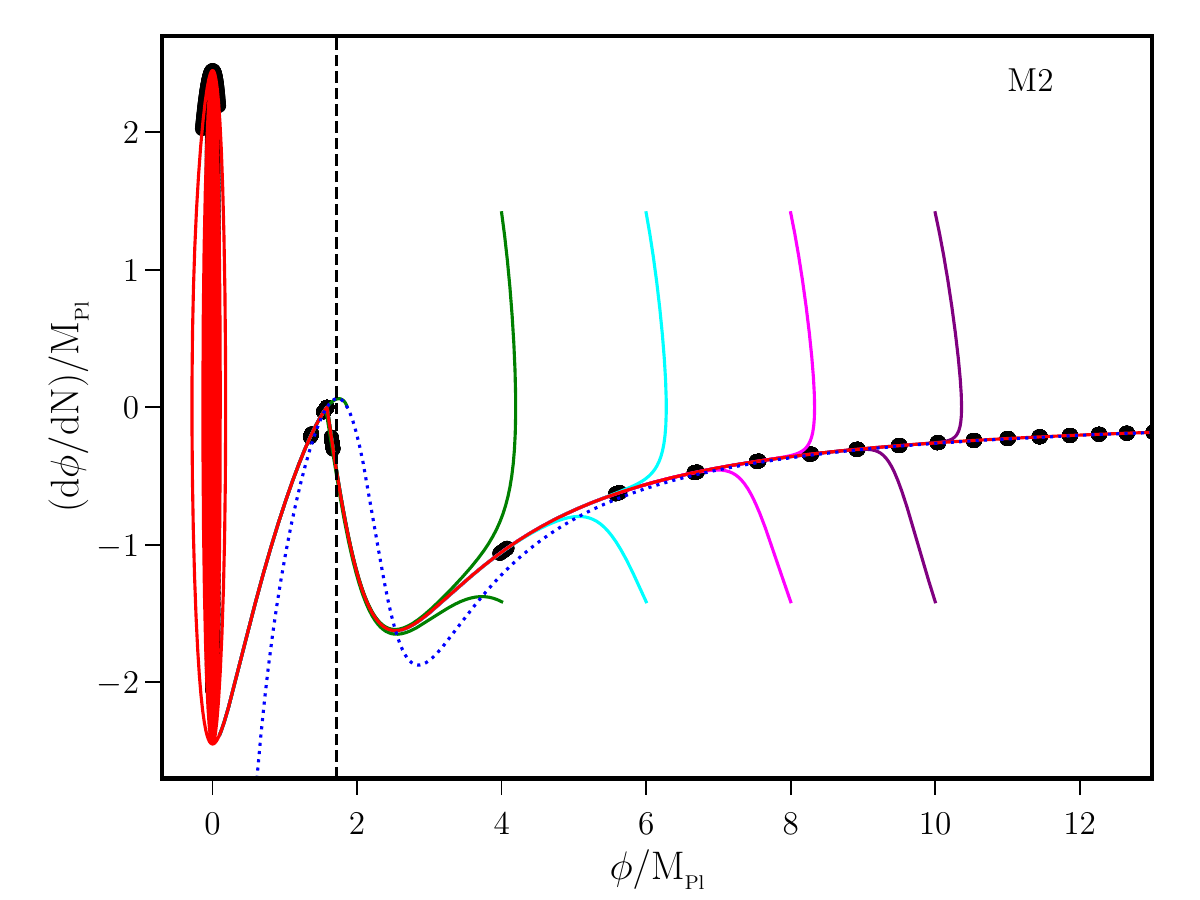}
\caption{The portrait of the scalar field in the phase 
space has been illustrated for the two models~M1 and~M2 (on the left and
the right, respectively).
We have plotted the trajectories for different initial conditions (as 
solid curves in different colors), along with the specific initial 
conditions (plotted in red) that we shall be focusing on in our later 
discussion. 
For the initial conditions apart from the ones of our interest, we have 
considered two values of $\phi_{N{\mathrm{i}}}$ for each value of 
$\phi_{\mathrm{i}}$ (plotted in same color). 
In the case of the primary trajectory, we have indicated the lapse in time 
every two~$e$-folds (as black dots on the red curves). 
We have identified the point of inflection (with black vertical lines) and we 
have also illustrated the evolution arrived at using the standard slow 
roll approximation (as dotted blue curves).}\label{fig:b-pp}
\end{figure}
Let us now understand the evolution of the background in the inflationary
models of interest. 
We can gain an overall perspective of the dynamics involved from the 
behavior of the field in the phase space $\phi$-$\phi_N$.
In Figure~\ref{fig:b-pp}, we have presented the phase portrait of the 
scalar field for the two models~M1 and~M2.
We should highlight three points regarding the figure.
Firstly, it is clear that the trajectories with different initial conditions 
quickly merge with the primary trajectory of interest (viz. the one evolved 
from the initial conditions mentioned above). 
Secondly, independent of the initial conditions, the speed of the scalar field 
reduces considerably as it approaches the point of inflection. 
It is this behavior that is responsible for the epoch of ultra slow roll
inflation in these models.
In fact, we find that, around the point of inflection, certain trajectories 
with insufficient velocities may stagnate and not evolve beyond the point.
Lastly, the slow roll approximation fails to capture the dynamics of the field 
near and beyond the point of inflection.

\begin{figure}[!t]
\centering
\hskip -46.2pt
\includegraphics[width=9.50cm]{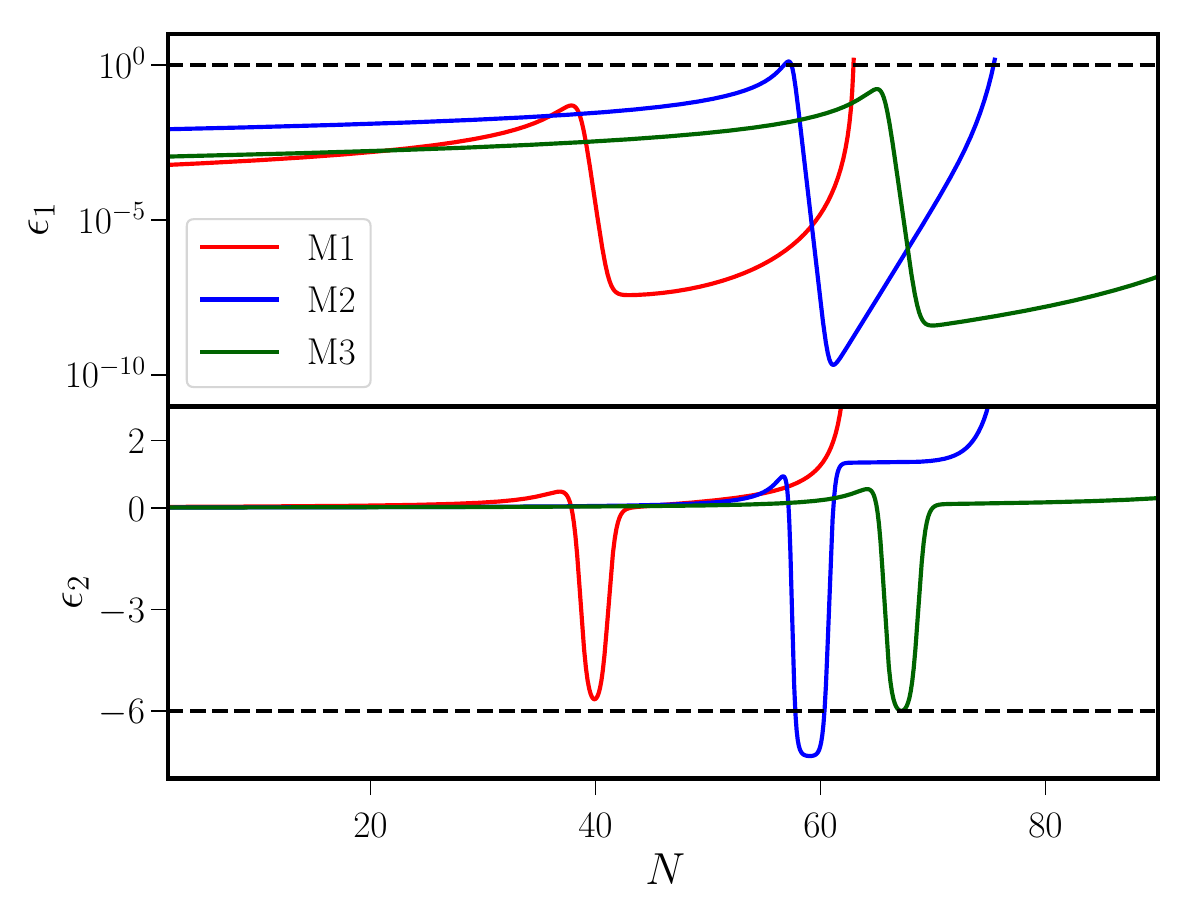}
\includegraphics[width=9.50cm]{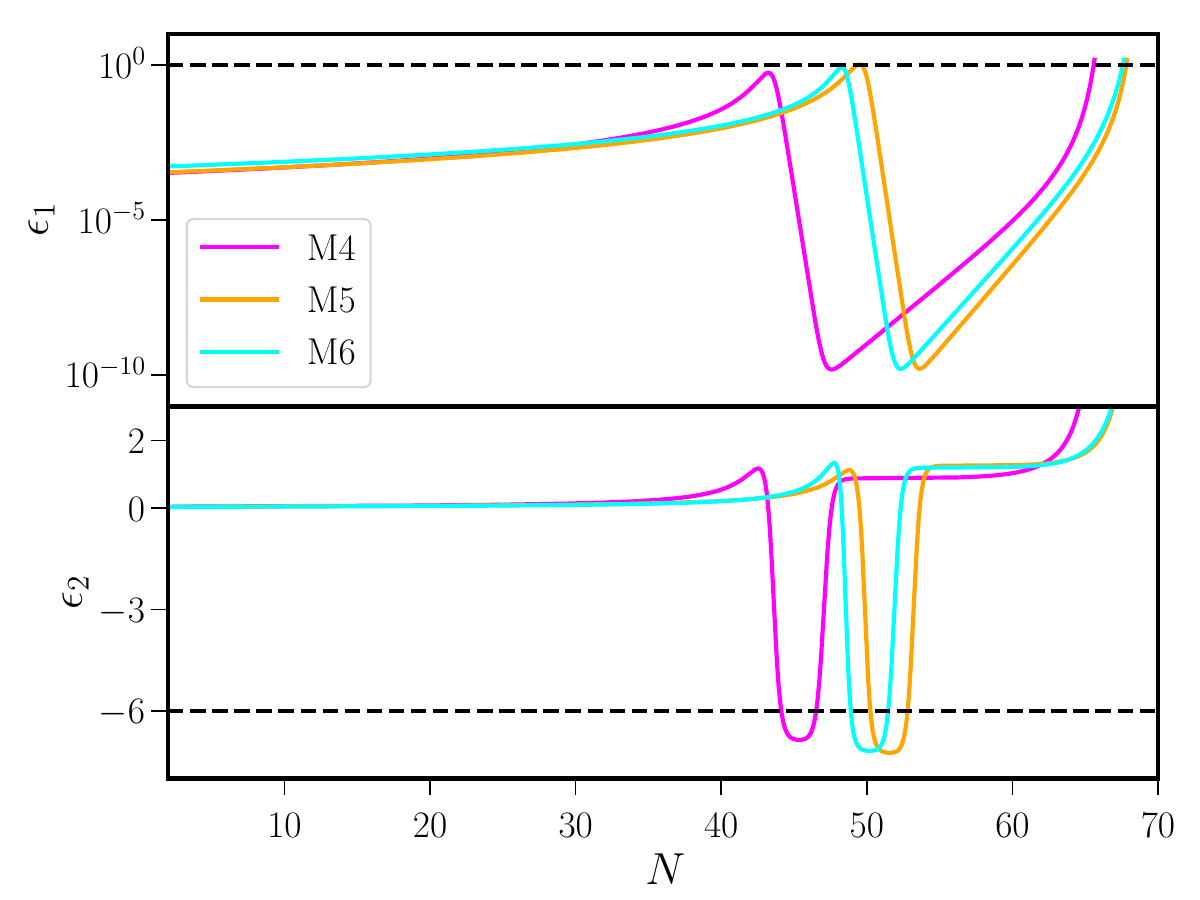}
\caption{The behavior of the first two slow roll parameters $\epsilon_1$ (on top) 
and $\epsilon_2$ (at the bottom) have been plotted for the models M1 to M3 (on the 
left) and for M4 to M6 (on the right).
In these plots, we have indicated the values of $\epsilon_1=1$ and $\epsilon_2
=-6$ (in dashed black).
We should point out that, when $\epsilon_1$ crosses unity, inflation is either 
interrupted (provided $\epsilon_1$ returns to a value smaller than one soon 
after) or terminated (if it does not).
Note that the value of $\epsilon_2=-6$ corresponds to the case wherein $V_\phi$ 
vanishes identically, a point we have discussed in the text.}\label{fig:b-srp}
\end{figure}
In Figure~\ref{fig:b-srp}, we have plotted the evolution of the first two 
slow roll parameters in the six models~M1 to~M6.
Note that all the models permit a short epoch wherein the first slow roll 
parameter~$\epsilon_1$ decreases rapidly and the second parameter $\epsilon_2$ 
remains nearly a constant.
In fact, from Eq.~\eqref{eq:eom-sf}, it is straightforward to establish that,
when a slowly rolling field approaches a regime wherein $V_\phi\simeq 0$, 
$\dot{\phi}$ will begin to behave as $a^{-3}$.
This implies that the first slow roll parameter behaves as $\epsilon_1
\propto a^{-6}$, thereby leading to $\epsilon_2 \simeq -6$ during the epoch.
It should be clear from the figure that the first two slow roll parameters 
indeed roughly exhibit such a behavior during the epoch of ultra slow roll 
in all the models of interest.
Interestingly, we find that, in some cases, inflation is briefly interrupted
for about an $e$-fold or so (a scenario dubbed as punctuated inflation; in
this context, see Refs.~\cite{Jain:2008dw,Jain:2009pm,Ragavendra:2020sop}) 
before the epoch of ultra slow roll sets in, and inflation is eventually 
terminated at a later stage when the first slow roll parameter crosses
unity again.


\subsection{Scalar and tensor power spectra in ultra slow roll inflation}

Let us now discuss the numerical evaluation of the scalar and tensor power 
spectra (in this context, see Refs.~\cite{Adams:2001vc,Mortonson:2009qv,
Hazra:2012yn,Agocs:2019vyk,Handley:2019anl,Ragavendra:2020old,
Ragavendra:2020sop}).
As in the case of the background, we shall work with $e$-folds $N$ as the 
independent time variable. 
In such a case, the differential equations~\eqref{eq:de-p} governing the 
perturbations can be expressed as follows:
\begin{subequations}\label{eq:de-p-N}
\begin{eqnarray}
f^{k}_{NN}+\l(3-\epsilon_1+\epsilon_2\r)\,f^{k}_{N} 
+ \l(\f{k}{a\,H}\r)^2\,f^k\!\!\! &=&\!\!\! 0,\\
g^{k}_{NN}+\l(3-\epsilon_1\r)\,g^{k}_{N} + \l(\f{k}{a\,H}\r)^2\,g^k\!\!\! 
&=&\!\!\! 0,
\end{eqnarray}
\end{subequations}
where, for convenience, we have denoted $(f_{k},g_k)$ as $(f^{k},g^k)$.

During inflation, the initial conditions on the perturbations are imposed 
when the physical wavelengths~$a/k$ are well inside the Hubble 
radius~$H^{-1}$, i.e. when $(a\, H)/k \ll 1$ or, equivalently, when
$k/(a\, H) \gg 1$.
In fact, to be precise, the conditions are to be imposed on the scalar and
tensor perturbations when $k \gg \sqrt{z''/z}$ and $k\gg \sqrt{a''/a}$,
respectively~\cite{Hazra:2012yn,Ragavendra:2020old}.
It should be clear from Eqs.~\eqref{eqs:ms} that, in such a limit, the $k^2$
term will dominate and, as a result, the Mukhanov-Sasaki variables $v_{k}$ 
and $u_{k}$ will exhibit oscillatory behavior, i.e. they behave in the same 
manner as they would in Minkowski spacetime.
Such a behavior should not be surprising.
The quantities $\sqrt{z''/z}$ and $\sqrt{a''/a}$ roughly determine the scale
of curvature of the inflationary background and the limits $k \gg \sqrt{z''/z}$ 
and $k\gg \sqrt{a''/a}$ correspond to the domain wherein the physical wavelengths 
are much smaller than the curvature scale.
In such a domain, the initial conditions imposed on the Mukhanov-Sasaki 
variables $v_{k}$ and $u_{k}$ are given by
\begin{subequations}
\begin{eqnarray}
v_k(\ei) \!\!\! &=& \!\!\! u_k(\ei) = \f{1}{\sqrt{2\,k}}\,\mathrm{e}^{-i\,k\,\ei},\\
v'_k(\ei) \!\!\! &=& \!\!\! u'_k(\ei) = -i\,\sqrt{\f{k}{2}}\,\mathrm{e}^{-i\,k\,\ei},
\end{eqnarray}
\end{subequations}
where $\ei$ denotes an adequately early time when the conditions are imposed.
The vacuum state that is associated with such initial conditions is popularly 
known as the Bunch-Davies vacuum~\cite{Bunch:1978yq}, as had mentioned earlier. 
In terms of the mode functions $f_k$ and $g_k$, the above initial conditions
correspond to 
\begin{subequations}\label{eq:ic-fg}
\begin{eqnarray}
f^k(\Ni) \!\!\! &=& \!\!\!  \f{1}{\sqrt{2\,k}\,z(\Ni)},\quad
f_N^k(\Ni) =  -\f{1}{\sqrt{2\,k}\;z(\Ni)}\,
\l[\f{i\,k}{a(\Ni)\,H(\Ni)}+\f{z_N(\Ni)}{z(\Ni)}\r],\\
g_k(\Ni) \!\!\! &=& \!\!\! \f{\sqrt{2}}{\Mpl}\,\f{1}{\sqrt{2\,k}\,a(\Ni)},\quad
g_N^k(\Ni) = -\f{\sqrt{2}}{\Mpl}\,\f{1}{\sqrt{2\,k}\;a(\Ni)}\,
\l[\f{i\,k}{a(\Ni)\,H(\Ni)}+1\r].\qquad\;
\end{eqnarray}
\end{subequations}
where $z_N=\d z/\d N$ and we have dropped the overall and unimportant phase
factor $\mathrm{exp}\,(-i\,k\,\ei)$ in all the expressions.

Numerically, one finds that it is often sufficient to impose the initial 
conditions when $k \simeq 10^2\,\sqrt{z''/z}$ and $k\simeq 10^2\,\sqrt{a''/a}$ 
on the scalar and tensor perturbations, respectively~\cite{Adams:2001vc,
Mortonson:2009qv,Hazra:2012yn,Agocs:2019vyk,Handley:2019anl,Ragavendra:2020old,
Ragavendra:2020sop}.
With the solutions to background at hand, we can evaluate the coefficients 
in the equations~\eqref{eq:de-p-N} governing the perturbations.
Starting with the initial conditions~\eqref{eq:ic-fg}, we use the fifth 
order Runge-Kutta method~\cite{Fehlberg:1969rk5} to evolve the scalar and 
tensor perturbations until late times.
In simple scenarios involving slow roll inflation, the amplitudes of the 
modes quickly approach a constant value soon after they leave the Hubble 
radius.
Therefore, typically, the spectrum of scalar and tensor perturbations are 
evaluated on super-Hubble scales, say, when $k \simeq 10^{-5}\,\sqrt{z''/z}$
and $k \simeq 10^{-5}\,\sqrt{a''/a}$.
However, in scenarios that permit a period of ultra slow roll, the amplitudes
of the modes which leave the Hubble radius just prior to or during the epoch 
of ultra slow roll inflation can be affected even after they leave the Hubble 
radius\footnote{In fact, if the duration of the ultra slow 
roll phase is, say, $\Delta N_{_{\mathrm{USR}}}$, then the range of wave numbers 
that are affected can be quantified as $\ln\,(k_2/k_1) 
\simeq \Delta N_{_{\mathrm{USR}}}$, where $k_1$ is the wave number that exits 
the Hubble radius at the onset of the ultra slow roll phase. 
This range corresponds to the region around the peak of the scalar power spectrum.
Besides, there is another range of wave numbers that are affected by the phase
of ultra slow roll.
These correspond to wave numbers which leave the Hubble radius a few $e$-folds prior 
to the onset of the ultra slow roll phase (for earlier discussions in this regard, 
see Refs.~\cite{Leach:2000yw, Leach:2001zf,Jain:2007au}; for a more recent discussion, 
see Ref.~\cite{Ragavendra:2020old}).
Over these range of wave numbers, there arises a sharp dip and a rise in the power 
spectra leading to the peak. 
The wave number at the dip, say, $k_\mathrm{dip}$, can be estimated to be 
$k_\mathrm{dip} \simeq \sqrt{3}\,k_1\,\exp\,(-3\,\Delta N_{_{\mathrm{USR}}}/2)$ and 
the range between the dip and the approach to the peak corresponds to $k_\mathrm{dip} 
\lesssim k \lesssim k_1$ (in this regard, see Ref.~\cite{Balaji:2022zur}).}
Due to these reasons, in the models of our interest that lead to an epoch 
of ultra slow roll inflation, we evaluate the scalar and tensor power
spectra close to the end of inflation, well past the epoch of ultra slow 
roll.

\begin{figure}[!t]
\centering
\includegraphics[width=9.5cm]{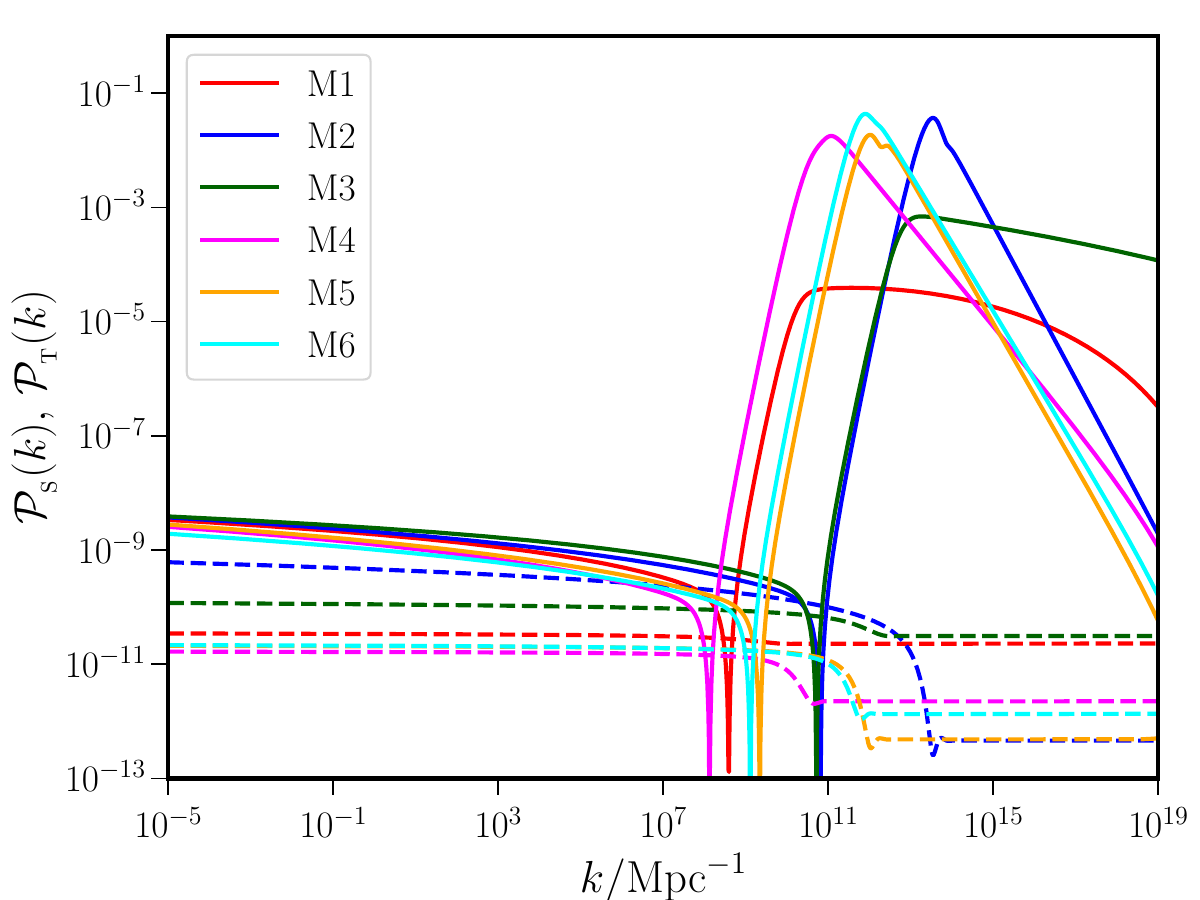}
\caption{The scalar power spectrum $\ps(k)$ (in solid lines) and the tensor
power spectrum $\pt(k)$ (in dashed lines) that arise in the models M1 to M6 
have been plotted for the parameters and initial conditions that we have 
discussed.
As expected, all the scalar power spectra exhibit a strong peak on small 
scales due to the epoch of ultra slow roll inflation that occurs in these 
models.}\label{fig:sps-tps}
\end{figure}
In Figure~\ref{fig:sps-tps}, we have plotted the scalar and tensor power
spectra that arise in the models M1 to M6. 
While the power spectra are nearly scale invariant over large scales, 
clearly, the scalar spectra exhibit a sharp rise in power at small 
scales.
It is interesting to note that, in contrast to the scalar spectra, the 
tensor spectra exhibit a suppression of power over small scales.
Evidently, it is the epoch of ultra slow roll inflation that is responsible 
for the enhancement in the scalar power at small scales.
Apart from the values of the parameters characterizing the potential,
the spectra are also determined by the choice of the quantity~$N_\ast$,
which is the time when the pivot scale of $k_\ast=0.05\,\mpcinv$ leaves 
the Hubble radius.
Our choices for the values of the parameters of the potential and $N_\ast$ 
were guided by the following conditions: (1)~there should be significant 
amplification of power on small scales and (2)~the scalar and tensor power 
spectra should be consistent with the constraints from the CMB on large 
scales.
In Table~\ref{tab:ns-r}, we have listed our choices of $N_\ast$ and the
values of the scalar spectral index~$\ns$ and the tensor-to-scalar 
ratio~$r$ at the pivot scale in the models M1 to M6. 
If one compares these values of $\ns$ and $r$ with the constraints from the CMB 
(viz. that $\ns = 0.9649 \pm 0.0042$ and $r < 0.036$~\cite{Planck:2015sxf,
Planck:2018jri,BICEP:2021xfz}, which we had quoted in the introductory section),
it is clear that, apart from the model~M3, all the other models are rather
inconsistent with the CMB data.
Either $\ns$ is at least 4-$\sigma$ away from the mean value and/or $r$ 
is larger than the strongest bound from the CMB observations.
Even in the case of M3, we had to choose a large value of $N_\ast$ (larger
than the typical value of $50 <N_\ast < 60$) to achieve a reasonable level 
of consistency with the CMB data. 
We should mention that the values of $r$ that we have obtained in these models
can be roughly understood by the behavior of the potentials over large values 
of the fields. 
For instance, in M1, the potential behaves as $V(\phi) \sim \phi^4/(\phi^2)^2$
over large~$\phi$, which is asymptotically flat. 
This leads to a low value of~$r$, which is typical for a plateau-type potential. 
On the other hand, in M2, we have $V(\phi) \sim \phi^2$ over large values of~$\phi$. 
Hence, in such a case, we obtain a value of $r \sim 0.2$, as can be expected from 
a quadratic potential.
\begin{table}[!t]
\centering
\begin{tabular}{|c|c|c|c|c|c|c|}
\hline
Models & M1 & M2 & M3 & M4 & M5 & M6\\ 
\hline
$N_\ast$ & $50$ & $55$ & $70$ & $50$ & $50$ & $50$ \\
\hline
$\ns$ & $0.945$ & $0.946$ & $0.956$ & $0.933$ & $0.936$ & $0.940$ \\
\hline
$r$ & $0.015$ & $0.244$ & $0.041$ & $0.011$ & $0.012$ & $0.017$\\
\hline
\end{tabular}
\vskip 5pt
\caption{The scalar spectral index~$\ns$ and tensor-to-scalar 
ratio~$r$, evaluated at the pivot scale of $k_\ast=0.05\,\mpcinv$, 
are tabulated for the models M1 to M6.
We have also listed the values of $N_\ast$ for the different models that we 
have worked with to achieve the $\ns$ and $r$ mentioned above.
Note that, apart from the case of M3, the other models lead to $\ns$ and $r$
that are fairly inconsistent with the CMB data.}\label{tab:ns-r}
\end{table}


\subsection{Reverse engineering desired potentials}\label{subsec:rcp}

We have seen that, while the models M1 to M6 lead to an epoch of ultra slow
roll inflation and to a strong peak in the scalar power spectrum on small 
scales, they generate primordial spectra that are not consistent with the CMB 
data on large scales.
Actually, the hurdle seems to crop up in most single field models involving 
the canonical scalar field wherein ultra slow roll inflation is achieved with
the aid of a (near) inflection point in the potential. 
When one attempts to modify the parameters so that the spectra are consistent
with the CMB data, two challenges are encountered.
Either there arises a prolonged duration of inflation or the power at small
scales is not enhanced significantly.
Moreover, in such models, it proves to be difficult to shift the location of 
the peak in the power spectrum.
In particular, when the peak is higher and is closer the CMB scales, the 
inconsistency with the CMB data turns out to be greater.

A method to overcome these difficulties in canonical, single field models of 
inflation is to reverse engineer potentials that simultaneously lead to 
spectra that are consistent with the constraints from the CMB on large scales 
and produce significant power on small scales (for discussions in this context, 
see Refs.~\cite{Hertzberg:2017dkh,Byrnes:2018txb,Motohashi:2019rhu,
Ragavendra:2020sop,Franciolini:2022pav}).
In order to do so, one begins with the desired functional form of the first 
slow roll parameter~$\epsilon_1(N)$, i.e. one that admits a brief period of 
ultra slow roll after an initial epoch of slow roll, before the termination
of inflation.
Given $\epsilon_1(N)$, from the definition~\eqref{eq:f-srp} of the first slow 
roll parameter, we can express the time evolution of the scalar field~$\phi(N)$ 
and the Hubble parameter~$H(N)$ in terms of the following integrals:
\begin{subequations}\label{eqs:phiHN}
\begin{eqnarray}
\phi(N)\!\!\! &=&\!\!\! \phi_\mathrm{i} 
-\Mpl\, \int^N_{N_\mathrm{i}}\d N\,\sqrt{2\,\epsilon_1(N)},\\
H(N)\!\!\!  
&=&\!\!\!  H_\mathrm{i}\;\mathrm{exp}\l[-\int^N_{N_\mathrm{i}}\d N\,
\epsilon_1(N)\r],
\end{eqnarray}
\end{subequations}
where~$\phi_\mathrm{i}$ and~$H_\mathrm{i}$ are the values of the scalar field
and the Hubble parameter specified at some initial $e$-fold~$\Ni$.
In other words, if the initial conditions~$\phi_\mathrm{i}$ and~$H_\mathrm{i}$ 
are provided, the functional form of~$\epsilon_1(N)$ completely determines~$\phi(N)$ 
and~$H(N)$.
It is easy to show that, using the Friedmann equation~\eqref{eq:ffe} and the 
definition of the first slow roll parameter, the potential $V(N)$ can be
expressed in terms of $H(N)$ and $\epsilon_1(N)$ as
\begin{equation}
V(N) = \Mpl^2\,H^2(N)\,\l[3-\epsilon_1(N)\r].\label{eq:VN}
\end{equation}
Therefore, using $\epsilon_1(N)$ and $H(N)$, we can construct $V(N)$ as well.
Also, as should be clear from the above equation, we require $H_\mathrm{i}$ to 
determine the overall amplitude of the potential.
With $\phi(N)$ and $V(N)$ at hand, we can then construct the potential $V(\phi)$ 
parametrically.
We should add that, once we have $\phi(N)$ and $H(N)$, all the other background 
quantities can be computed using them.

We shall consider the following form for $\epsilon_1(N)$ which leads to an 
intermediate epoch of ultra slow roll inflation for suitable choice of the
parameters involved~\cite{Ragavendra:2020sop}:
\begin{equation}
\epsilon_1(N) 
= \l[{\epsilon_{1a}\,\l(1+\epsilon_{2a}\,N\r)}\r]\,
\l[1 - {\mathrm{tanh}}\l(\f{N - N_1}{\Delta N_1}\r)\r] 
+ \epsilon_{1b} + \mathrm{exp}\l(\f{N - N_2}{\Delta N_2}\r).\label{eq:rs1}
\end{equation}
Given such a form for $\epsilon_1$, apart from arriving at the potential describing
the background, we can determine the coefficients in Eqs.~\eqref{eq:de-p-N} and evolve 
the perturbations from the initial conditions~\eqref{eq:ic-fg} to eventually arrive
at the scalar and tensor power spectra.
Let us clarify a few points regarding the parameters $(\epsilon_{1a}, \epsilon_{2a}, 
N_1,  \Delta N_1, \epsilon_{1b}, N_2, \Delta N_2)$ that appear in the above expression 
for $\epsilon_1(N)$.
The first term within the square brackets containing the parameters $\epsilon_{1a}$ 
and $\epsilon_{2a}$ leads to an initial epoch of slow roll inflation.
We can choose these two parameters suitably so that the resulting scalar and tensor
power are consistent with the CMB data on large scales. 
The hyperbolic tangent function  in the first term helps us achieve the epoch of
ultra slow roll, which sets in at~$N_1$ when counted from an initial $e$-fold 
$N_{\mathrm{i}}$ and the transition from slow roll to ultra slow roll inflation
occurs over a duration~$\Delta N_1$.
The parameter $\epsilon_{1b}$ corresponds to the value of the first slow roll parameter 
at the end of the regime of ultra slow roll.
The exponential function in the last term leads to a rise in the first slow parameter
leading to an end of inflation.
Inflation ends at the $e$-fold $N_2$ and the parameter $\Delta N_2$ regulates the 
duration between the end of ultra slow roll and the termination of inflation.
Apart from the values of these parameters, we need to provide the initial value of 
$H_\mathrm{i}$ which influences the overall amplitude of the potential and hence 
the power spectra.
We work with the following values of the parameters involved: $H_{\mathrm{i}} = 
8.5 \times 10^{-6}\,\Mpl$, $\epsilon_{1a} = 7.38\times10^{-5}$, $\epsilon_{2a} 
= 9 \times 10^{-2}$, $\epsilon_{1b} = 1.7\times 10^{-10}$, $N_2 = 72$ and 
$\Delta N_2 = 5.5 \times 10^{-1}$.
We vary $N_1$ and $\Delta N_1$ over the ranges $[41,55]$ and $[0.31,0.32]$, 
respectively, to illustrate their effects on the background quantities and 
hence on the observables.

\begin{figure}[!t]
\centering
\hskip -46.2pt
\includegraphics[width=9.50cm]{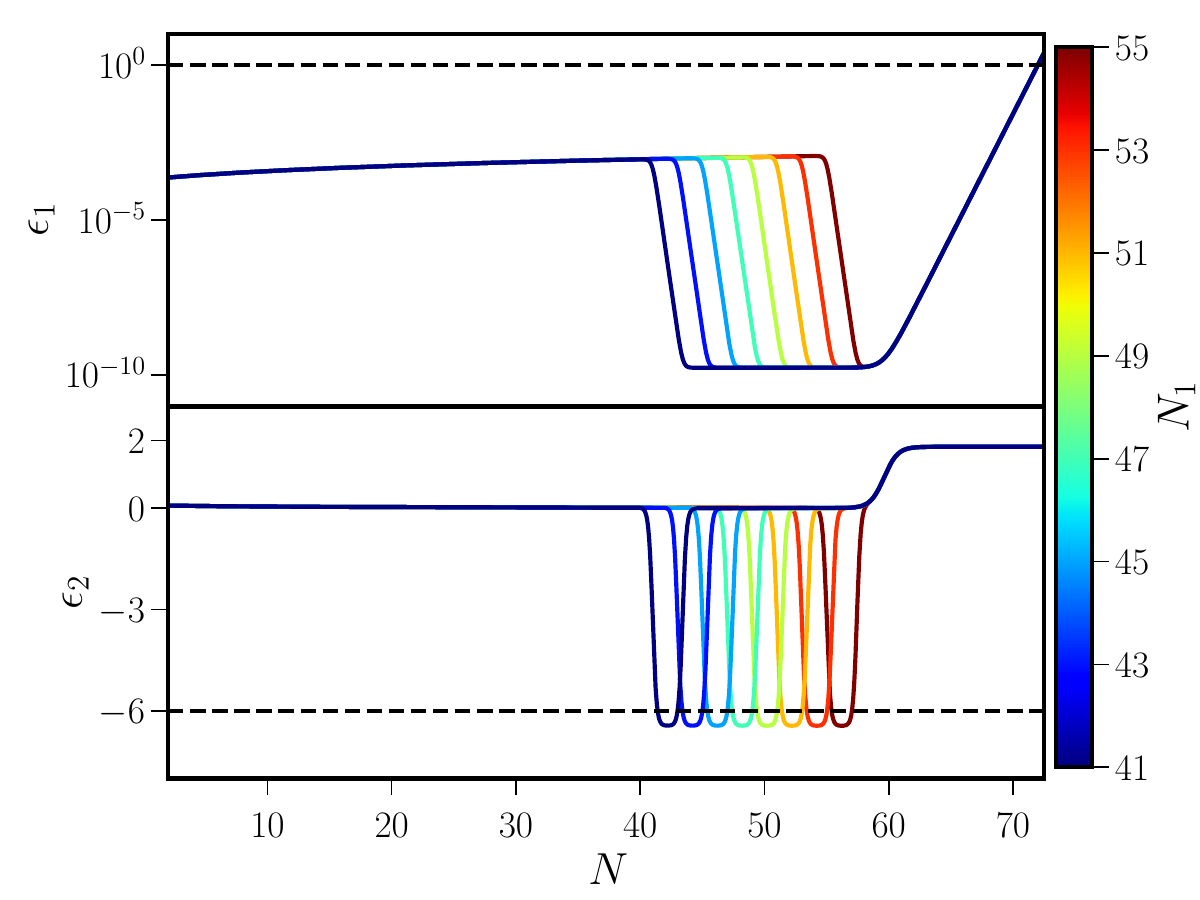}
\includegraphics[width=9.50cm]{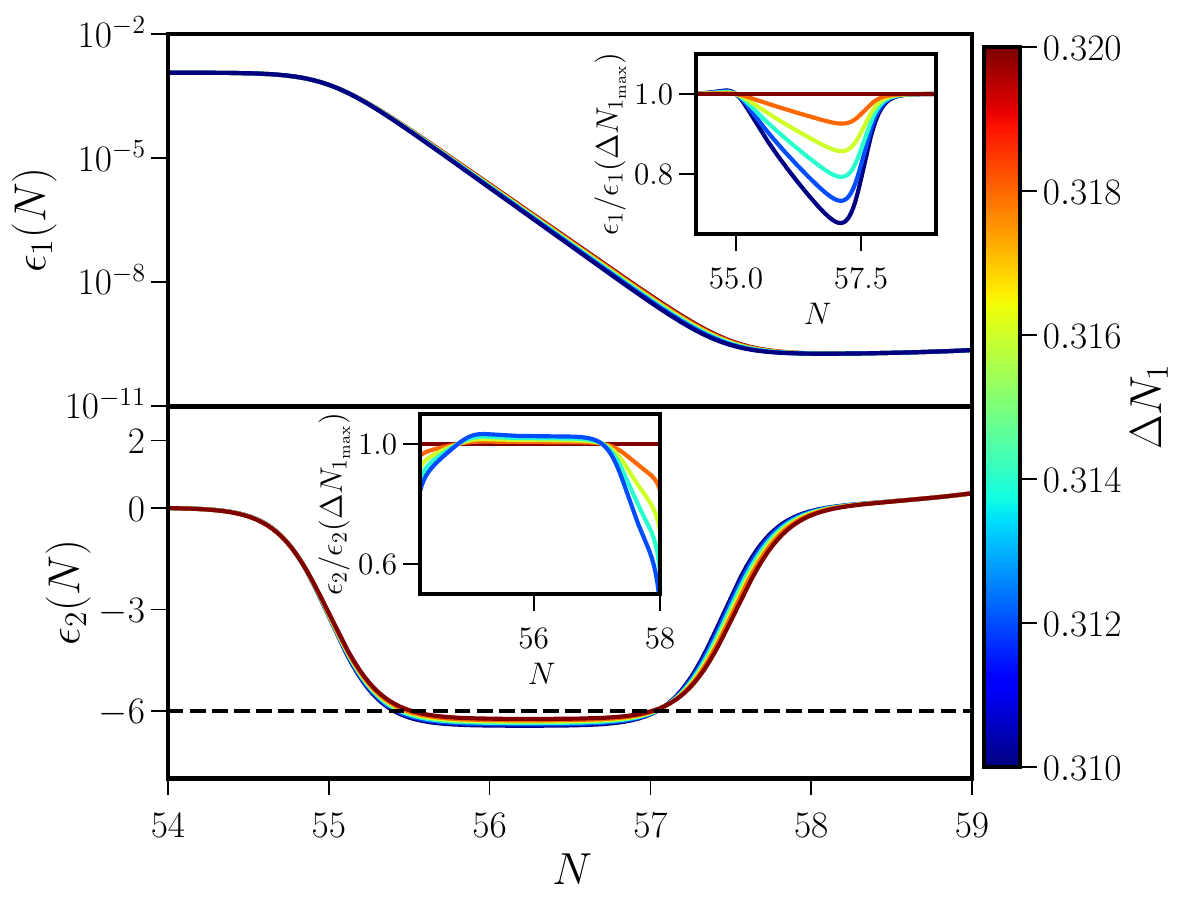}
\caption{The behavior of the first two slow roll parameters $\epsilon_1$ 
(on top) and $\epsilon_2$ (at the bottom) are presented for the scenario 
wherein $\epsilon_1(N)$ is given by Eq.~\eqref{eq:rs1}. 
We have plotted the slow roll parameters over a range of $N_1$ varied in steps 
of two $e$-folds (on the left) and $\Delta N_1$ (on the right).
While the effects of the variation in $N_1$ are evident in the figure, the effects 
of the variation of $\Delta N_1$ are not so visible. 
Hence, we have included insets (on the right) to highlight the minor differences 
that arise in the parameters.
Clearly, the slow roll parameters in the reconstructed scenario behave broadly 
in the same fashion as in the six, specific inflationary models we have considered.}
\label{fig:b-srp-rs}
\end{figure}
\begin{figure}[!t]
\centering
\hskip -46.2pt
\includegraphics[width=9.5cm]{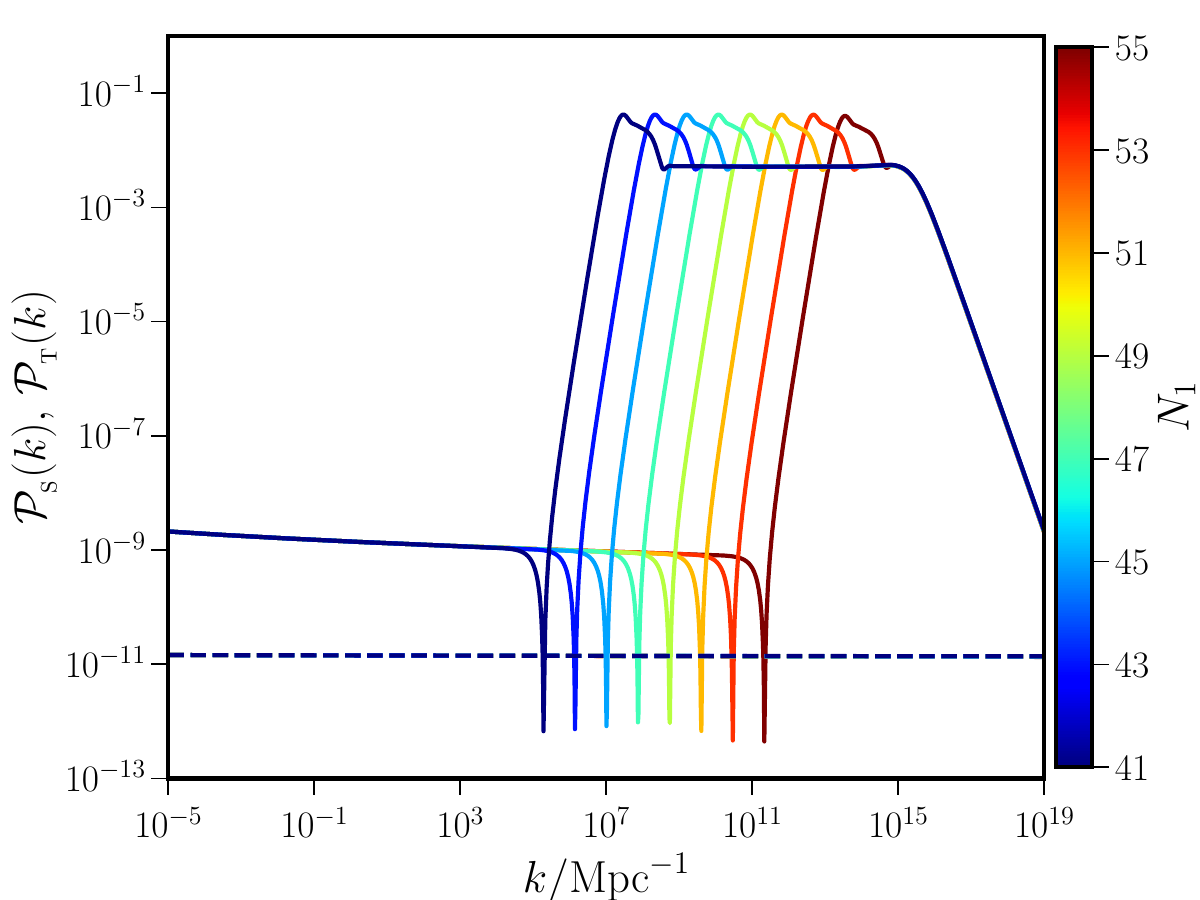}
\includegraphics[width=9.5cm]{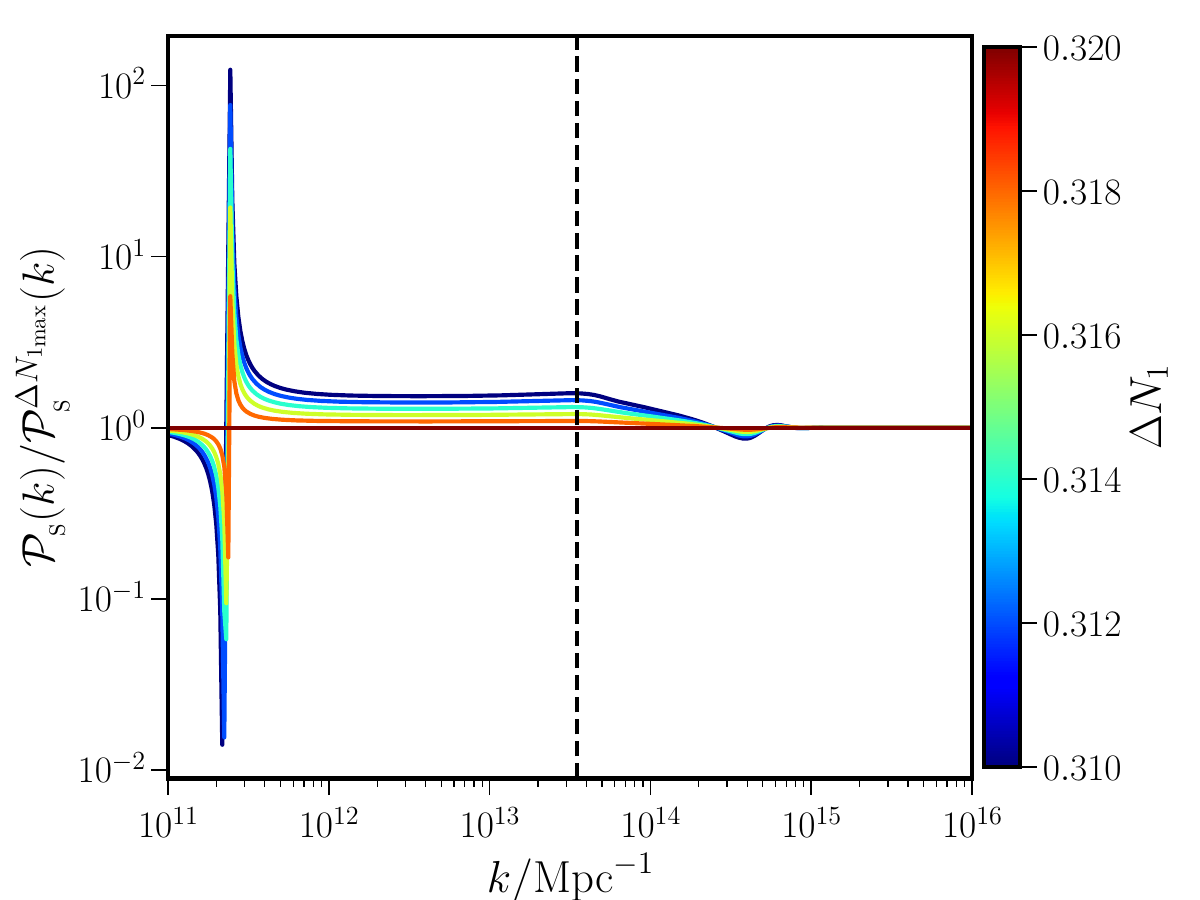}
\caption{The scalar power spectra $\ps(k)$ (in solid lines) and the tensor
power spectra $\pt(k)$ (in dashed lines) that arise in the reconstructed 
scenario have been plotted for a range of~$N_1$ (on top) and~$\Delta N_1$ (at 
the bottom).
It should be clear (from the figure on top) that, earlier the onset of ultra 
slow roll, broader is the peak in the scalar power spectra.
Also, for the effects due to the variation in $\Delta N_1$ to be distinguishable, 
we have illustrated (in the figure at the bottom) the relative change in the scalar 
power spectrum with respect to the spectrum corresponding to the maximum value of 
$\Delta N_1$ that we have considered (viz. $\Delta N_1 = 0.320$). 
We have also indicated the wave number where the peak is located in the spectra
(in dashed black).
We find that, for a given value of~$N_1$ (in this case, $N_1=41$), a reduction 
in $\Delta N_1$ leads to an increase in the amplitude of power over the range 
containing the rise and the peak in the spectrum.}\label{fig:ps-rs}
\end{figure}

In Figures~\ref{fig:b-srp-rs} and~\ref{fig:ps-rs}, we have plotted the evolution of 
the first two slow roll parameters and the resulting scalar and tensor power spectra.
In fact, we have plotted these quantities over the range of $N_1$ and $\Delta N_1$
that we mentioned above.
Note that the first two slow roll parameters broadly behave in the same manner as 
they did in the case of the models M1 to M6.
For the values of the parameters we have worked with, the amplitude of the scalar
power spectrum at the pivot scale of $k_\ast = 0.05\,\mpcinv$ proves to be 
$A_{_{\mathrm{S}}} = 2.10\times 10^{-9}$.
Also, we find that the scalar spectral index and the tensor-to-scalar ratio are
given by $\ns = 0.968$ and $r = 1.18\times 10^{-3}$, which are consistent with the 
CMB data.
It is the additional parameters that are available in the parametrization of 
$\epsilon_{1}(N)$ which permit us to arrive at spectra that are consistent with
the CMB observations.
We should also point out that, surprisingly, in complete contrast to the scalar
power spectrum, the tensor power spectrum is nearly scale invariant and does
not contain any feature at all.

\section{Formation of PBHs in the radiation dominated epoch}\label{sec:pbhs}

In this section, we shall discuss the extent of PBHs that are formed due to
the enhanced power in the scalar spectra on small scales.
Specifically, we shall be interested in calculating the function $\fpbh(M)$,
which describes the fractional contribution of PBHs to the dimensionless 
parameter~$\Omega_\mathrm{c}$ that describes the density of cold matter 
{\it today},\/ as a function of the mass~$M$ of the PBHs.

Recall that scales with wave numbers $k \gtrsim 10^{-2}\,{\rm Mpc}^{-1}$ 
renter the Hubble radius during the radiation dominated epoch.
Once these scales are inside the Hubble radius, the perturbations in the matter
density at the corresponding scales collapse to form structures.
Let the density contrast in matter be characterized by the quantity~$\delta$.
The matter power spectrum~$P_\delta(k)$ during the radiation dominated epoch
is related to the scalar power spectrum $\ps(k)$ generated during inflation
though the relation~\cite{Chongchitnan:2006wx}
\begin{equation}
P_\delta(k) 
= \f{16}{81}\,\l(\f{k}{aH}\r)^4\,\ps(k).\label{eq:pm-k}
\end{equation}
As we shall see, the fraction of PBHs formed when matter collapses after the
modes reenter the Hubble radius is determined by the quantity $\sigma^2(R)$, 
which represents the variance in the spatial density fluctuations that has 
been smoothed over a length scale~$R$.
The variance~$\sigma^2(R)$ smoothed with the aid of the window function
$W(k\,R)$ is defined as~\cite{Chongchitnan:2006wx}
\begin{equation}
\sigma^2(R) 
= \int_{0}^{\infty} \f{\d k}{k}\,P_\delta(k)\, W^2(k\,R).\label{eq:sigma2}
\end{equation}
In our discussion below, we shall work with the following Gaussian form for
the window function:~$W(k\,R) = \mathrm{exp}\,[-(k^2\,R^2)/2]$.

\begin{figure}[!t]
\centering
\includegraphics[width=9.5cm]{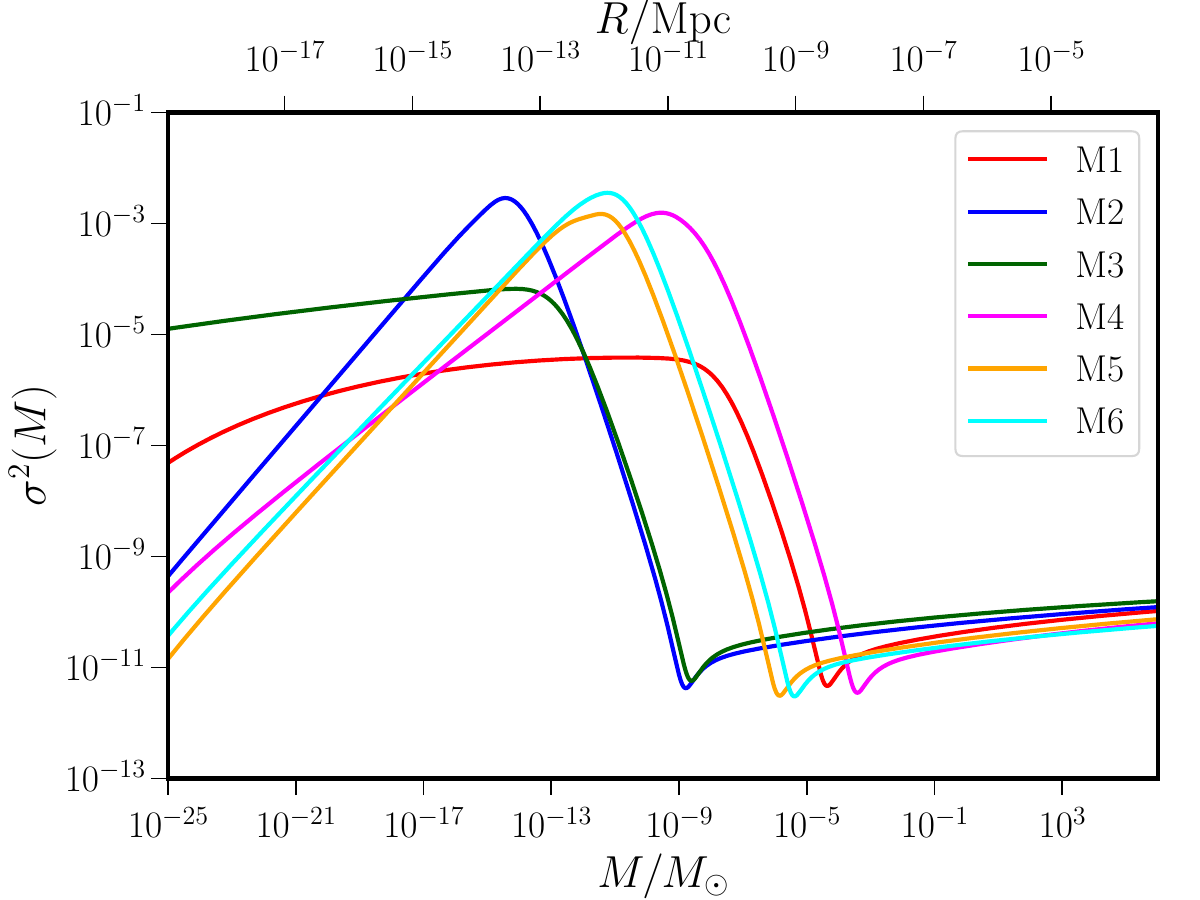}
\caption{The variance $\sigma^2(M)$ of the density fluctuations has been plotted 
as a function of the mass~$M$ of the PBHs for the inflationary models~M1 to~M6.}\label{fig:variance}
\end{figure}

As we mentioned, our aim will be to arrive at the number of PBHs formed as 
a function of their mass.
To do so, we need to relate their mass $M$ of the PBHs to the smoothing 
scale~$R$ that we have introduced through the window function.
If~$M_{_\mathrm{H}}$ represents the mass within the Hubble radius~$H^{-1}$ 
at a given time, it seems reasonable to assume that a certain fraction of the
total mass goes on to form PBHs.
Let the parameter $\gamma$ reflect the efficiency of the collapse of the 
density contrast to form PBHs.
In such a case, when a scale with wave number~$k$ reenters the Hubble radius,
we can express the mass of the PBHs formed to be $M=\gamma\, M_\mathrm{H}$. 
Since no other scale is present, it seems reasonable to set $k = R^{-1}$ and
use the fact that $k=a\,H$ when the perturbation with wave number~$k$ reenters 
the Hubble radius, to obtain the relation between~$R$ and~$M$.
It can be easily shown that $R$ and $M$ are related as follows:
\begin{equation}
R=\f{2^{1/4}}{\gamma^{1/2}}\,
\l(\f{g_{\ast,k}}{g_{\ast,\mathrm{eq}}}\r)^{1/12}\,
\l(\f{1}{k_\mathrm{eq}}\r)\,
\l(\f{M}{M_\mathrm{eq}}\r)^{1/2},\label{eq:R-M}
\end{equation}
where $k_\mathrm{eq}$ denotes the wave number that reenters the Hubble radius 
at the time of radiation-matter equality, and the quantity $M_\mathrm{eq}$ 
represents the mass within the Hubble radius at equality.
Moreover, the quantities $g_{\ast,k}$ and $g_{\ast,\mathrm{eq}}$ denote the 
effective number of relativistic degrees of freedom at the times of PBH 
formation and radiation-matter equality, respectively.
One finds that $M_\mathrm{eq} = 5.83\times 10^{47}\, \mathrm{kg}$ and using
this result, the above relation between $R$ and $M$ can be expressed in terms 
of the solar mass $M_\odot$ as follows:
\begin{equation}
R=4.72\times10^{-7}\,\l(\f{\gamma}{0.2}\r)^{-1/2}\,
\l(\f{g_{\ast,k}}{g_{\ast,\mathrm{eq}}}\r)^{1/12}\,
\l(\f{M}{M_\odot}\r)^{1/2}\,\mathrm{Mpc}.\label{eq:R-Ms}
\end{equation}
Given an inflationary scalar power spectrum $\ps(k)$, we can make use of the
relations~\eqref{eq:pm-k}, \eqref{eq:sigma2} and \eqref{eq:R-Ms} to compute
the quantity~$\sigma^2(M)$.
In Figure~\ref{fig:variance}, we have illustrated the variance $\sigma^2(M)$
corresponding to the scalar power spectra that arise in the inflationary 
models~M1 to~M6.
As can be expected, the variance exhibits peaks over smoothing scales corresponding 
to the wave numbers containing the peaks in the scalar power spectra, i.e. at
$R \simeq k^{-1}$ (cf. Figure~\ref{fig:sps-tps}).

To calculate the number of PBHs produced, we shall assume that the density 
contrast~$\delta$ is a Gaussian random variable described by the probability 
density
\begin{equation}
{\cal P}_M(\delta) 
= \f{1}{\sqrt{2\,\pi\,\sigma^2(M)}}\; 
\mathrm{exp}{\l[-\f{\delta^2}{2\,\sigma^2(M)}\r]}.\label{eq:P-d}
\end{equation}
The quantity~$\sigma^2(M)$ in this expression is the variance of the density
fluctuations smoothed over the scale~$R$ that we introduced above, with $R$ 
and $M$ being related by Eq.~\eqref{eq:R-Ms}.
Let us further assume that perturbations with a density contrast beyond a 
certain threshold, say, $\delta_\mathrm{c}$, go on to form PBHs.
In such a case, the fraction, say, $\beta(M)$, of the density fluctuations 
that collapse to form PBHs is described by the integral (in this context, 
see the reviews~\cite{Carr:2016drx,Carr:2018rid,Sasaki:2018dmp,Carr:2020xqk})
\begin{equation}
\beta(M) = \int^{1}_{_{\delta_\mathrm{c}}} \d\delta\, \cP_M(\delta)
\simeq \f{1}{2}\,\l\{1-\mathrm{erf}\,\l[\f{\delta_\mathrm{c}}
{\sqrt{2\,\sigma^2(M)}}\r]\r\},\label{eq:b-pbh}
\end{equation}
where $\mathrm{erf}(z)$ denotes the error function.
We should stress here that the quantity $\beta(M)$ is exponentially 
sensitive to the choice of the threshold value of the density 
contrast~$\delta_\mathrm{c}$, as much as it is to variance~$\sigma^2(M)$.
Importantly, the choice of $\delta_\mathrm{c}$ is not unique and, actually,
it can depend on the amplitude of the perturbation at a given scale 
as well as on the epoch of formation of the PBHs (for 
early discussions in this context, see Refs.~\cite{Carr:1975qj,Green:2004wb};
for some recent discussions, see Refs.~\cite{Sasaki:2018dmp,Germani:2018jgr,
Germani:2019zez,Escriva:2019nsa,Escriva:2019phb,Escriva:2020tak}).
In our results that we present below, we shall work with $\delta_{\mathrm{c}}
=0.45$ (for further details in this regard, see 
Refs.~\cite{Nakama:2013ica,Musco:2004ak,Young:2019osy}, especially 
Ref.~\cite{Young:2019osy}).

\begin{figure}[!t]
\centering
\includegraphics[width=9.50cm]{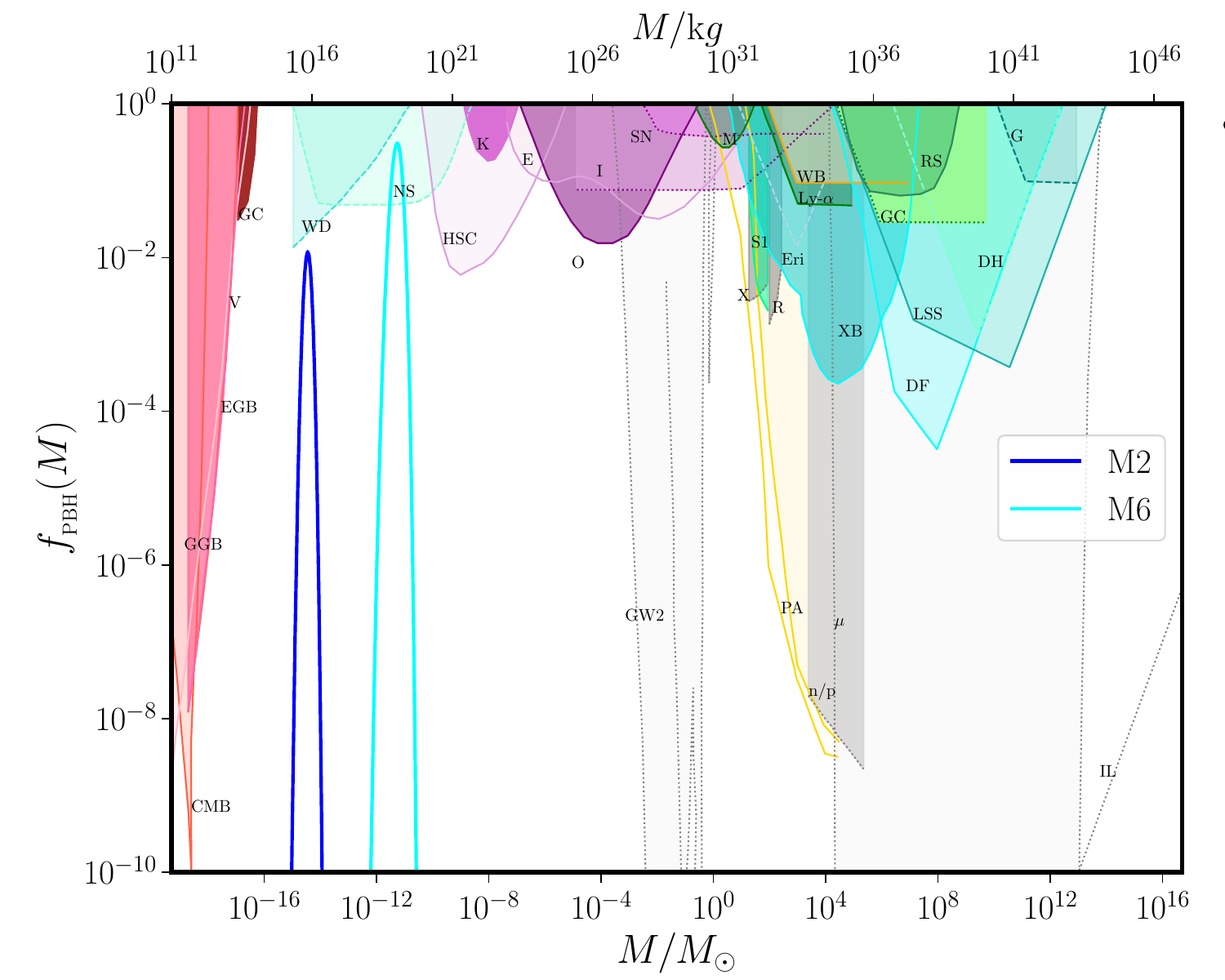}
\caption{The fraction of PBHs~$\fpbh(M)$ that constitute the cold dark 
matter density in the current universe has been plotted for the 
inflationary models of interest.
We had mentioned that the quantity $\fpbh(M)$ is very sensitive to the 
value of~$\delta_{\mathrm{c}}$.
We find that, for the choice of~$\delta_{\mathrm{c}}=0.45$, it is only the
inflationary models M2 and M6 that lead to substantial formation of PBHs.
In the figure, we have also included the constraints from the different 
observations, corresponding to a monochromatic spectrum
of PBHs.}\label{fig:fpbh-im}
\end{figure}
\begin{figure}[!t]
\centering
\hskip -46.2pt
\includegraphics[width=9.50cm]{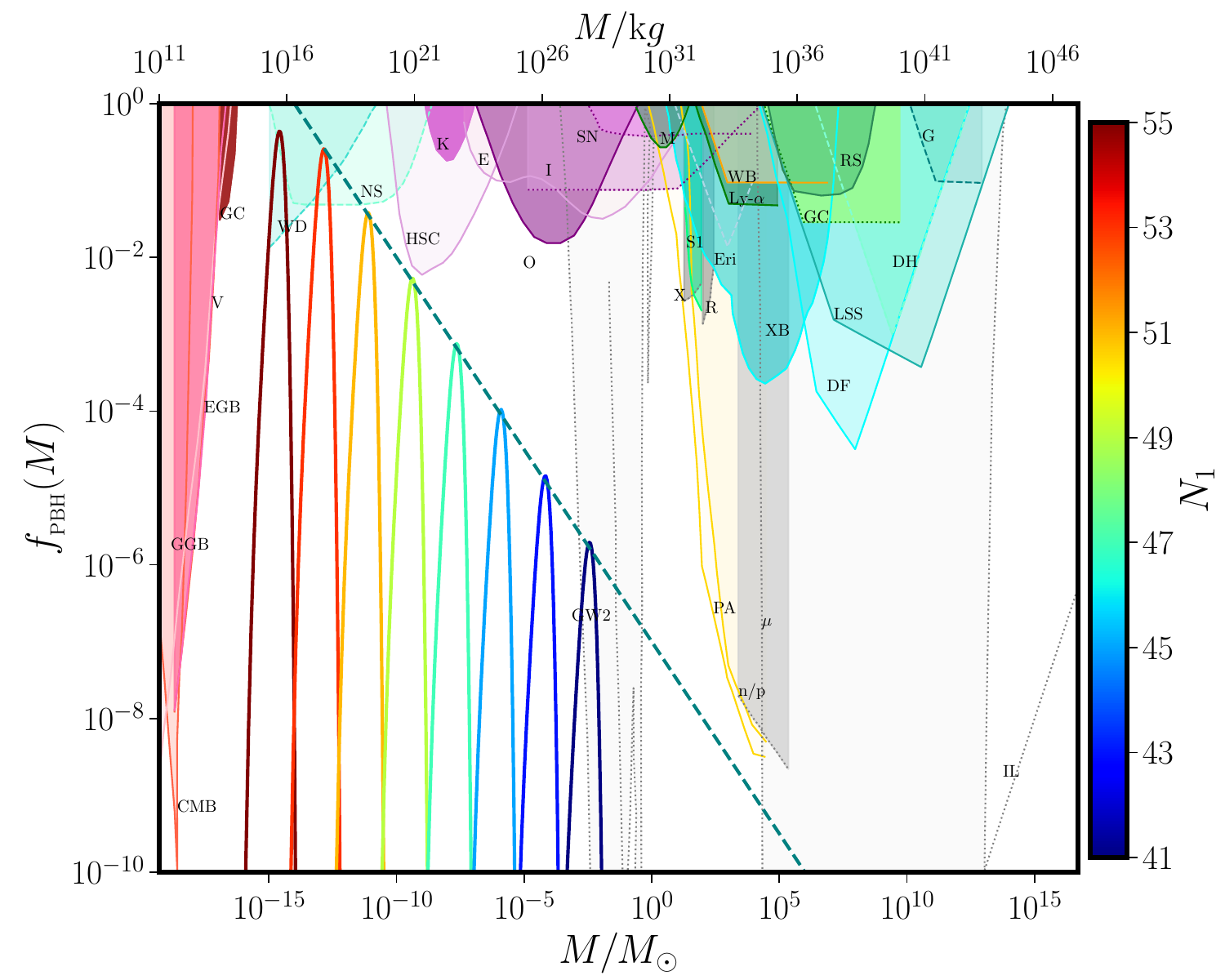}
\includegraphics[width=9.50cm]{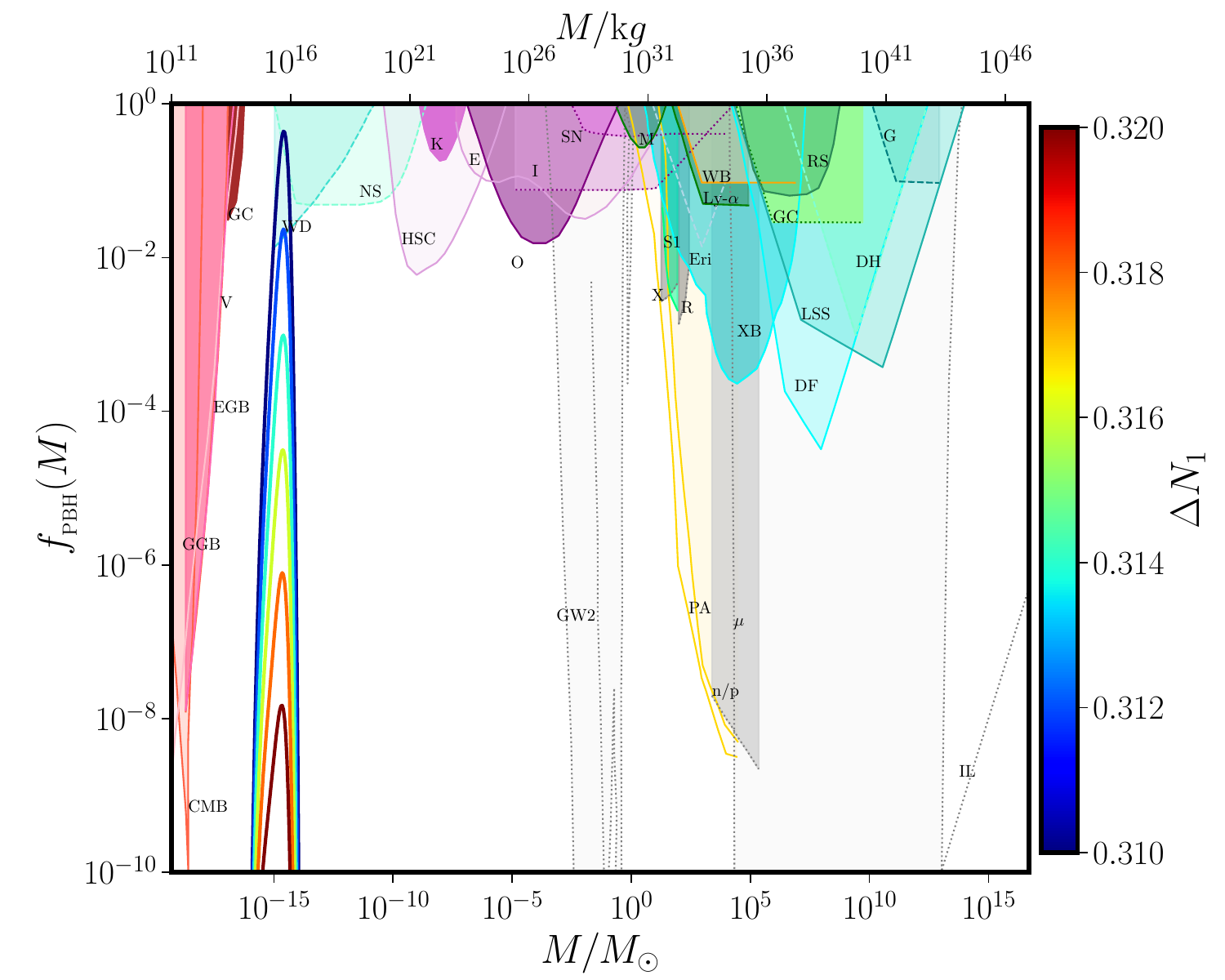}
\caption{The quantity $\fpbh(M)$ that arises in the reconstructed 
scenario has been plotted for different values of $N_1$ (on the left) 
and $\Delta N_1$ (on the right).
As $N_1$ is varied, we find that the peaks of $\fpbh(M)$ behave as~$M^{-1/2}$
(indicated in dashed teal) for reasons explained in the text.
The increase in $\Delta N_1$, with a fixed value of $N_1=41$, results in drastic 
reduction in the amplitude of $\fpbh$ (as shown in the figure on the right).
This can be attributed to the fact that an increase in $\Delta N_1$ leads to 
a gentler transition from slow roll to ultra slow roll inflation, and hence
to a smaller height of the peak in the power spectrum.
We have also included the constraints from the different observations, as 
in the previous figure.
Note that the quantity $\fpbh(M)$ depends exponentially 
on the amplitude of the scalar power spectrum.
Hence, despite the wide peaks in the scalar power spectra [cf. 
Fig.~\ref{fig:ps-rs}], the shape of $\fpbh(M)$ proves to be rather narrow. 
Due to this reason, we believe that the constraints for the monochromatic
power spectra (which we have indicated in the figure) also apply well to 
the situations of our interest.}
\label{fig:fpbh-rc}
\end{figure}

On using the arguments presented above and propagating the density of PBHs 
produced to the current epoch, we find that the quantity $\fpbh(M)$ can be 
written as  
\begin{equation}
\fpbh(M) = 2^{1/4}\;\gamma^{3/2}\,\beta(M)\,
\l(\f{\Omega_\mathrm{m}\,h^2}{\Omega_\mathrm{c}\,h^2}\r)\,
\l(\f{g_{\ast,k}}{g_{\ast,\mathrm{eq}}}\r)^{-1/4}\,
\l(\f{M}{M_\mathrm{eq}}\r)^{-1/2}, 
\end{equation}
where $\Omega_\mathrm{m}$ and $\Omega_\mathrm{c}$ are the dimensionless 
parameters describing the matter and cold matter densities, with the Hubble 
parameter expressed as $H_0=100\,h\,\mathrm{km}\,\mathrm{sec}^{-1}\,
\mathrm{Mpc}^{-1}$.
In our estimates of $\fpbh(M)$, we shall choose $\gamma = 0.2$, $g_{\ast,k} 
= 106.75$ and $g_{\ast,\mathrm{eq}} = 3.36$.
We shall also set $\Omega_\mathrm{m}\,h^2= 0.14$ and $\Omega_{\mathrm{c}}\,h^2 
= 0.12$, which are the best fit values from the recent Planck 
data~\cite{Ade:2015xua,Aghanim:2018eyx}.
On substituting these values, we arrive at the following expression 
for~$\fpbh(M)$:
\begin{equation}
\fpbh(M) 
=\l(\f{\gamma}{0.2}\r)^{3/2}\,
\l(\f{\beta(M)}{1.46\times 10^{-8}}\r)\, 
\l(\f{g_{\ast,k}}{g_{\ast,\mathrm{eq}}}\r)^{-1/4}\,
\l(\f{M}{M_\odot}\r)^{-1/2}.\label{eq:fpbh-f}
\end{equation}
As we have discussed, given an inflationary scalar power spectrum $\ps(k)$, 
the relations~\eqref{eq:pm-k}, \eqref{eq:sigma2} and \eqref{eq:R-Ms} can be
utilized to arrive at the variance~$\sigma^2(M)$.
Once we have obtained~$\sigma^2(M)$, we can evaluate the quantity~$\beta(M)$ 
using the expression~\eqref{eq:b-pbh}.
Finally, we can use the relation~\eqref{eq:fpbh-f} to arrive at~$\fpbh(M)$. 
In Figures~\ref{fig:fpbh-im} and~\ref{fig:fpbh-rc}, we have plotted 
the quantity $\fpbh(M)$ in the different inflationary models and the 
reconstructed scenario that we discussed in the previous section.
We have also included the constraints from the different observations in 
the figure, which correspond to a monochromatic spectrum
of PBHs (in this regard, see the discussions in Refs.~\cite{Carr:2020gox,
Green:2020jor,Escriva:2022duf,Franciolini:2022htd} and references therein).
In the case of the reconstructed scenario, we have plotted the quantity 
$\fpbh(M)$ for a range of values of $N_1$ and $\Delta N_1$. 
Recall that, while $N_1$ denotes the $e$-fold at which the phase of 
ultra slow roll sets in, $\Delta N_1$ determines the duration of the 
transition from the initial slow roll phase to the ultra slow roll
epoch.
We find that, among the six inflationary models we have considered, it is 
only the inflationary models M2 and M6 that produce a significant number
of PBHs.
In the reconstructed scenario with varied values of $N_1$, the peaks in 
$\fpbh(M)$ roughly behave as $M^{-1/2}$, as can be expected from the 
relation~\eqref{eq:fpbh-f} for a constant amplitude of $\beta(M)$ that is 
shifted only by the mass~$M$.
This is a direct consequence of delaying the onset of ultra slow roll to 
later and later stages of inflation while retaining a fixed amplitude and 
shape of the scalar power spectrum~$\ps(k)$.


\section{Generation of secondary GWs in the radiation dominated 
epoch}\label{sec:sgws}

In Section~\ref{sec:qo}, we had discussed the evolution of the scalar and
tensor perturbations during inflation.
We had seen that, at the linear order in the perturbations (or, equivalently,
at the quadratic order in the action), the scalar and tensor perturbations 
evolve independently [cf. Eqs.~\eqref{eq:a-p} and~\eqref{eq:de-p}], a result
that is often referred to as the decomposition theorem.
But, when the perturbations at the second order are taken into account, one
finds that the second order scalar perturbations can source the tensor 
perturbations (for the original discussions in this context, see 
Refs.~\cite{Ananda:2006af,Baumann:2007zm,Saito:2008jc,Saito:2009jt}).
Such a phenomenon becomes important particularly in the scenarios involving
ultra slow roll inflation that we have considered.
The enhanced scalar power on small scales can source the tensor perturbations 
to such an extent that the strength of the induced, secondary GWs can be 
significantly larger than the amplitude of the primary GWs generated during
inflation.
In this section, our aim will be calculate the dimensionless spectral energy
density of the secondary GWs, say, $\ogw$, induced by the scalar perturbations
in the inflationary models and reconstructed scenario of interest.
Specifically, we shall focus on the situation wherein the secondary GWs are 
generated when the scales of interest have reentered the Hubble radius during
the radiation dominated era.
As we shall see, interestingly, in many situations, the spectral energy density 
of the secondary GWs generated in such a manner can be comparable to the sensitivity
curves of some of the ongoing as well as forthcoming 
GW observatories (in this regard, see, for instance, Refs.~\cite{Moore:2014lga,
NANOGrav:2021flc,Mu:2022dku} and references therein).

We shall first sketch the essential arguments for calculating the quantity~$\ogw(f)$, 
where~$f$ is the frequency associated with the wave number~$k$ that can be determined 
by the relation
\begin{equation}
f = \frac{k}{2\,\pi} 
= 1.55\times10^{-15}\,\l(\f{k}{1\, \mathrm{Mpc}^{-1}}\r)\,\mathrm{Hz}.
\end{equation}
For simplicity, we shall assume that the anisotropic stresses are absent
during the era of radiation domination.
In such a case, the scalar perturbations at the first order can be described 
by the Bardeen potential, say, $\Phi$.
Recall that, earlier, we had represented the first order tensor perturbations 
as~$\gamma_{ij}$ [cf. Eq.~\eqref{eq:metric}]. 
In order to distinguish the first and the second order tensor perturbations,
we shall denote the second order tensor perturbations as~$h_{ij}$\footnote{The 
second order tensor perturbations~$h_{ij}$ should not be confused with the
quantity~$\sf{h}_{ij}$ which had denoted the spatial components of the metric 
in the ADM form of the line-element~\eqref{eq:adm-m}.}.
On taking into account the first order scalar and the second order tensor
perturbations, the FLRW line-element can be written as
\begin{equation}
\d s^2 = a^2(\eta)\, \l\{-\l(1+2\,\Phi\r)\, \d\eta^2
+ \l[\l(1-2\,\Phi\r)\,\delta_{ij}+\f{1}{2}\,h_{ij}\r]\, \d x^i \d x^j\r\}.
\end{equation}

Let $h_{\bm k}$ denote the Fourier modes associated with the second
order tensor perturbations.
In terms of the mode functions~$h_{\bm k}$, the tensor perturbations~$h_{ij}$ 
can be decomposed as follows:
\begin{equation}
h_{ij}(\eta,\vx)
=\int\f{\d^3\vk}{(2\,\pi)^{3/2}}\,
\l[e_{ij}^+(\vk)\, h_{\vk}^+(\eta)
+e_{ij}^\times(\vk)\, h_{\vk}^\times(\eta)\r]\, {\rm e}^{i\,\vk\cdot\vx},
\end{equation}
where quantities $e_{ij}^+(\vk)$ and $e_{ij}^\times(\vk)$ denote the 
polarization tensors.
As in the case of the first order tensor perturbations~$\gamma_{ij}$, the 
second order tensor perturbations~$h_{ij}$ too are transverse and traceless,
i.e. they satisfy the conditions $\delta^{ij}\,k_i\,e^{\lambda}_{jl}=
\delta^{ij}\, e^{\lambda}_{ij}=0$.
The transverse nature of the tensor perturbations implies that the polarization
tensors have  non-zero components only in the plane perpendicular to the 
direction of propagation~$\hat{\vk}$.
The polarization tensors $e_{ij}^+(\vk)$ and $e_{ij}^\times(\vk)$ can be 
expressed in terms of the set of orthonormal unit vectors $(e(\vk),
{\bar e}(\vk),\hat{\vk})$ in the following manner (for a discussion on 
this point, see, for example, the review~\cite{Maggiore:1999vm}):
\begin{subequations}
\begin{eqnarray}
e_{ij}^+(\vk)\!\!\! 
&=&\!\!\! \f{1}{\sqrt{2}}\,\l[e_{i}(\vk)\,e_{j}(\vk)
-{\bar e}_{i}(\vk)\, {\bar e}_{j}(\vk)\r],\\
e_{ij}^\times(\vk)\!\!\! 
&=&\!\!\! \f{1}{\sqrt{2}}\,\l[e_{i}(\vk)\,{\bar e}_{j}(\vk)
+{\bar e}_{i}(\vk)\, e_{j}(\vk)\r].
\end{eqnarray}
\end{subequations}
The orthonormal nature of the vectors $e(\vk)$ and ${\bar e}(\vk)$ lead to 
the normalization condition:~$\delta^{il}\,\delta^{jm}\, e_{ij}^{\lambda}(\vk)\,
e^{\lambda'}_{lm}(\vk)=\delta^{\lambda \lambda'}$, where $\lambda$ and 
$\lambda'$ can represent either of the two states of polarization~$+$ or~$\times$.

Let us now turn our attention to the equation of motion governing the modes
functions~$h_\vk$.
The equation of motion can be arrived at using the second order Einstein equations 
describing the tensor perturbations~$h_{ij}$ and the Bardeen equation describing 
the scalar perturbation~$\Phi$ at the first order (for the initial discussions, 
see Refs.~\cite{Ananda:2006af,Baumann:2007zm}; for some of the recent discussions, 
see Refs.~\cite{Bartolo:2016ami,Bartolo:2018evs,Bartolo:2018rku,Espinosa:2018eve}).
It can be shown that, during the radiation dominated epoch, the equation 
governing~$h_\vk$ can be written as
\begin{equation}
{h_\vk^\lambda}''+ \f{2}{\eta}\, {h_\vk^\lambda}'+ k^2\, h_\vk^\lambda
=S_\vk^\lambda,\label{eq:sgw}
\end{equation}
where the quantity $S_\vk^\lambda$ denotes the source due to the scalar 
perturbations.
The source term~$S_\vk^\lambda$ is given by
\begin{eqnarray}
S_\vk^\lambda(\eta)\!\!\! 
&=&\!\!\!  4\, \int\frac{\d^3 \vp}{(2\,\pi)^{3/2}}\, e^\lambda(\vk,\vp)\,
\Biggl\{2\,\Phi_\vp(\eta)\,\Phi_{\vk-\vp}(\eta)\nn\\
& &\!\!\! +\, \l[\Phi_\vp(\eta)+\eta\,\Phi_\vp'(\eta)\r]\,
\l[\Phi_{\vk-\vp}(\eta)+\eta\,\Phi_{\vk-\vp}'(\eta)\r]\,\Biggr\},
\end{eqnarray}
where $\Phi_\vk$ represents the Fourier modes associated with the Bardeen potential 
and, for convenience, we have defined the quantity $e^{\lambda}(\vk,\vp)=
e^{\lambda}_{ij}(\vk)\,p^i\,p^j$.
As is well known, during the epoch of radiation domination, we can express the 
Fourier modes~$\Phi_\vk$ of the Bardeen potential in terms of the Fourier 
modes~$\cR_\vk$ of the curvature perturbations generated during inflation through 
the relation
\begin{equation}
\Phi_\vk(\eta)=\f{2}{3}\,\cT(k\,\eta)\, \cR_\vk,
\end{equation}
where $\cT(k\,\eta)$ is the transfer function given by
\begin{equation}
\cT(k\,\eta)=\f{9}{\l(k\,\eta\r)^2}\,
\l[\f{{\rm sin}\l(k\,\eta/\sqrt{3}\r)}{k\,\eta/\sqrt{3}}
-{\rm cos}\l(k\,\eta/\sqrt{3}\r)\r].\label{eq:tf}
\end{equation}
If we make use of the Green's function corresponding to the tensor modes 
during radiation domination, we find that we can express the inhomogeneous
contribution to $h_\vk^\lambda$ as~\cite{Espinosa:2018eve}
\begin{eqnarray}
h_\vk^\lambda(\eta)\!\!\! 
&=&\!\!\! \frac{4}{9\,k^3\,\eta}\,
\int \f{\d^3 \vp}{(2\,\pi)^{3/2}}\, 
e^{\lambda}(\vk,\vp)\, \cR_\vk\,\cR_{\vk-\vp}\nn\\
& &\!\!\! 
\times\,\l[\cI_c\l(\f{p}{k},\f{\vert\vk-\vp\vert}{k}\r)\,{\rm cos}\l(k\,\eta\r)
+\cI_s\l(\f{p}{k},\f{\vert\vk-\vp\vert}{k}\r)\,{\rm sin}\l(k\,\eta\r)\r],
\label{eq:hkl}
\end{eqnarray}
where the quantities $\cI_c(v,u)$ and $\cI_s(v,u)$ are described by the integrals
\begin{subequations}
\begin{eqnarray}
\cI_c(v,u)\!\!\! 
&=&\!\!\! -4\,\int_{0}^{\infty}\,\d \tau\,\tau\,{\rm sin}\,\tau\,
\biggl\{2\,\cT(v\,\tau)\,\cT(u\,\tau)\nn\\
& &\!\!\! +\,\l[\cT(v\,\tau)+v\,\tau\,\cT_{v\tau}(v\,\tau)\r]\,
\l[\cT(u\,\tau)+u\,\tau\,\cT_{u\tau}(u\,\tau)\r]\biggr\},\\
\cI_s(v,u)\!\!\! 
&=&\!\!\!  4\,\int_{0}^{\infty}\,\d \tau\,\tau\,{\rm cos}\,\tau\,
\biggl\{2\,\cT(v\,\tau)\,\cT(u\,\tau)\nn\\
& &\!\!\! +\,\l[\cT(v\,\tau)+v\,\tau\,\cT_{v\tau}(v\,\tau)\r]\,
\l[\cT(u\,\tau)+u\,\tau\,\cT_{u\tau}(u\,\tau)\r]\biggr\},
\end{eqnarray}
\end{subequations}
with $\cT_z=\d\cT/\d z$.
Upon utilizing the transfer function~\eqref{eq:tf}, these integrals can be
calculated analytically to obtain that (see, for example,
Refs.~\cite{Kohri:2018awv,Espinosa:2018eve})
\begin{subequations}\label{eq:cI}
\begin{eqnarray}
\cI_c(v,u)\!\!\!  
&=&\!\!\!  -\f{27\,\pi}{4\,v^3\,u^3}\,
\Theta\l(v+u-\sqrt{3}\r)\, (v^2+u^2-3)^2,\\
\cI_s(v,u)\!\!\!  
&=&\!\!\!  -\f{27}{4\,v^3\,u^3}\, (v^2+u^2-3)\,
\l[4\,v\,u+ (v^2+u^2-3)\;{\rm log}\,
\biggl\vert\f{3-(v-u)^2}{3-(v+u)^2}\biggr\vert\r],\qquad
\end{eqnarray}
\end{subequations}
where $\Theta(z)$ denotes the theta function.
It is useful to note that $\cI_{c,s}(v,u) =\cI_{c,s}(u,v)$.

The power spectrum of the secondary GWs, say, $\ph(k,\eta)$, generated
due to the second order scalar perturbations can be defined through 
the relation
\begin{eqnarray}
\langle h_{\vk}^{\lambda}(\eta)\,h_{\vk'}^{\lambda'}(\eta)\rangle
=\f{2\,\pi^2}{k^3}\,\cP_h(k,\eta)\,\delta^{(3)}(\vk+\vk')\,
\delta^{\lambda\lambda'}.\label{eq:phk}
\end{eqnarray}
It should be evident that, because the quantity $h_\vk^\lambda$ involves products 
of the Fourier modes $\cR_\vk$ and $\cR_{\vk-\vp}$ of the curvature perturbations 
generated during inflation [see Eq.~\eqref{eq:hkl}], the power spectrum $\ph(k)$ 
of the secondary GWs will involve products of four such variables.
If we assume that the Fourier modes of the curvature perturbations are Gaussian 
random variables, we can express the four-point function involving $\cR_\vk$ in
terms of the two-point functions, i.e. in terms of the the inflationary scalar
power spectrum~$\ps(k)$ [cf. Eq.~\eqref{eq:sps-d}].
Equivalently, it can be said that, since the expectation value in the 
definition~\eqref{eq:phk} of the secondary tensor power spectrum has to
be evaluated in the Bunch-Davies vacuum, the four-point function of the
curvature perturbations can be expressed in terms of the two-point 
functions using Wick's theorem.
Upon doing so, we can arrive at the following expression for the secondary 
tensor power spectrum:
\begin{eqnarray}
\ph(k,\eta)\!\!\! 
&=&\!\!\!  \f{4}{81\,k^2\,\eta^2}
\int_{0}^{\infty}\d v\,\int_{\vert 1-v\vert}^{1+v}\d u\,
\l[\f{4\,v^2-(1+v^2-u^2)^2}{4\,u\,v}\r]^2\,\ps(k\,v)\,\ps(k\,u)\nn\\
& &\!\!\! \times\,
\l[\cI_c(u,v)\,{\rm cos}\l(k\,\eta\r)+\cI_s(u,v)\,{\rm sin}\l(k\,\eta\r)\r]^2.
\label{eq:ph}
\end{eqnarray}
The trigonometric functions in this expression arise because of the form of
the transfer function~$\cT(k\,\eta)$ [cf. Eq.~\eqref{eq:tf}].
They reflect the fact that the Bardeen potentials~$\Phi_{\bm k}$ oscillate
when the corresponding scales are inside the Hubble radius during the 
radiation dominated epoch.
On averaging the secondary tensor power $\ph(k,\eta)$ over small time 
scales, we can replace the trigonometric functions in the above expression
by their average over a time period.
In such a case, only the overall time dependence remains, leading 
to~\cite{Kohri:2018awv,Espinosa:2018eve}
\begin{eqnarray}
\overline{\ph(k,\eta)}\!\!\! 
&=&\!\!\! \f{2}{81\,k^2\,\eta^2}
\int_{0}^{\infty}\d v\,\int_{\vert 1-v\vert}^{1+v}\d u\,
\l[\f{4\,v^2-(1+v^2-u^2)^2}{4\,u\,v}\r]^2\,\ps(k\,v)\,\ps(k\,u)\nn\\
& &\times\,
\l[\cI_c^2(u,v)+\cI_s^2(u,v)\r],\label{eq:phf}
\end{eqnarray}
where the line over $\ph(k,\eta)$ implies that we have averaged over 
small time scales.
The energy density of GWs associated with a Fourier mode corresponding 
to the wave number~$k$ (i.e. the spectral energy density of GWs) at a 
time~$\eta$ is given by~\cite{Maggiore:1999vm}
\begin{equation}
\rho_{_{\mathrm{GW}}}(k,\eta)
= \f{\Mpl^2}{8}\,\l(\frac{k}{a}\r)^2\;\overline{\ph(k,\eta)}.
\end{equation}
We can define the corresponding dimensionless density parameter~$\ogw(k,\eta)$ 
in terms of the critical density $\rho_{\mathrm{cr}}(\eta)$ 
as~\cite{Espinosa:2018eve}
\begin{equation}
\ogw(k,\eta)
=\f{\rho_{_{\mathrm{GW}}}(k,\eta)}{\rho_{\mathrm{cr}}(\eta)}
=\f{1}{24}\,\l(\f{k}{\mathcal{H}}\r)^2\; \overline{\ph(k,\eta)}
=\f{k^2\,\eta^2}{24}\; \overline{\ph(k,\eta)},\label{eq:ogw-rd}
\end{equation}
where, in the final expression, we have made use of the fact that 
$\mathcal{H}=1/\eta$ in radiation domination era.
Note that, since $\ph(k,\eta)\propto \eta^{-2}$, the dimensionless
spectral energy density $\ogw(k,\eta)$ is actually independent of 
time.

\begin{figure}[!t]
\centering
\includegraphics[width=9.50cm]{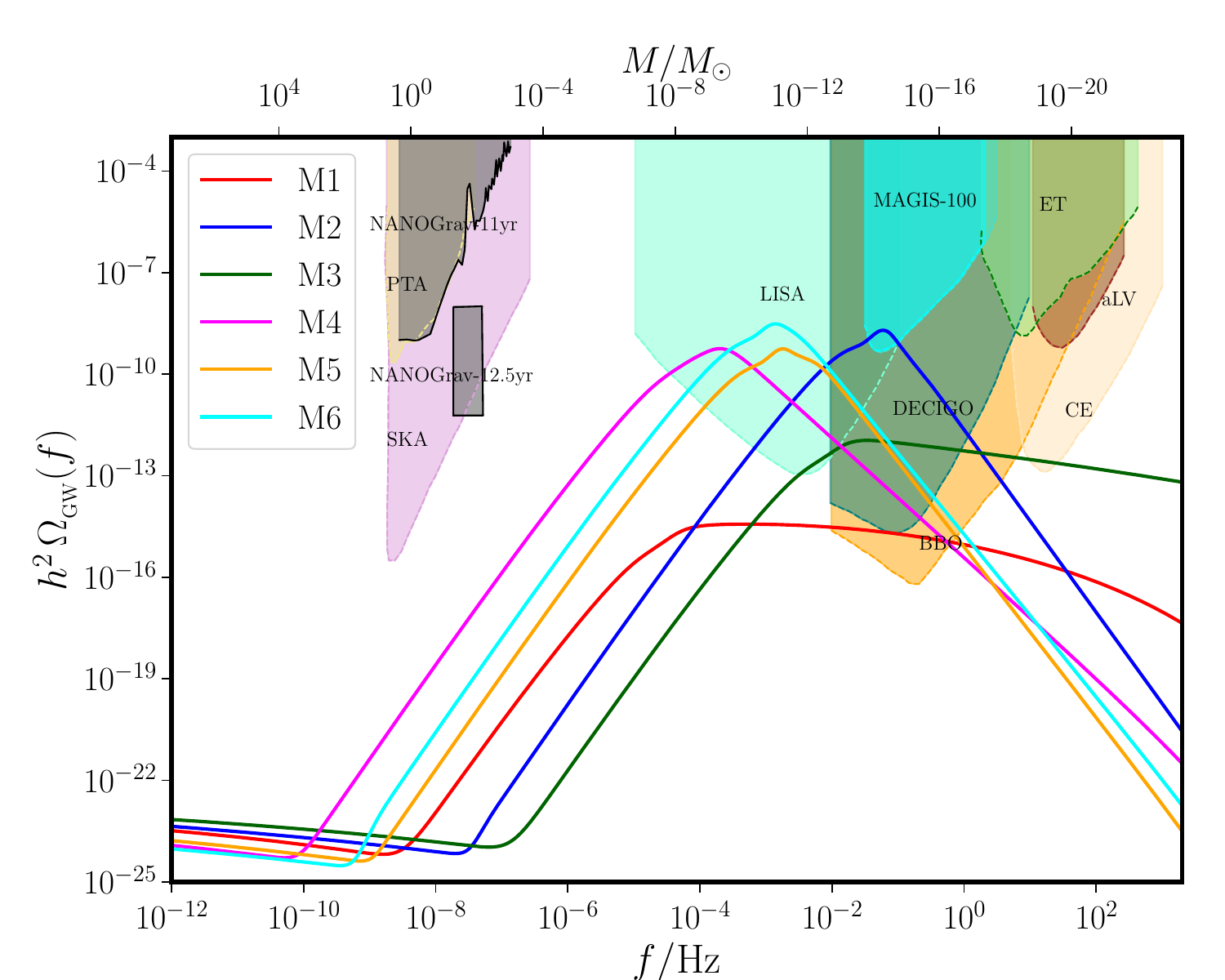}
\caption{The dimensionless spectral density of secondary gravitational waves 
{\it today}, viz. $\ogw(f)$, arising in the inflationary models M1 to M6 has 
been plotted as function of the frequency~$f$.
On the top part of the figure, we have also included the sensitivity curves of 
the various ongoing and forthcoming GW observatories.
It is clear that, in all the models we have considered, the strengths of 
the secondary GWs are comparable to the sensitivity curves of one or more 
of the observatories suggesting that it should be possible to detect such
signals in the future.}\label{fig:ogw-im}
\end{figure}
\begin{figure}[!t]
\hskip -46.2pt
\includegraphics[width=9.50cm]{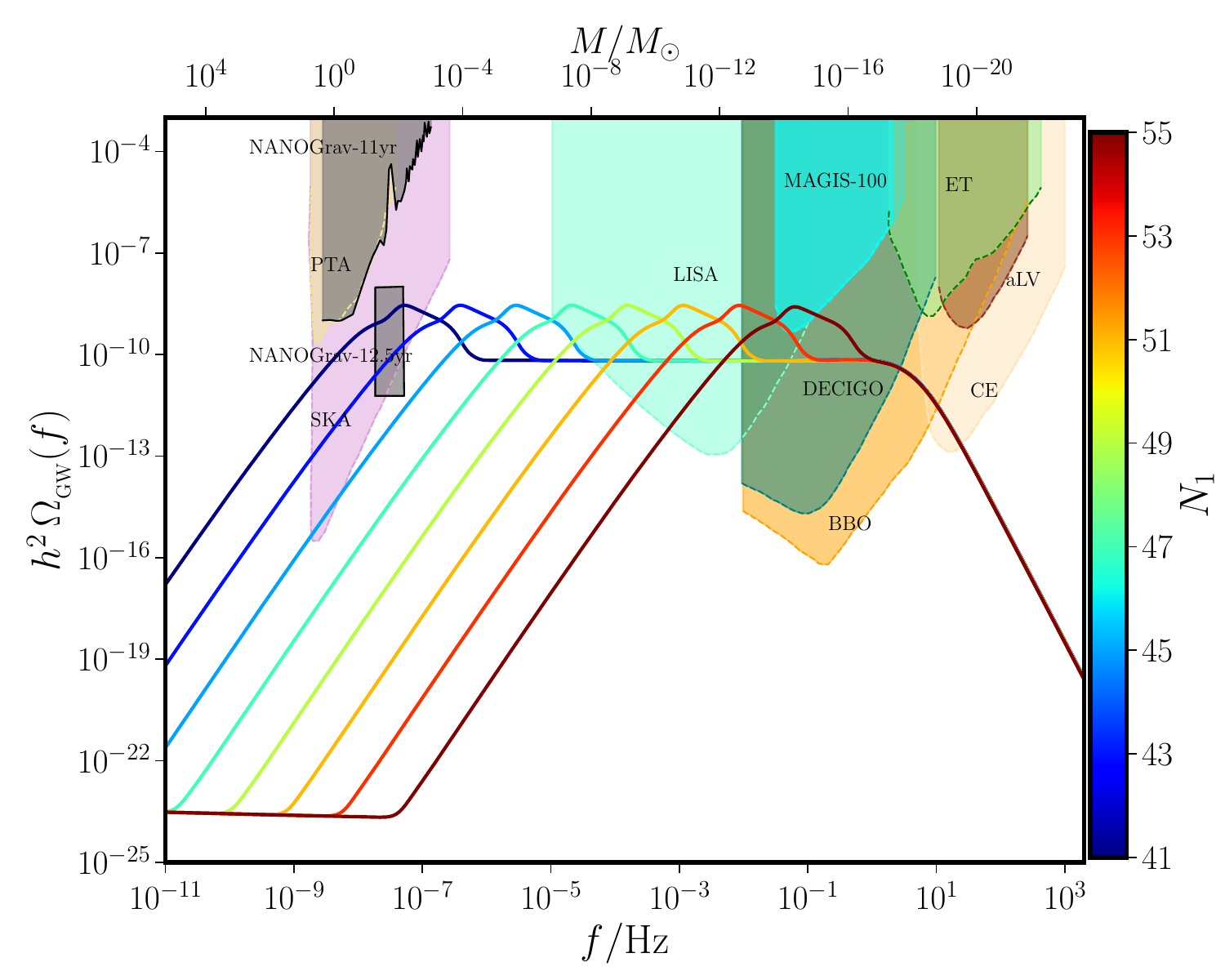}
\includegraphics[width=9.50cm]{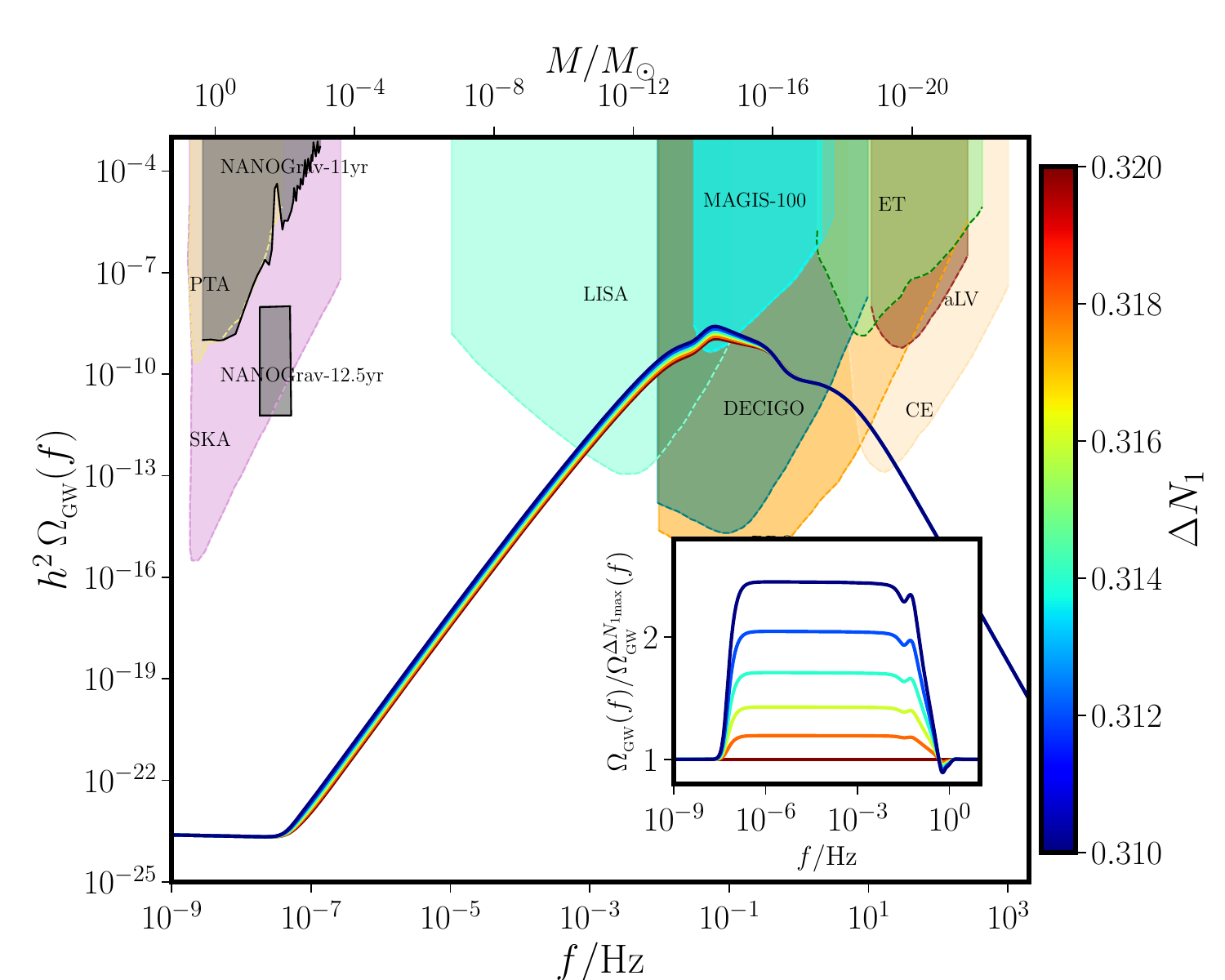}
\caption{The dimensionless spectral density of secondary GWs $\ogw(f)$ that
arise in the reconstructed scenario has been plotted for a range of $N_1$ 
(on the left) and $\Delta N_1$ (on right).
Clearly, smaller the $N_1$, longer is the epoch of ultra slow roll and wider 
is the peak in~$\ogw(f)$. 
In contrast, the variation in $\Delta N_1$ (at least over the range we have 
considered) does not alter the shape of $\ogw(f)$ appreciably.
We should mention that these spectra of different shapes are consistent
with the current bounds on on $h^2\,\ogw$ due to BBN.
Even for the widest of spectra, upon integration over all frequencies, the 
dimensionless density of GWs turns out to be about $1.40\times 10^{-8}$. 
This value is substantially lower than the corresponding BBN bound of 
around~$10^{-6}$ (in this context, see, for instance,
Refs.~\cite{Guzzetti:2016mkm,Caprini:2018mtu}.}
\label{fig:ogw-rc}
\end{figure}

The dimensionless spectral density of GWs above has been calculated 
during the late stages of the epoch of radiation domination, when 
all the scales of interest are inside the Hubble radius.
In such a domain, the energy density of GWs decreases in the same 
fashion as the energy density of radiation (i.e. as $a^{-4}$).
Utilizing this behavior, we can express the dimensionless spectral
energy density of GWs {\it today},\/ i.e. $\ogw(k)$, in terms of the 
$\ogw(k,\eta)$ above in the following manner
\begin{eqnarray}
h^2\,\ogw(k)\!\!\! 
&=&\!\!\! \l(\f{g_{\ast,k}}{g_{\ast,0}}\r)^{-1/3}\,\Omega_{\mathrm{r}}\,h^2\;
\ogw(k,\eta)\nn\\
& \simeq &\!\!\!  
1.38\times10^{-5}\, 
\l(\f{g_{\ast,k}}{106.75}\r)^{-1/3}\,
\l(\f{\Omega_{r}\,h^2}{4.16\times10^{-5}}\r)\,\ogw(k,\eta).\label{eq:ogw0}
\end{eqnarray}
In this expression, $\Omega_\mathrm{r}$ and $g_{\ast,0}$ denote the 
dimensionless energy density of radiation and the number of relativistic 
degrees of freedom today.
In Figures~\ref{fig:ogw-im} and~\ref{fig:ogw-rc}, we have plotted the quantity
$\ogw$ as a function of the frequency $f$ in the six inflationary models 
and the reconstructed scenario we have considered.
We have also included the sensitivity curves of the different ongoing as 
well as forthcoming GW observatories in the figures (in this context, 
see, for example, Refs.~\cite{Moore:2014lga,Arzoumanian:2018saf,NANOGrav:2020bcs,
NANOGrav:2021flc,Mu:2022dku}).
It is clear that all the models and the reconstructed scenario lead to GW
spectral densities that are comparable to the sensitivity curves of the
different observatories.
This gives us hope that the imprints of non-trivial dynamics during the 
late stages of inflation can either be detected or, at the least, strongly
constrained with the aid of GW observations set to emerge over the coming 
decade or two. 


\section{Non-Gaussianities on small scales}\label{sec:sbs}

As we have seen, the onset of ultra slow roll leads to strong departures 
from slow roll inflation with the second slow roll parameter $\epsilon_2$ 
(as well as the higher order slow roll parameters) attaining rather large 
values.
It is the strong departure from slow roll that results in sharp features
in the scalar power spectrum, such as the peak with significantly enhanced 
power that we have discussed earlier.
We should clarify that, we have chosen the parameters of the inflationary
potentials and the reconstructed scenario so that the ultra slow roll phase
sets in during the latter stages of inflation (after the wave numbers 
corresponding to the CMB scales have left the Hubble radius) and the peak 
occurs at small scales. 

In fact, there has been a constant effort in the literature to investigate 
if certain features in the inflationary scalar power spectrum provide a 
better fit to the CMB data than the more standard, nearly scale invariant
spectrum (see the recent efforts~\cite{Ragavendra:2020old,Hazra:2022rdl,
Braglia:2022ftm} and references therein).
Often, these features are generated due to moderate departures from slow 
roll inflation.
However, we should point out that strong departures such as those occur in 
ultra slow roll inflation have been considered to suppress the power on the 
largest scales (comparable to the Hubble radius today) so as to improve the
fit to the lowest multipoles in the CMB data~\cite{Jain:2008dw,Jain:2009pm,
Ragavendra:2020old}.
In slow roll inflation involving the canonical scalar field, typically, 
the non-Gaussianities generated are rather small with the dimensionless
parameter $\fnl$ 
[defined below in Eq.~\eqref{eq:fnl}] that reflects 
the amplitude of the scalar bispectrum being of the order of the first slow
roll parameter $\epsilon_1$~\cite{Maldacena:2002vr,Seery:2005wm,Chen:2006nt,
Hazra:2012yn,Ragavendra:2020old}.
But, when departures from slow roll occur, it is known that the amplitude
of the scalar bispectrum and the associated non-Gaussianity parameter can
be considerably larger~\cite{Chen:2006nt,Hazra:2012yn,Ragavendra:2020old}.
Moreover, while the scalar bispectrum has an equilateral shape in slow
roll inflation, the shape can be considerably different when deviations
from slow roll occur.
These suggest that the epoch of ultra slow roll inflation that we have 
considered to enhance power on small scales can also be expected to
generate significant levels of non-Gaussianities with characteristic
shapes~\cite{Ragavendra:2020old}.

In this section, we shall compute the scalar bispectrum and the associated
non-Gaussianity parameter for two of the inflationary models that we have 
discussed earlier.
However, before we go on to present these results, we shall first describe
the third order action that governs the curvature perturbations and the 
numerical method we shall adopt to compute the scalar bispectrum.


\subsection{The complete third order action governing the scalar 
bispectrum}\label{sec:toa}

Let us begin by recalling a few necessary points regarding the scalar bispectrum.
Just as the power spectrum characterizes the two-point function of the perturbations
in Fourier space, the bispectrum describes corresponding three-point function.
The scalar bispectrum, say, $\cB_{_{\mathrm{S}}}(\vka,\vkb,\vkc)$, is defined in 
terms of the operator~$\hat{\cR}_\vk$ [that we had introduced earlier in 
Eq.~\eqref{eq:s-m-dc}] as follows~\cite{Ade:2015ava,Planck:2019kim}:
\begin{equation}
\langle \hat{\cR}_{\vka}(\ee)\, 
\hat{\cR}_{\vkb}(\ee)\, \hat{\cR}_{\vkc}(\ee)\rangle 
=(2\,\pi)^3\, \cB_{_{\rm S}}(\vka,\vkb,\vkc)\,
\delta^{(3)}(\vka+\vkb+\vkc),\label{eq:bi-s}
\end{equation}
where $\ee$ is a time close to the end of inflation and the expectation value 
on the left hand side is to be evaluated in the {\it perturbative}\/ 
vacuum~\cite{Maldacena:2002vr,Seery:2005wm,Chen:2010xka}.
Note that the delta function that appears in the above definition implies that 
the three wave vectors $(\vka,\vkb,\vkc)$ form the edges of a triangle.
Hence, it is only two of the vectors that are truly independent and, it is for
this reason the quantity $\cB_{_{\mathrm{S}}}(\vka,\vkb,\vkc)$ is referred to 
as the {\it bi}\/-spectrum.
Hereafter, for convenience, we shall set
\begin{equation}
\cB_{_{\rm S}}(\vka,\vkb,\vkc)
=(2\,\pi)^{-9/2}\, G(\vka,\vkb,\vkc)
\end{equation}
and refer to $G(\vka,\vkb,\vkc)$ as the scalar bispectrum.

Conventionally, in quantum field theory, the correlation functions beyond 
the two-point functions that describe the fields are often calculated using 
perturbative methods.
As is well known, the three-point function associated with a field can be 
expected to be non-zero if the action governing the field of interest contains
a cubic order term.  
The same approach can be utilized to calculate the scalar bispectrum generated 
during inflation (for the original discussions in this context, see 
Refs.~\cite{Maldacena:2002vr,Seery:2005wm,Chen:2006nt}).
Evidently, in order to do so, one first requires the action describing the 
curvature perturbation at the third order. 
With such an action at hand, one can use the standard methods of perturbative
quantum field theory to arrive at the scalar bispectrum.

Recall that, in Section~\ref{sec:qo}, we had arrived at the action and the 
equations of motion governing the background as well as the scalar and tensor 
perturbations using the ADM formalism. 
Starting from the original action~\eqref{eq:adm-a} that governs the system of 
the gravitational and scalar fields and the line-element~\eqref{eq:metric}, 
the third order action describing the curvature perturbation can arrived at in 
the same manner~\cite{Maldacena:2002vr,Seery:2005wm,Chen:2006nt,Martin:2011sn}.
In fact, a set of temporal and spatial boundary terms arise in the process,
when the action is repeatedly integrated by parts to simplify its form.
One can easily establish that, due to the triangularity condition on the wave 
vectors $(\vk_1,\vk_2,\vk_3)$, the spatial boundary terms do not contribute
to the scalar bispectrum under any condition.
However, some of the temporal boundary terms can contribute to the scalar 
bispectrum even in the simple case of slow roll inflation~\cite{Arroja:2011yj}.
It can be shown that, at the third order, the action governing the
curvature perturbation $\cR$ can be expressed as (see, for instance, 
Refs.~\cite{Maldacena:2002vr,Seery:2005wm,Martin:2011sn,Arroja:2011yj})
\begin{eqnarray}
\mathcal{S}_3[\cR]\!\!\!   
&=&\!\!\!  \Mpl^2\, \int \d\eta \int \d^3{\bm x}\, 
\biggl[a^2\, \epsilon_1^2\, \cR\,\cR'^2
+ a^2\, \epsilon_1^2\, \cR\,(\pa\cR)^2
- 2\,a\,\epsilon_1\, \cR^{\prime}\, (\pa\cR)\,(\partial\chi)\nn\\ 
& &\!\!\!  +\, \f{a^2}{2}\,\epsilon_1\,\epsilon_2'\,\cR^2\,\cR'
+ \frac{\epsilon_1}{2}\,(\pa\cR)\, (\pa\chi)\, \partial^2\chi
+ \frac{\epsilon_1}{4}\,\pa^2\cR\,(\partial\chi)^2 
+ 2\,{\cal F}(\cR)\, \frac{\delta {\cal L}_2}{\delta \cR}\biggr],\qquad 
\label{eq:S3}
\end{eqnarray}
where $\epsilon_1$ and $\epsilon_2$ are the slow roll parameters we have 
repeatedly encountered, while $\pa^2\chi= a\,\epsilon_1\,\cR'$.
Moreover, the quantity ${\cal F}$($\cR$) is given by
\begin{eqnarray}
{\cal F}(\cR)\!\!\!   
&=&\!\!\!   \frac{\epsilon_2}{4}\,\cR^2 
+ \f{1}{a\,H}\,\cR\,\cR'
+ \f{1}{4\,a^2\,H^2} \biggl\{-(\pa\cR)\, (\pa\cR)
+ \pa^{-2}[\pa_i\,\pa_j\,(\pa_i\cR\,\pa_j\cR)]\biggr\} \nn \\
& &\!\!\!  +\, \f{1}{2\,a^2\,H} \biggl\{(\pa\cR)\, (\pa\chi) 
- \partial^{-2}[\pa_i\,\partial_j\,(\pa_i\cR\,\pa_j\chi)]\biggr\}
\end{eqnarray}
and ${\mathcal{L}}_2$ denotes the Lagrangian density associated with the 
action~\eqref{eq:a-rr} that governs the curvature perturbation at the 
second order.
Further, the temporal boundary terms are given by~\cite{Arroja:2011yj}
\begin{eqnarray}
\mathcal{S}^{\mathrm{B}}_3[\cR]\!\!\!  
&=&\!\!\!  \Mpl^2\, \int\d\eta \int\d^3{\bm x}\;
\f{\d}{\d\eta}\biggl\{-9\,a^3H\,\cR^3
+\f{a}{H}\,(1-\epsilon_1)\,\cR\,(\pa\cR)^2
- \f{1}{4\,a\,H^3}\,(\pa\cR)^2\,\pa^2\cR\nn\\ 
& &\!\!\!  -\, \f{a\,\epsilon_1}{H}\,\cR\,\cR'^2
-\,\f{a\,\epsilon_2}{2}\,\cR^2\,\pa^2\chi
+ \frac{1}{2\,a\,H^2}\,\cR\,\l(\pa_i\pa_j\cR\,\pa_i\pa_j\chi 
- \partial^2\cR\,\partial^2\chi\r)\nn\\
& &\!\!\! -\, \frac{1}{2\,aH}\,\cR\,
\l[\pa_i\pa_j\chi\,\pa_i\pa_j\chi - (\pa^2\chi)^2\r]\biggr\}.\label{eq:S3B}
\end{eqnarray}
In most of the situations of interest, including the scenarios that we are
considering here, one finds that it is only the term involving $\epsilon_2$ 
(in the above expression) that contributes to the scalar bispectrum.
Usually, the contribution due to this term is taken into account through a
field redefinition (for a discussion in this context, see, for instance,
Refs.~\cite{Maldacena:2002vr,Arroja:2011yj}).
Instead, apart from the calculating the contributions to the bispectrum due 
to the bulk terms in the action~\eqref{eq:S3}, we shall explicitly evaluate
contribution due to the term containing~$\epsilon_2$ in Eq.~\eqref{eq:S3B}.


\subsection{Numerical computation of the scalar bispectrum and the associated 
non-Gaussianity parameter}\label{subsec:sbs}

Upon taking into account the contributions due to the bulk and the boundary 
terms we discussed above, it can be shown that the scalar bispectrum, evaluated
at a time $\ee$ close to the end of inflation, can be written as (in this context, 
see, for instance, Refs.~\cite{Martin:2011sn,Hazra:2012yn,Ragavendra:2020old,
Ragavendra:2020sop})
\begin{eqnarray}
G(\vka,\vkb,\vkc)\!\!\!   
&=&\!\!\!  \Mp^2\; \sum_{C=1}^{6}\; 
\l[f_{k_1}(\ee)\, f_{k_2}(\ee)\,f_{k_3}(\ee)\,
\cG_{_{C}}(\vka,\vkb,\vkc)+\mathrm{complex~conjugate}\r]\nn\\
& & +\, G_{7}(\vka,\vkb,\vkc),\label{eq:sbs}
\end{eqnarray}
where $f_k$ are the mode functions associated with the curvature perturbation 
[cf. Eq.~(\ref{eq:s-m-dc})].
The quantities $\cG_{_{C}}(\vka,\vkb,\vkc)$ that appear in the above expression
represent six integrals that involve the scale factor, the slow roll parameters, 
the mode functions~$f_k$ and their time derivatives~$f_k'$.
They correspond to the six bulk terms appearing in the cubic order 
action~(\ref{eq:S3}) and are described by the following expressions:
\begin{subequations}\label{eqs:cG}
\begin{eqnarray}
\cG_1(\vka,\vkb,\vkc)\!\!\!  
&=&\!\!\!   2\,i\,\int_{\ei}^{\ee} \d\eta\; a^2\, 
\epsilon_{1}^2\, \biggl(f_{k_1}^{\ast}\,f_{k_2}'^{\ast}\,
f_{k_3}'^{\ast}
+{\rm two~permutations}\biggr),\label{eq:cG1}\\
\cG_2(\vka,\vkb,\vkc)\!\!\!  
&=&\!\!\!  -2\,i\;\l(\vka\cdot \vkb +\,{\rm two~permutations}\r)\, 
\int_{\ei}^{\ee} \d\eta\; a^2\, 
\epsilon_{1}^2\, f_{k_1}^{\ast}\,f_{k_2}^{\ast}\,
f_{k_3}^{\ast},\label{eq:cG2}\\
\cG_3(\vka,\vkb,\vkc)\!\!\!  
&=&\!\!\!  -2\,i\,\int_{\ei}^{\ee} \d\eta\; a^2\,\epsilon_{1}^2\,
\biggl(\f{\vka\cdot\vkb}{k_2^2}\,
f_{k_1}^{\ast}\,f_{k_2}'^{\ast}\, f_{k_3}'^{\ast}
+{\rm five~permutations}\biggr),\label{eq:cG3}\\
\cG_4(\vka,\vkb,\vkc)\!\!\!  
&=&\!\!\!   i\,\int_{\ei}^{\ee} \d\eta\; a^2\,\epsilon_{1}\,\epsilon_{2}'\, 
\biggl(f_{k_1}^{\ast}\,f_{k_2}^{\ast}\,f_{k_3}'^{\ast}
+ {\rm two~permutations}\biggr),\label{eq:cG4}\\
\cG_5(\vka,\vkb,\vkc)\!\!\!  
&=&\!\!\!  \frac{i}{2}\,\int_{\ei}^{\ee} \d\eta\; 
a^2\, \epsilon_{1}^{3}\;\biggl(\f{\vka\cdot\vkb}{k_2^2}\,
f_{k_1}^{\ast}\,f_{k_2}'^{\ast}\, f_{k_3}'^{\ast} 
+ {\rm five~permutations}\biggr),\label{eq:cG5}\\
\cG_6(\vka,\vkb,\vkc)\!\!\!   
&=&\!\!\!  \frac{i}{2}\,\int_{\ei}^{\ee}\d\eta\, a^2\, \epsilon_{1}^{3}\,
\biggl(\f{k_1^2\,\l(\vkb\cdot\vkc\r)}{k_2^2\,k_3^2}\, 
f_{k_1}^{\ast}\, f_{k_2}'^{\ast}\, f_{k_3}'^{\ast} 
+ {\rm two~permutations}\biggr).\qquad\label{eq:cG6}
\end{eqnarray}
\end{subequations}
These integrals are to be calculated from an early time ($\ei$) when the 
scales of interest are well inside the Hubble radius, until a time towards 
the end of inflation~($\ee$).
We should also clarify that the last term in the action~\eqref{eq:S3} 
involving ${\cal F}(\cR)\, ({\delta {\cal L}_2}/{\delta \cR})$ actually 
vanishes when we assume that the curvature perturbation satisfies the 
linear equation of motion [cf. Eq.~\eqref{eq:de-fk}].
The contribution $G_{7}(\vka,\vkb,\vkc)$ is due to the term 
containing $\epsilon_2$ in the boundary terms~(\ref{eq:S3B}),
and it can be expressed as
\begin{eqnarray}
G_{7}(\vka,\vkb,\vkc)\!\!\!  
&=&\!\!\!  -i\,\Mpl^2\,(f_{k_1}(\ee)\,f_{k_2}(\ee)\,f_{k_3}(\ee)) \nn \\
& &\!\!\!  \times\, \biggl[a^2\epsilon_1\epsilon_{2}\,
f_{k_1}^{\ast}(\eta)\,f_{k_2}^{\ast}(\eta)\,f_{k_3}'^{\ast}(\eta) 
+ {\rm two~permutations} \biggr]_{\eta_i}^{\ee}\nn\\
& &\!\!\! +\, {\mathrm{complex~conjugate}}.\label{eq:G7}
\end{eqnarray} 
When one imposes the initial conditions when the scales are well inside the 
Hubble radius, the contribution due to $\ei$ in the above expression for
$G_{7}(\vka,\vkb,\vkc)$ vanishes with the introduction of a regulator, and 
it is only the term evaluated towards end of inflation (i.e. at $\ee$) that
contributes.
Usually, instead of the scalar bispectrum, it is the dimensionless 
non-Gaussianity parameter~$\fnl(\vka,\vkb,\vkc)$ that is often 
quoted and constrained.
The non-Gaussianity parameter corresponding to the scalar bispectrum 
$G(\vka,\vkb,\vkc)$ is defined as (see, for instance, 
Refs.~\cite{Martin:2011sn,Hazra:2012yn})
\begin{eqnarray}
\fnl(\vka,\vkb,\vkc)
& =&-\frac{10}{3}\,\frac{1}{\l(2\,\pi\r)^4}\;k_1^3\, k_2^3\, k_3^3\;
G(\vka,\vkb,\vkc)\nn\\
& & \times\, \biggl[k_1^3\,\ps(k_2)\,\ps(k_3) 
+ {\rm two~permutations}\biggr]^{-1},\label{eq:fnl}
\end{eqnarray}
where $\ps(k)$ denotes the scalar power spectrum [cf. Eq.~\eqref{eq:sps}].

When departures from slow roll inflation occur, as in the case of the power 
spectrum, in general, it proves to be difficult to evaluate the scalar
bispectrum analytically.
Hence, one has to construct methods to compute the scalar bispectrum 
numerically~\cite{Chen:2008wn,Hazra:2012yn,Ragavendra:2020old,Ragavendra:2020sop}.
As we pointed out, the quantities $\cG_C(\vk_1,\vk_2,\vk_3)$ are described by 
integrals which involve the background quantities as well as the mode functions 
$f_k$ and their time derivatives~$f_k'$ [cf. Eqs.~\eqref{eqs:cG}].
We have already discussed the numerical evaluation of the background quantities 
and the scalar mode functions~$f_k$.
It is now a matter of utilizing them and carrying out the integrals describing 
the quantities~$\cG_C(\vk_1,\vk_2,\vk_3)$.
In analytical calculations, to evaluate these integrals, one assumes that 
$\ei\to-\infty$ and $\ee\to 0^-$.
But, evidently, it is not possible to achieve these extreme limits in 
numerical computations.
Actually, it does not seem to be necessary either.
We had seen earlier that, to evaluate the scalar power spectrum, it is often
adequate to evolve the mode functions $f_k$ from $k\simeq 10^2\,\sqrt{z''/z}$ 
until $k\simeq 10^{-5}\,\sqrt{z''/z}$.
The reason being that, in most situations, the mode functions only oscillate 
in the sub-Hubble regime and their amplitudes quickly freeze once they leave 
the Hubble radius.
Interestingly, one finds that, since the amplitudes of the mode functions
freeze, the super-Hubble contributions to the bispectrum prove to be 
negligible~\cite{Hazra:2012yn}.
However, in contrast to the power spectrum wherein we needed to focus on a 
single wave number, the bispectrum depends on three wave numbers.
Therefore, we need to carry out the integrals from an early time when the 
smallest of the three wave numbers (in the range of our interest) satisfies  
the condition $k\simeq 10^2\,\sqrt{z''/z}$ until a late time when the 
largest of them satisfies the condition $k\simeq 10^{-5}\,\sqrt{z''/z}$.
Also, in order to choose the correct perturbative vacuum, one has to 
regulate the integrals $\cG_C(\vk_1,\vk_2,\vk_3)$ by imposing a cut-off 
in the sub-Hubble regime~\cite{Seery:2005wm}.
Often, in the integrals, one introduces a cut-off function that is democratic 
in wave number and is of the form $\mathrm{exp}\,\l[-\kappa\,(k_1+k_2+k_3)/
(3\,\sqrt{z''/z})\r]$, where~$\kappa$ is a suitably chosen, positive definite and 
small quantity (for a discussion in this regard, see Refs.~\cite{Chen:2008wn,
Hazra:2012yn,Ragavendra:2020old}).
Numerically, such a cut-off proves to be convenient as it aids in the 
efficient computation of the integrals.
In scenarios involving ultra slow roll inflation, we had pointed out that,
due to the non-trivial evolution of the scalar mode functions $f_k$ at late 
times, it is safer to evaluate the power spectra at the end of inflation.
For the same reason, we also evaluate the scalar bispectrum 
close to the end of inflation
in the models and scenarios of our interest 
here\footnote{A clarification is in order at this stage of 
the discussion.
Note that computing the scalar and tensor power spectra only require the
evaluation of the corresponding Fourier mode functions~$f_k$ and~$g_k$ at 
the end of inflation [cf. Eqs.~\eqref{eqs:stps}]. 
These can be calculated numerically without difficulty.
However, as we have seen, the calculation of the scalar bispectrum 
also involves carrying out integrals over quantities that describe 
the background, the mode functions~$f_k$ and their time derivatives
[cf. Eqs.~\eqref{eqs:cG}]. 
As we mentioned, in slow roll inflation, the super-Hubble contributions 
to the integrals can be shown to be negligible~\cite{Hazra:2012yn}.
But, when there arise departures from slow roll, particularly at late
times as in the ultra slow roll scenarios of our interest here, it 
becomes important to calculate the integrals until after the epoch of 
ultra slow roll and as close to the end of inflation as possible. 
In some models, computing the integrals right until the end of inflation 
(for a wide range of scales) becomes numerically taxing and it can also 
induce some numerical inaccuracies at large wave numbers. 
In such situations, we calculate the integrals until as close to the end 
of inflation as numerically feasible. 
We should hasten to add that, in these cases, we have checked that the late 
time contributions to the scalar bispectra are indeed insignificant.
It is for this reason we have said that we evaluate the power and bi-spectra
close to the end of inflation rather than at the end of inflation.}.
Lastly, we should mention that the non-trivial boundary 
term~$G_{7}(\vka,\vkb,\vkc)$ is easier to compute as it does not involve
any integral and depends only the background parameters, the mode function
$f_k$ and its time derivative $f_k'$, {\it evaluated at the end of inflation}.\/

\begin{figure}[!t]
\centering
\hskip -46.2pt
\includegraphics[width=9.50cm]{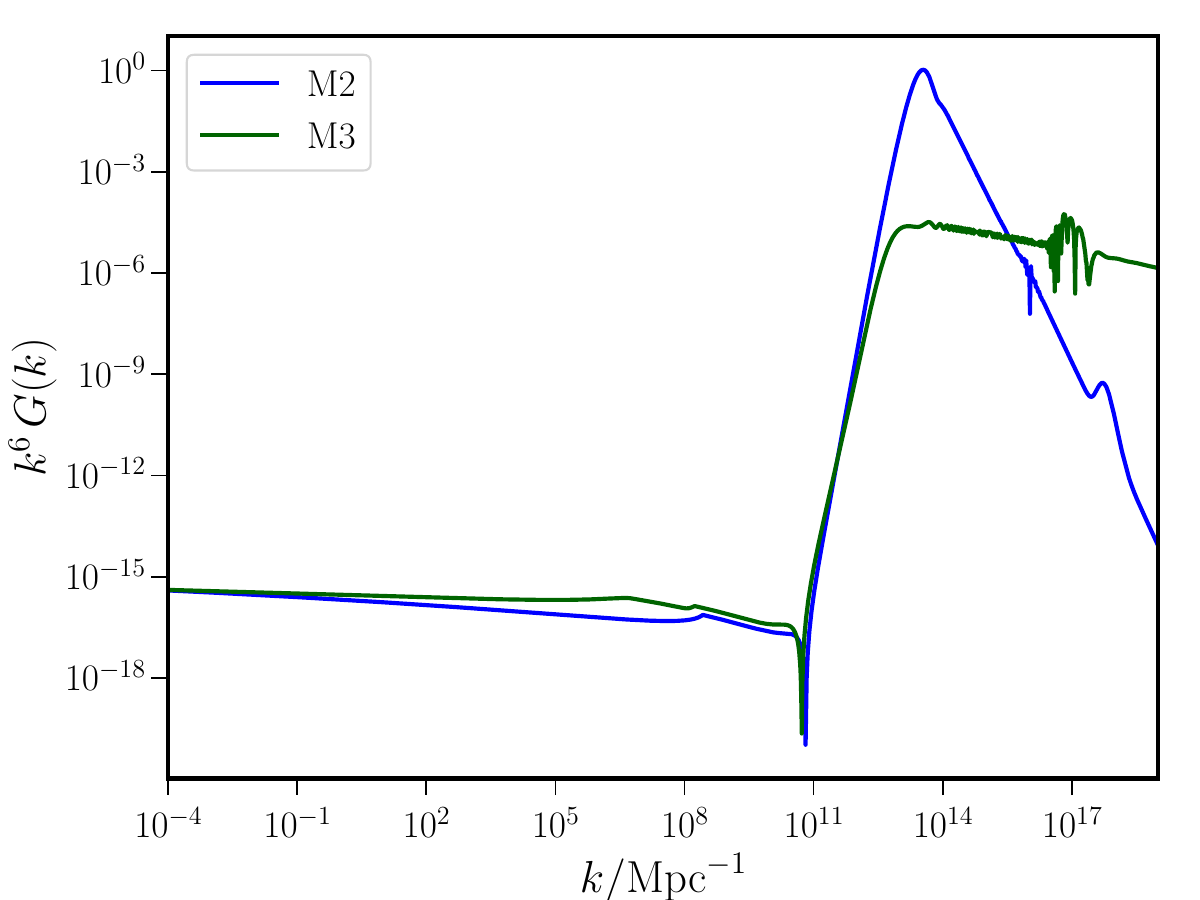}
\includegraphics[width=9.50cm]{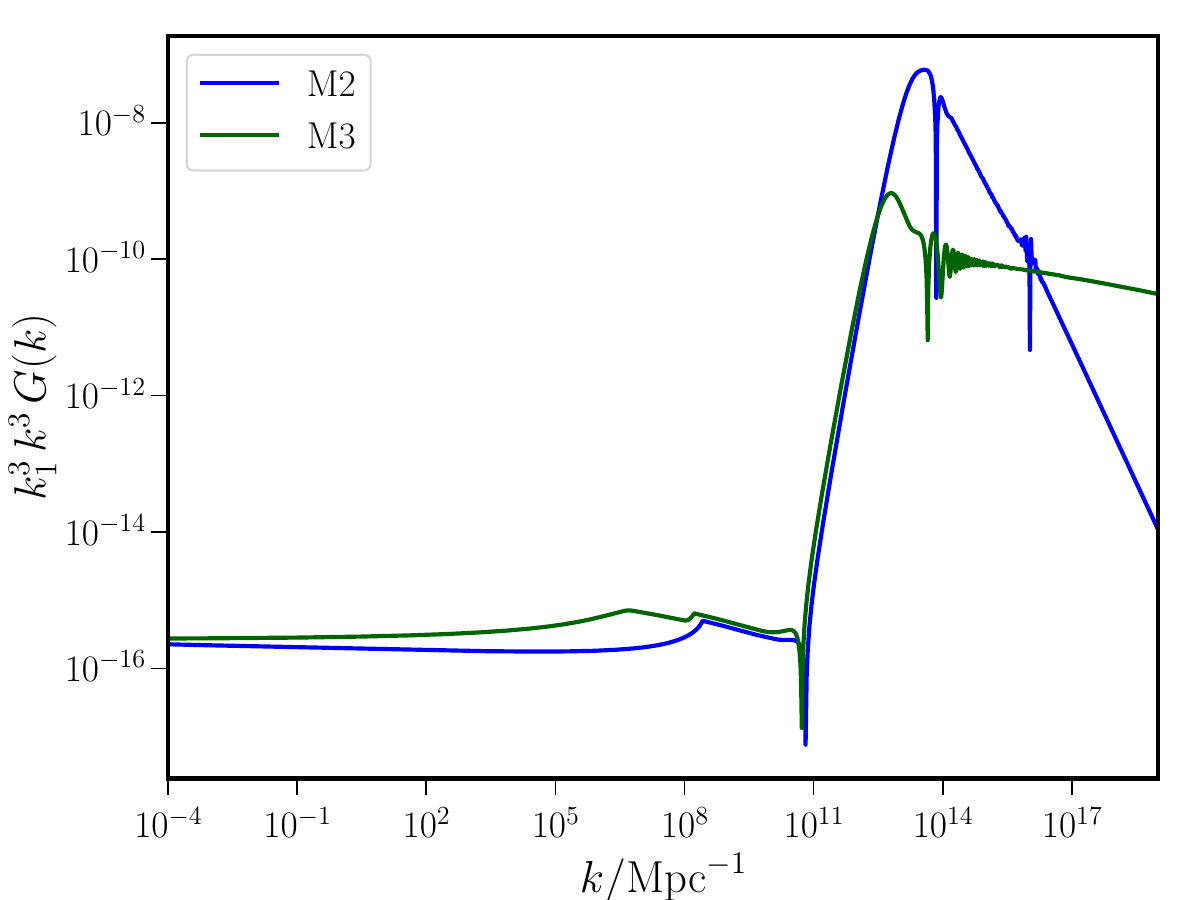}
\caption{The behavior of the bispectrum~$G(\vka,\vkb,\vkc)$ in the equilateral 
and squeezed limits have been plotted (on the left and right, respectively) for 
the two inflationary models M2 and M3 (in blue and green).
We have chosen these models since they exhibit a sharp or 
a broad peak in their power spectra. 
Upon comparison with Figure~\ref{fig:sps-tps}, it is evident that the bispectra 
closely mimic the shapes of the corresponding power spectra. 
Note that, at small scales, the amplitudes of the bispectra are considerably 
higher in the equilateral limit than in the squeezed limit.}\label{fig:G-eq-sq}
\end{figure}
In Figure~\ref{fig:G-eq-sq}, we have illustrated suitable dimensionless 
combinations of the wave numbers $(k_1,k_2,k_3)$ and the bispectrum 
$G(\vk_1,\vk_2,\vk_3)$ in the equilateral (i.e. when $k_1=k_2
=k_3=k)$ and the squeezed limits (i.e. when $k_1\to 0, k_2\simeq k_3 \simeq k)$
for the inflationary models M2 (which has a narrow peak in the scalar power 
spectrum and M3 (which has a rather broad peak).
Remarkably, the shape of the bispectra in these limits closely resemble 
the shape of the corresponding power spectra.
However, note that, on small scales, around the peak, the amplitude of the 
bispectra are considerably higher in the equilateral limit than in the 
squeezed limit.
Also, it is interesting to note that the dip and the peak in the bispectra 
occur at the same wave numbers as observed in the power spectra. 
Such a sharp dip actually implies the vanishing of the mode function $f_k$ 
for a specific wave number.
As a result, it can be expected that any higher order correlation function
involving the wave number also identically vanishes (for a detailed discussion
regarding the dip, see Refs.~\cite{Goswami:2010qu,Tasinato:2020vdk,
Ozsoy:2021qrg,Ozsoy:2021pws,Balaji:2022zur}).
We should add that the bispectra also contain additional features that are unique
to the nature of the terms in the cubic order action and the corresponding integrals 
involved in the computation.

\begin{figure}[!t]
\centering
\hskip -46.2pt
\includegraphics[width=9.50cm]{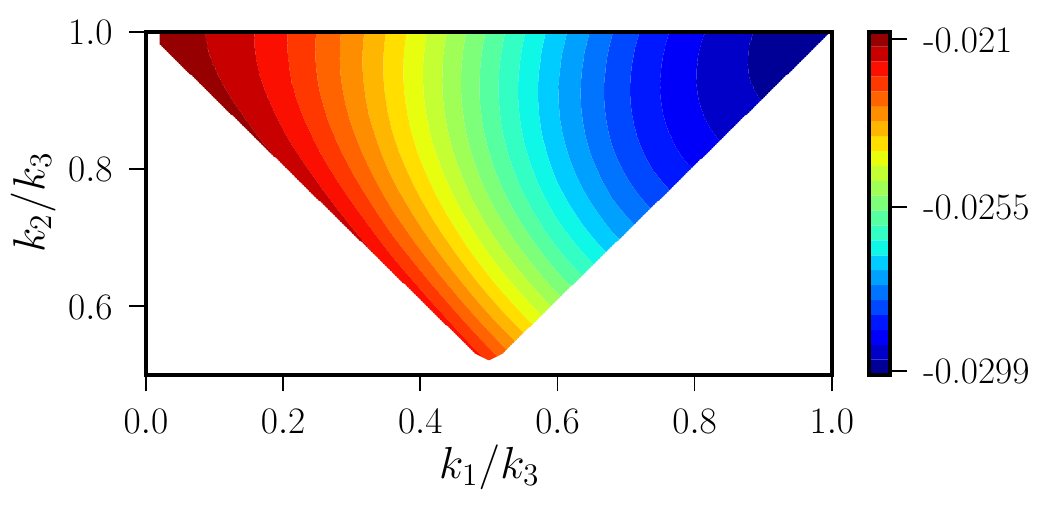}
\includegraphics[width=9.50cm]{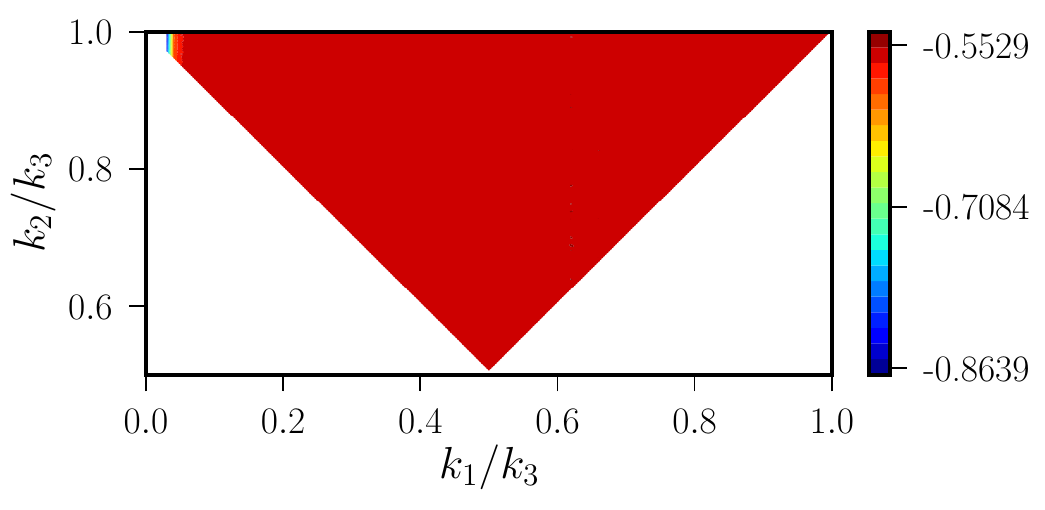}\\
\hskip -46.2pt
\includegraphics[width=9.50cm]{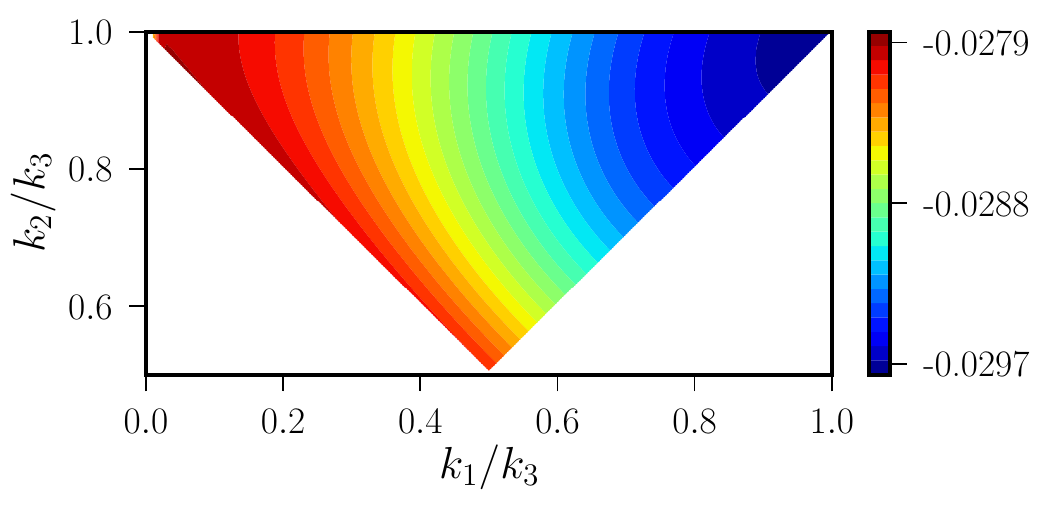}
\includegraphics[width=9.50cm]{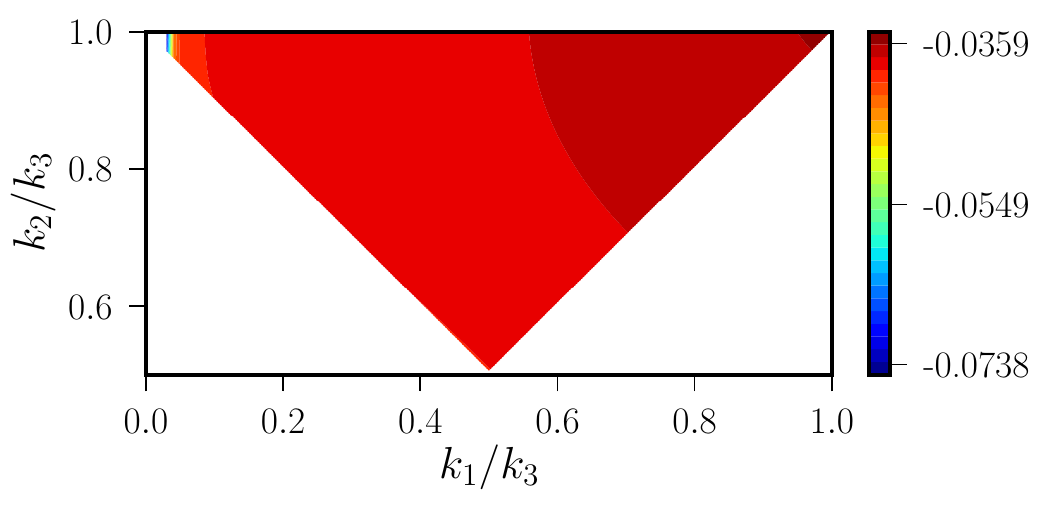}
\caption{The behavior of the scalar non-Gaussianity 
parameter~$\fnl(\vk_1,\vk_2,\vk_3)$ has been presented as a density plot for 
the inflationary models~M2 (in the top row) and~M3 (in the bottom row). 
In arriving at these figures, we have chosen the value of $k_3$ to be $k_\ast$ 
(i.e. the pivot scale, on the left) or $k_{\mathrm{peak}}$ (viz. the location 
of the peak in the scalar power spectrum, on the right). 
It should be clear from the density plots that the non-Gaussianity parameter
is equilateral in shape near~$k_\ast$ (i.e. over the CMB scales, indicating 
the slow roll evolution of the background during the early stages of inflation), 
whereas it is highly local in shape around $k_{\mathrm{peak}}$ (i.e. on small 
scales, reflecting the ultra slow roll behavior of the background during the
later stages of inflation).}\label{fig:fnl-2d}
\end{figure}
In Figure~\ref{fig:fnl-2d}, we have presented the density plots of the scalar
non-Gaussianity parameter $\fnl(\vk_2,\vk_2,\vk_3)$ for the two inflationary
models~M2 and~M3.
We have illustrated the density of the non-Gaussianity parameter around two 
wave numbers chosen in two distinct regimes---$k_\ast$ (i.e. the pivot
scale) over large scales and $k_{\mathrm{peak}}$ (i.e. the wave number 
corresponding to the peak in the scalar power spectra) over small scales.
We find that, for M2, $k_{\mathrm{peak}}=3.5\times 10^{13}\,{\rm Mpc}^{-1}$
and, for M3, $k_{\mathrm{peak}}=1.8\times 10^{13}\,{\rm Mpc}^{-1}$. 
Clearly, $\fnl$ is equilateral in shape around $k_\ast$ with values of order 
$\mathcal{O}(10^{-2})$ as expected from perturbations evolving over a slow 
roll regime. 
But, the shape of $\fnl$ turns out to be local around $k_{\mathrm{peak}}$ and
the amplitude becomes model dependent, with values close to $0.56$ for~M2 and 
$0.05$ in case of~M3.
An analytical understanding of the amplitude and shape of the scalar bispectrum 
in models permitting a brief epoch of ultra slow roll inflation is still elusive
and requires attention~\cite{Ragavendra:2021usr}.
Moreover, the behavior of bispectrum in the reconstructed scenario and the
effect of the parameters such as $N_1$ and $\Delta N_1$ on the associated 
$\fnl$ is an interesting topic of exploration. 
We are currently investigating these issues.


\section{Outlook}\label{sec:outlook}

In this review, we have considered ultra slow roll inflation driven by 
a single, canonical scalar field which leads to enhanced scalar power 
on small scales.
We have examined the corresponding effects on the extent of formation of 
PBHs and the production of secondary GWs during the radiation dominated epoch.
We have also computed the shape and the strengths of non-Gaussianities
generated on small scales in such situations.
We should mention that the numerical codes used to arrive at the results 
presented in this review are available at the 
following URL: \href{https://gitlab.com/ragavendrahv/pbs-pbh-sgw.git}{\tt 
https://gitlab.com/ragavendrahv/pbs-pbh-sgw.git}.
The package computes the scalar and tensor power spectra as well as the scalar 
bispectrum for a given canonical, single field model of inflation.
Further, it can compute the corresponding~$\fpbh(M)$ and $\ogw(f)$ arising 
from such spectra, as discussed in this review\footnote{Users making use of 
the code in part or whole can cite this manuscript in their publications.}

There are many related scenarios and effects that we could not include
in this review. While a few of these effects have been investigated already, 
some of them require further study. In this concluding section, we shall 
describe them briefly.

$\bullet$~{\bf Effects of non-Gaussianities on the formation of PBHs:}~In
our discussion, we have restricted our attention to the effects of the 
increased scalar power (due to the epoch of ultra slow roll) on the number
of PBHs produced.
Since the amplitude of the bispectrum generated due to ultra slow roll is 
significantly higher than the slow roll values, the non-Gaussianities can 
be expected to boost the extent of PBHs formed (for earlier discussions on 
this point, see, for instance, Refs.~\cite{Atal:2018neu,Passaglia:2018ixg,
Atal:2019cdz,Yoo:2019pma}).
There has been recent efforts to account for a skewness in the probability 
distribution describing the density contrast [cf. Eq.~\eqref{eq:P-d}],
arising due to increased strengths of the scalar bispectrum on small scales, 
and calculate the corresponding effects on the number of PBHs 
produced~\cite{Taoso:2021uvl,Riccardi:2021rlf,Matsubara:2022nbr}.
We should point out that alternative methods have also been proposed to 
account for the scalar non-Gaussianity in such calculations (see 
Refs.~\cite{Cai:2022erk,Gow:2022jfb}; for a brief summary of the different 
methods, see Refs.~\cite{Ferrante:2022mui}).

$\bullet$~{\bf Effects of non-Gaussianities on secondary GWs:}~It has been
argued that large amplitudes of~$\fnl$, as arising in ultra slow roll models,
can considerably influence the strengths of secondary GWs that are
generated during the radiation dominated epoch~\cite{Unal:2018yaa,
Pi:2020otn,Adshead:2021hnm}. 
However, rather than calculate the bispectrum arising in specific inflationary 
models, these attempts often assume certain well motivated amplitudes and shapes 
of $\fnl$ to calculate the corresponding contributions to~$\ogw$. 
There have also been efforts to compute such non-Gaussian contributions 
to~$\ogw$, while accounting for complete scale dependence of~$\fnl$, arising 
from the bispectrum in specific models of ultra slow roll (in this regard, see 
Ref.~\cite{Ragavendra:2021qdu}). 
These computations suggest that the non-Gaussian contributions to $\ogw$ are
highly model dependent and can, in principle, alter the shape and amplitude of 
$\ogw$ around the peak of the spectra.

$\bullet$~{\bf Loop corrections to the primordial power spectrum:}~There is 
a gathering interest in the literature towards computing the contributions 
due to the loops to the scalar and tensor power spectra generated during
inflation (for related early efforts, see, for example 
Refs.~\cite{Seery:2007wf,Yokoyama:2008by,Cogollo:2008bi,Rodriguez:2008hy}). 
These contributions capture the effects of the higher order correlations on 
the power spectra and can lead to characteristic signatures on the predicted
observables.
There have been attempts to investigate such effects on observables such 
as $\ogw$ and the 21-cm signals from neutral hydrogen of the Dark 
Ages~\cite{Yamauchi:2022fri,Chen:2022dah,Ota:2022xni}.
There have also been efforts to theoretically restrict models of ultra slow 
roll inflation based on the amplitude of the corrections due to the loops and
the associated consequences for the validity of perturbative treatment of the
correlations (in this regard, see Refs.~\cite{Cheng:2021lif,Kristiano:2021urj,
Kristiano:2022zpn,Kristiano:2022maq}).

We are presently investigating different issues in these directions.


\section*{Acknowledgement}
{HVR acknowledges support from the Indian Institute of Science Education and 
Research Kolkata through postdoctoral fellowship.
LS wishes to acknowledge support from the Science and Engineering Research Board, 
Department of Science and Technology, Government of India, through the Core Research 
Grant CRG/2018/002200.}

\appendix

\section{Determining the locations of the point of inflection}\label{app:pi}

\begin{figure}[!t]
\centering
\vskip -1.8cm
\hskip -46.2pt
\includegraphics[width=9.00cm]{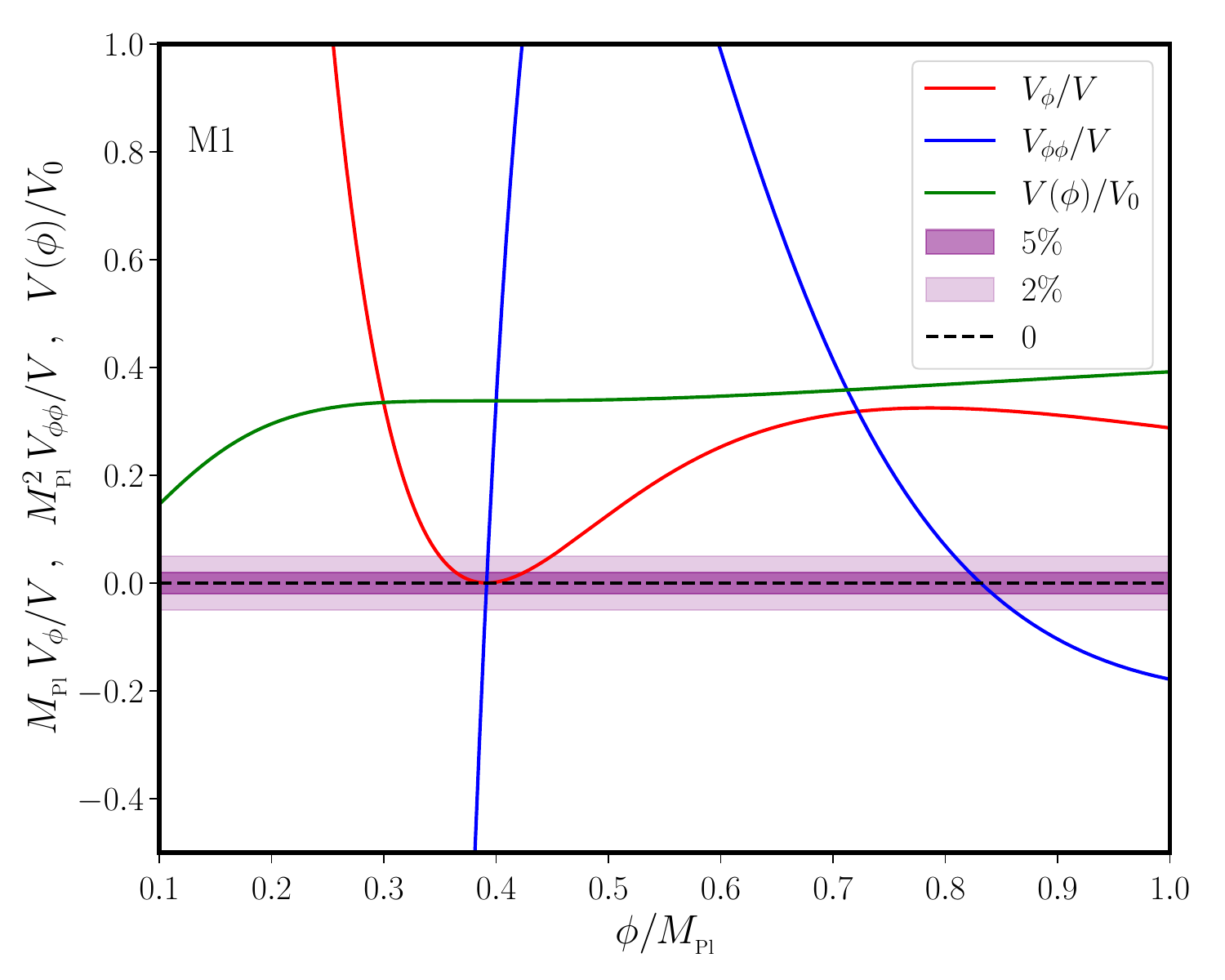}
\includegraphics[width=9.00cm]{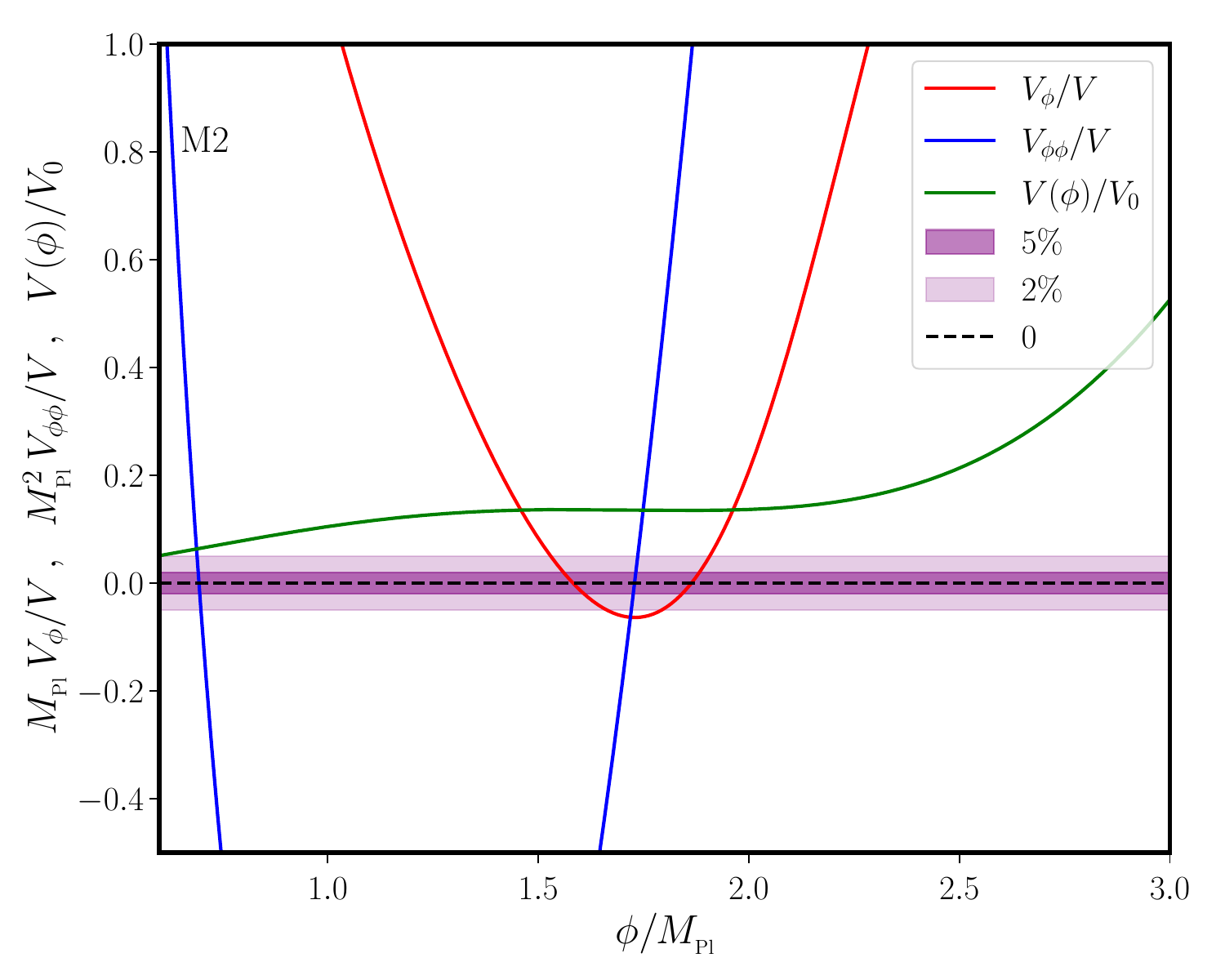}\\
\hskip -46.2pt
\includegraphics[width=9.00cm]{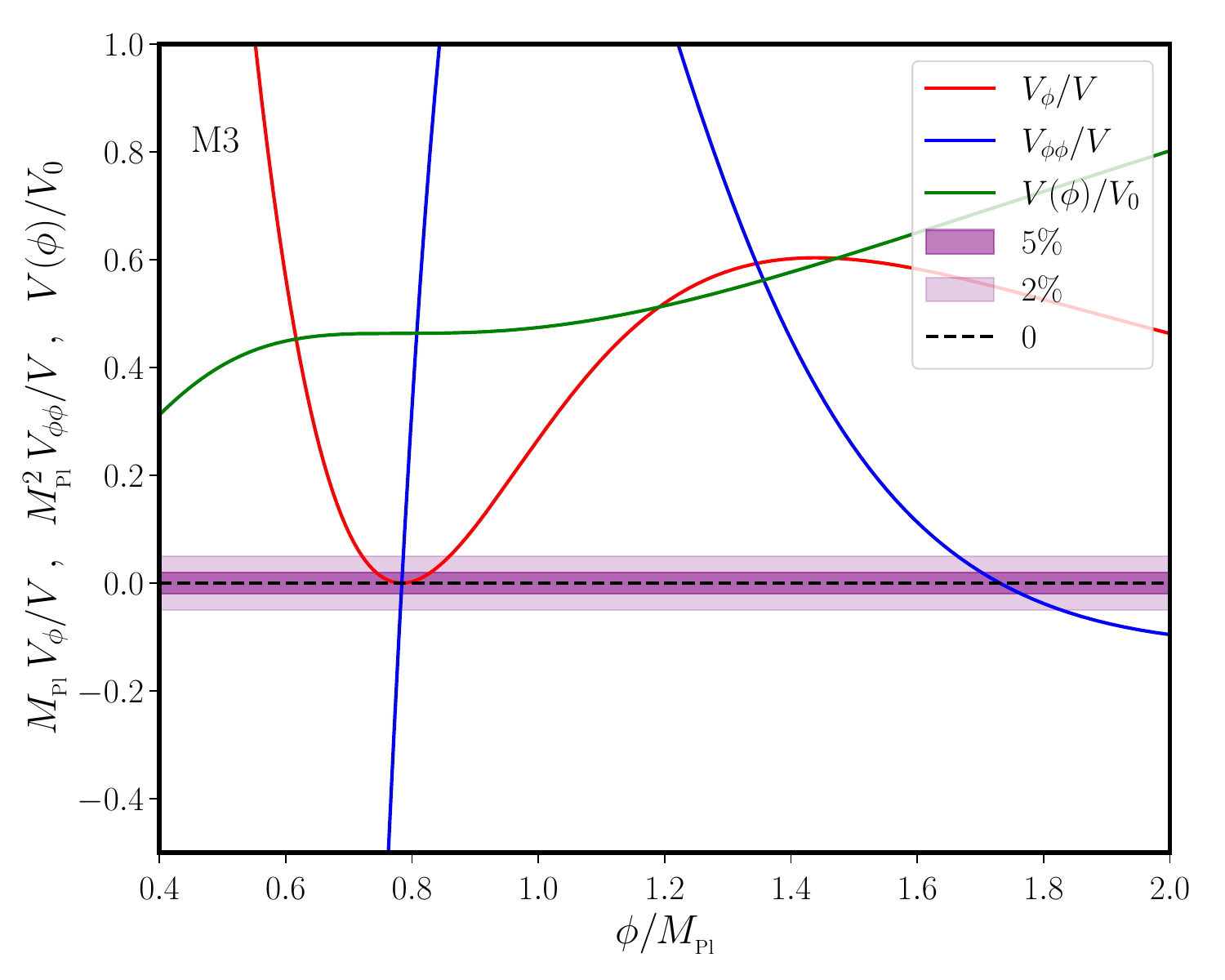}
\includegraphics[width=9.00cm]{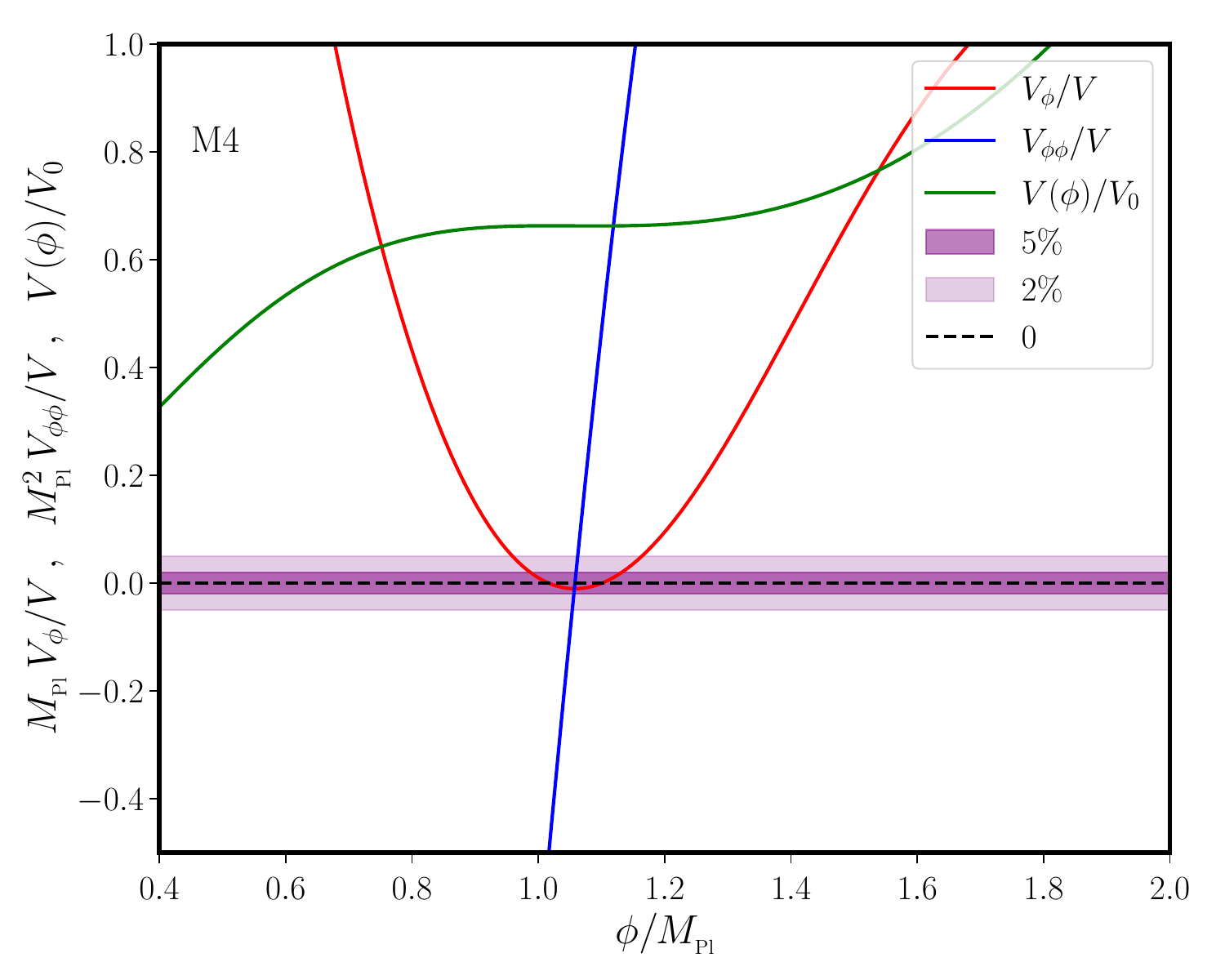}\\
\hskip -46.2pt
\includegraphics[width=9.00cm]{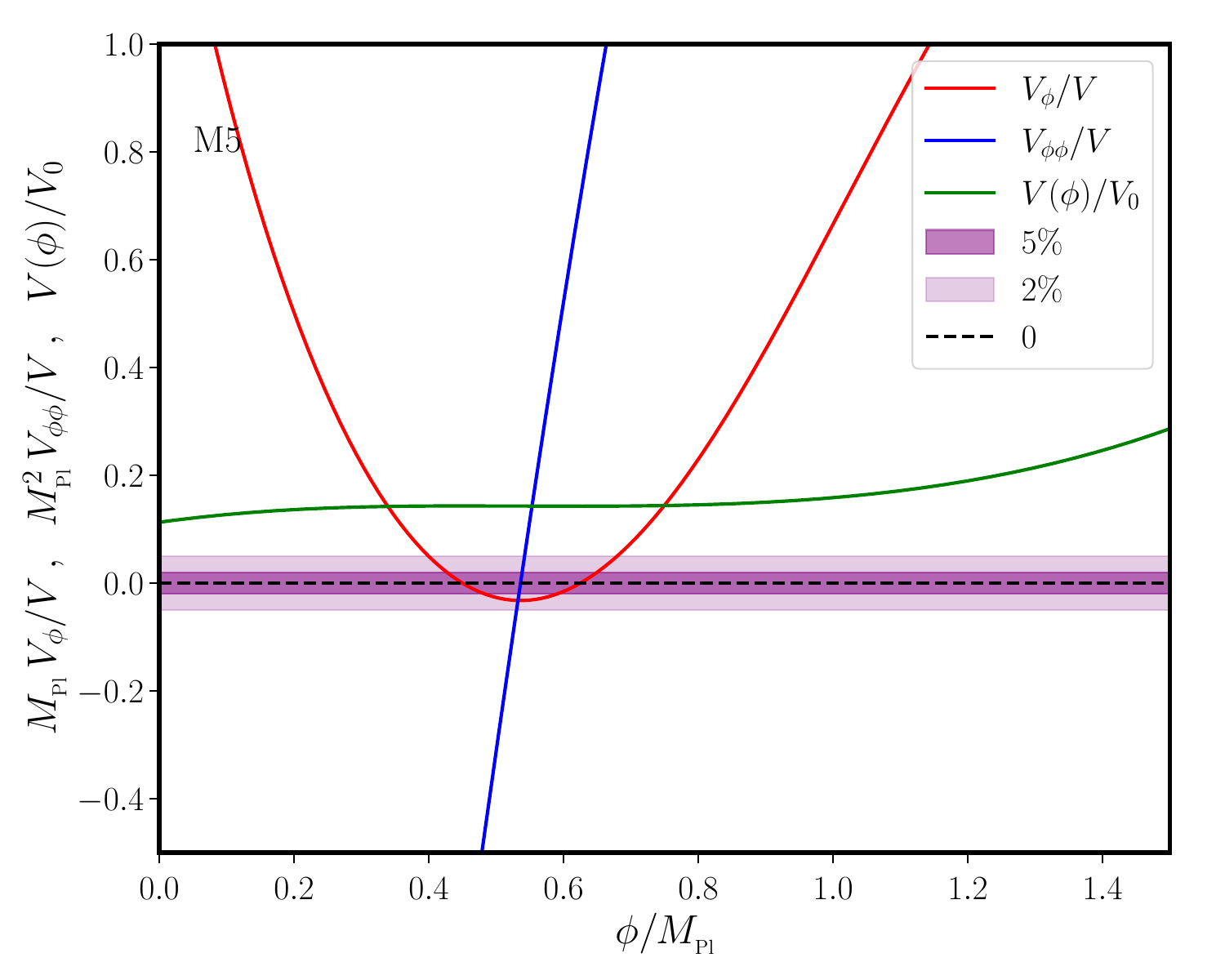}
\includegraphics[width=9.00cm]{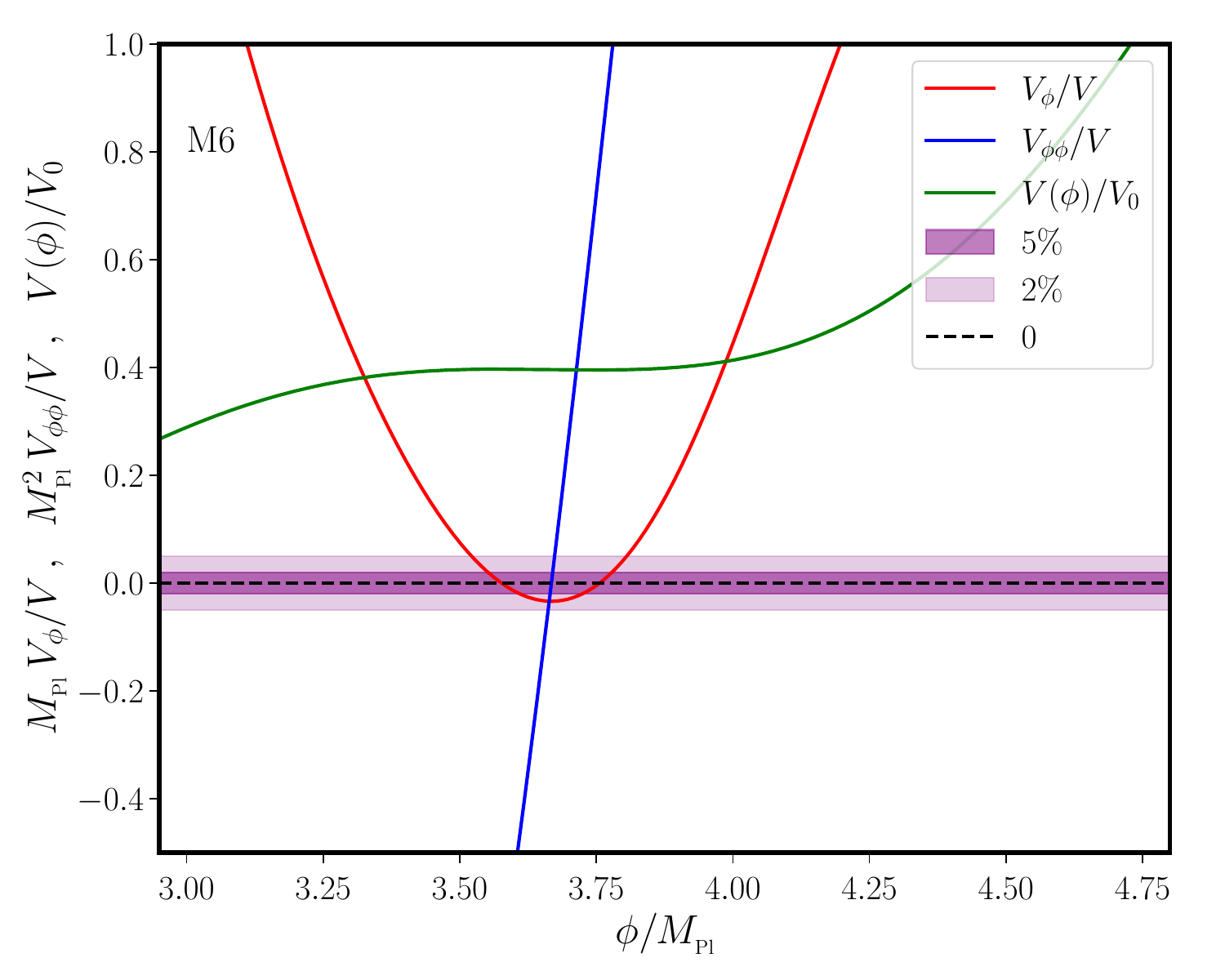}
\caption{We have illustrated the manner in which we have numerically determined 
the points of inflection in the inflationary models of interest.
We have plotted the quantities $V_\phi/V$ and $V_{\phi\phi}/V$ as well as a 
suitably rescaled potential $V$ in the models M1 to M6 (from the top left to 
the bottom right corner).
In the figures, we have included bands of $2\%$ and $5\%$ around zero.
The location of the point of inflection is determined by the condition that 
both $V_\phi/V$ and $V_{\phi\phi}/V$ lie within~$5\%$ of zero.
While, in some of the models, these quantities lie even within~$2\%$ of zero, 
we find that, in the model M2, the quantities deviate by as much as~$7\%$.}
\label{fig:ip}
\end{figure}

In our discussion in Section~\ref{subsec:im}, we had mentioned the locations 
of the point of inflection (viz. the value of~$\phi_0$) in the inflationary 
models M1 to M6.
We should clarify that some of these are actually {\it near}\/ inflection 
points, where the first and the second derivatives of the potential $V_\phi$ 
and $V_{\phi\phi}$ {\it almost}\/ vanish.
In Figure~\ref{fig:ip}, we have illustrated the method by which we have identified 
the points of inflection.
In the figure, we have plotted the behavior of the inflationary potential $V$ 
as well as the quantities $V_\phi/V$ and $V_{\phi\phi}/V$ in the six models of
interest.
Note that, since the first two derivatives of the potentials almost vanish at 
these (near) inflection points, the potentials have a plateau around the point.
As the field enters the flat region of the potentials, it considerably slows down 
resulting in an epoch of ultra slow roll inflation in these models.

\begin{figure}[!t]
\centering
\vskip -45pt
\hskip -46.2pt
\includegraphics[width=9.0cm]{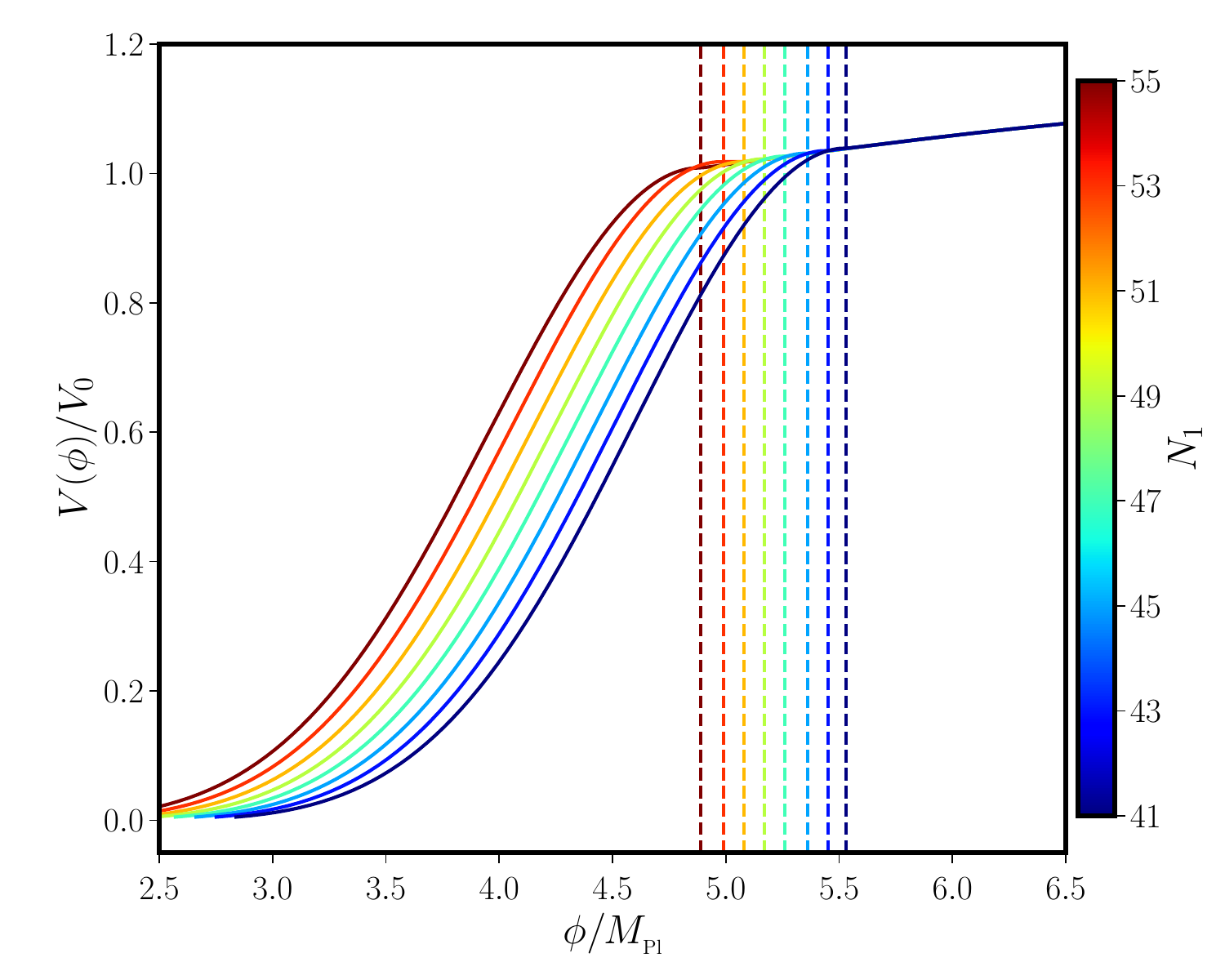}
\includegraphics[width=9.0cm]{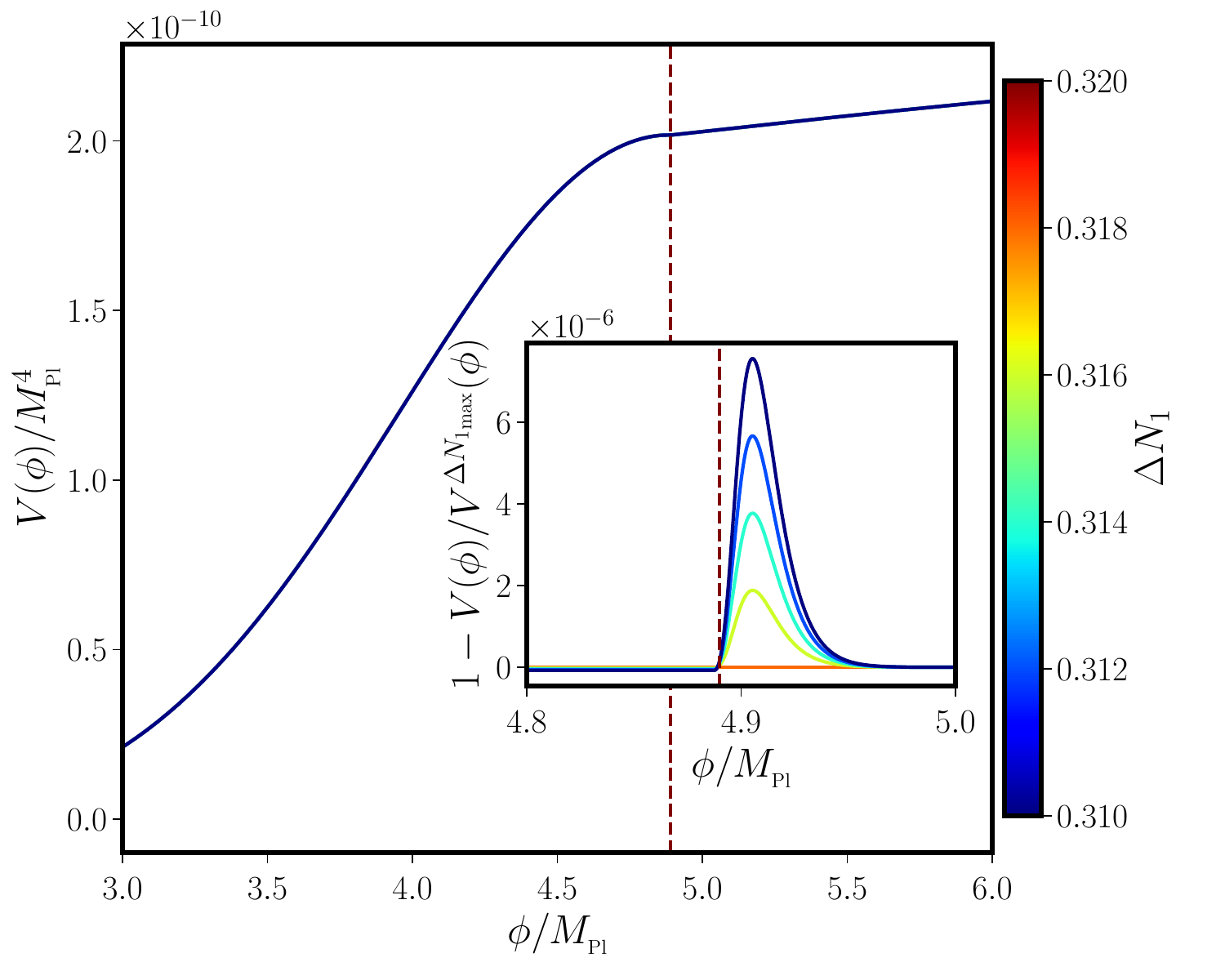}
\vskip -10pt
\caption{The shape of the potential obtained from reconstruction is presented 
for a range of values of $N_1$ (on the left) and $\Delta N_1$ (on the right). 
Understandably, the change in $N_1$ moves the location of the point of inflection
(marked with dashed lines of respective colors). 
The variation in $\Delta N_1$ leads to imperceptibly small change in the shape 
of the potential around the point of inflection. 
This can be better observed in the relative difference between a given $V(\phi)$ 
and the one corresponding to the maximum value of $\Delta N_1$, as plotted in the 
inset.}\label{fig:rcp}
\end{figure}

In Figure~\ref{fig:rcp}, we have plotted the potentials that correspond to 
the scenario described by the first slow parameter $\epsilon_1(N)$ in 
Eq.~\eqref{eq:rs1}.
As described in Section~\ref{subsec:rcp}, from the form of $\epsilon_1(N)$, 
we have computed $\phi(N)$ and $V(N)$ using Eqs.~\eqref{eqs:phiHN} 
and~\eqref{eq:VN}, and plotted the potential $V(\phi)$ parametrically. 
We have illustrated the numerically calculated potentials for the range of 
$N_1$ and $\Delta N_1$ discussed earlier.

\begin{figure}[H]
\centering
\vskip -20pt
\hskip -46.2pt
\includegraphics[width=9.5cm]{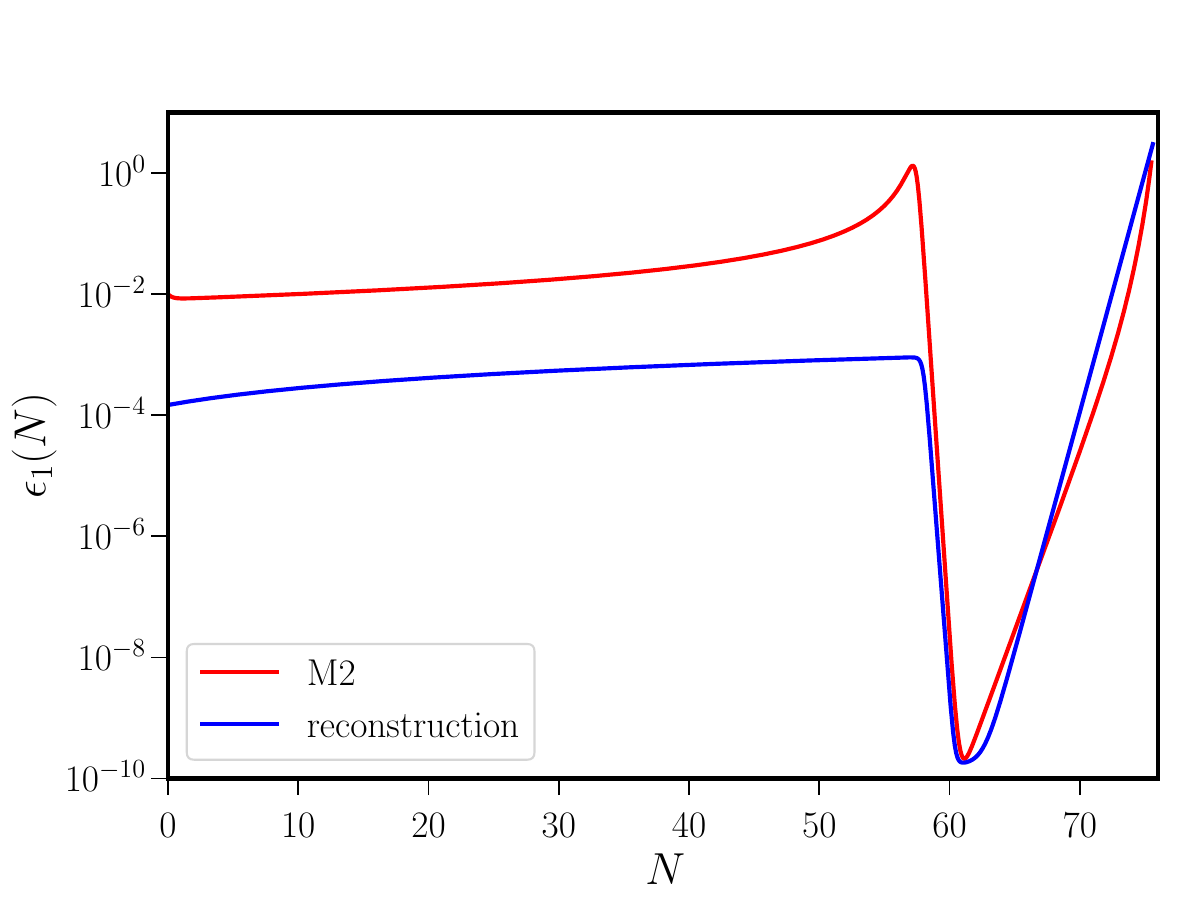}
\includegraphics[width=9.5cm]{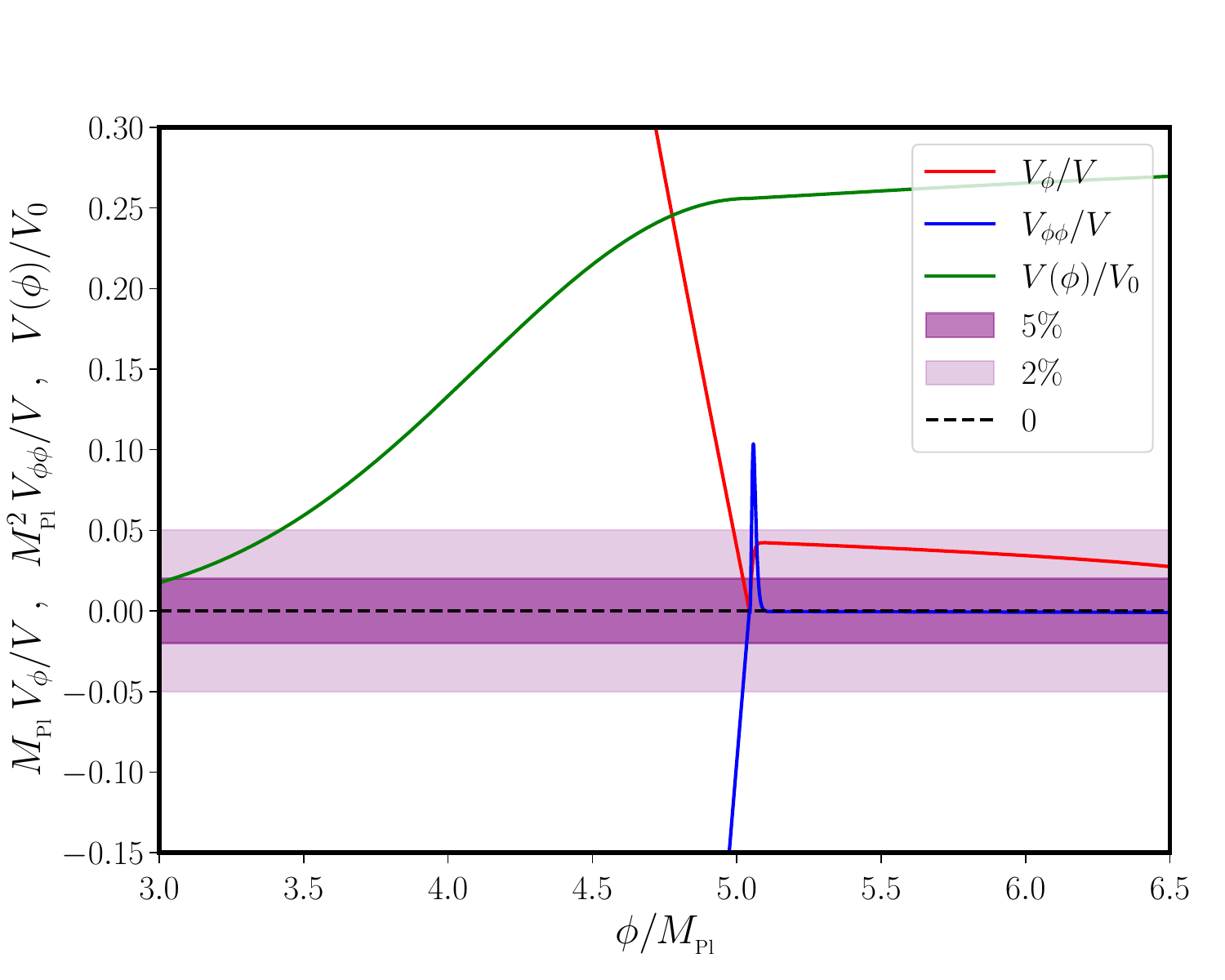}
\vskip -10pt
\caption{We have presented the behavior of slow roll parameter 
$\epsilon_1(N)$ as described by Eq.~\eqref{eq:rs1} (on the 
left) and the corresponding potential arrived at numerically, 
along with its first and second derivatives (on the right), 
for a representative set of parameters. 
The values of the parameters have been chosen to arrive at these behavior 
are $N_1=58$, $N_2 = 75$, $\Delta N_1 = 0.31$ and $\Delta N_2 = 0.55$.
The other related parameters are set to the values mentioned in the main text.
We have chosen these parameters to illustrate the manner in which the reconstructed 
$\epsilon_1$ closely mimics the behavior arising in a specific model described 
by a potential, say, M2 (plotted on the left). 
It is clear that the reconstructed $\epsilon_1(N)$ contains all the relevant 
features of an inflationary model that permits a brief epoch of ultra slow roll.
Further, we should note that the value of $\epsilon_1$ in the reconstructed
scenario over the initial phase of slow roll is much smaller value than that 
in~M2.
This leads to a tensor to scalar ratio which is within the observational bound, 
unlike M2 [cf.~Table~\ref{tab:ns-r}].
Note that, in the case of the reconstructed scenario, there exists a point where 
the first two derivatives of the potential vanish (as illustrated in the figure on
the right), implying a point of inflection. 
This plot has to be compared with the second plot of Figure~\ref{fig:ip} 
that illustrates the behavior of the corresponding quantities in model M2, 
which only contains a near-inflection point.}
\label{fig:eps-recon-comp}
\end{figure}

Further, in Figure~\ref{fig:eps-recon-comp}, we have presented the behavior of 
quantity~$\epsilon_1(N)$ in the reconstructed scenario [cf. Eq.~\eqref{eq:rs1}] 
for a specific set of parameters that have been chosen to mimic the behavior of
the first slow roll parameter in the model M2. 
This is to illustrate the manner in which our reconstruction captures the 
essential features of a typical model that permits a phase of ultra slow roll,
while at the same time providing better handle on the associated dynamics. 
We achieve the onset and duration of the phase of ultra slow roll at $e$-folds
similar to that of M2, and also terminate inflation around the same time as 
in~M2. 
Moreover, in the reconstructed scenario, we are able to work with a lower value 
of $\epsilon_1$ during the initial stage of slow roll, which ensures a viable
tensor-to-scalar ratio over the CMB scales, unlike the case of M2 
[cf.~Table~\ref{tab:ns-r}].
In the figure, we have also illustrated the behavior of the potential (as
well as its first and second derivatives) associated with the reconstructed
scenario, in the same manner as we had presented for the models M1--M6 in
Figure~\ref{fig:ip}. 
We have plotted them parametrically against the field, and have focused on
the behavior around the point of inflection. 
We notice that the point of inflection occurs quite rapidly, unlike the 
models M1--M6 which show a smoother behavior. 
The vanishing of the derivatives are rather sharp and highly coincident, in
contrast to the model M2 which only contains a near-inflection point.

\bibliographystyle{unsrt}
\bibliography{references}
\end{document}